\documentclass[fleqn,usenatbib]{mnras}

\usepackage{newtxtext,newtxmath}
\usepackage{float}
\usepackage{graphicx}
\usepackage{amsmath}
\usepackage{color}
\usepackage[table]{xcolor}

\title[Dynamics of spinning SMBH binaries]{Dynamics and spin alignment in massive, gravito-turbulent circumbinary discs around supermassive black hole binaries}

\author[Bourne et al.]{Martin A. Bourne$^{1, 2, 3}$\thanks{E-mail: m.bourne@herts.ac.uk}, Davide Fiacconi$^{1, 2}$, Debora Sijacki$^{1, 2}$, Joanna M. Piotrowska$^{2,4}$\newauthor and Sophie Koudmani$^{1,2,5,6}$\\
$^{1}$ Institute of Astronomy, University of Cambridge, Madingley Road, Cambridge, CB3 0HA, UK\\ 
$^{2}$ Kavli Institute for Cosmology, University of Cambridge, Madingley Road, Cambridge, CB3 0HA, UK\\
$^{3}$ Centre for Astrophysics Research, Department of Physics, Astronomy and Mathematics, University of Hertfordshire, College Lane, Hatfield AL10 9AB, UK \\
$^{4}$ Cahill Center for Astronomy and Astrophysics, California Institute of Technology, Pasadena, CA, USA\\
$^{5}$ Center for Computational Astrophysics, Flatiron Institute, 162 5$^{th}$ Avenue, New York, NY 10010, USA\\
$^{6}$ St Catharine’s College, University of Cambridge, Trumpington Street, Cambridge CB2 1RL, UK\\
}

\begin{document}

\date{\today}

\pagerange{\pageref{firstpage}--\pageref{lastpage}} \pubyear{2023}

\maketitle

\label{firstpage}

\begin{abstract}
Parsec-scale separation supermassive black hole binaries in the centre of gas-rich galaxy merger remnants could be surrounded by massive circumbinary discs (CBDs). Black hole mass and spin evolution during the gas-rich binary inspiral are crucial in determining the direction and power of relativistic jets that radio observations with LOFAR and SKAO will probe, and for predicting gravitational wave (GW) emission that IPTA and LISA will measure. We present 3D hydrodynamic simulations capturing gas-rich, self-gravitating CBDs around a $2\times 10^6$~M$_{\odot}$ supermassive black hole binary, that probe different mass ratios, eccentricities and inclinations. We employ a sub-grid Shakura-Sunyaev accretion disc to self-consistently model black hole mass and spin evolution together with super-Lagrangian refinement techniques to resolve gas flows, streams and mini-discs within the cavity, which play a fundamental role in torquing and feeding the binary. We find that higher mass ratio and eccentric binaries result in larger cavities, while retrograde binaries result in smaller cavities. All of the simulated binaries are expected to shrink with net gravitational torques being negative. Unlike previous simulations, we do not find preferential accretion onto the secondary black hole. This implies smaller chirp masses at coalescence and hence a weaker GW background. Critically this means that spin-alignment is faster than the binary inspiral timescale even for low mass ratios. When considering initially misaligned systems, the orientation of the mini-discs around each black hole can vary significantly. We discuss the implications of this behaviour for black hole spin alignment and highlight the need for broader parameter space studies of misaligned systems to understand the impact on black hole recoil velocities.
\end{abstract}


\section{Introduction}
\label{sec:intro}

Within the standard $\Lambda$-cold dark matter ($\Lambda$CDM) cosmological model of our Universe \citep{WhiteRees78, Peebles82, BlumethalEtAl84, DavisEtAl85}, cosmic structure formation proceeds hierarchically in a bottom-up fashion. Small mass structures form first and then grow in mass via ``smooth'' inflows and mergers to form larger mass objects as cosmic time progresses \citep{NavarroEtAl95, MoWhite02, FakhouriMa08, FakhouriEtAl10, GuoWhite08, GenelEtAl09, O'LearyEtAl21}. Mergers of cosmic structures have been studied extensively \citep{ToomreToomre72, Barnes92, Boylan-KolchinEtAl05, CoxEtAl06, NaabTrujillo06, Conselice14, FiacconiEtAl15, TalbotEtAl23}, and are a crucial mechanism in galaxy evolution. Not only does this lead to the mass assembly of dark matter halos and galaxies within but, in some cases significant amounts of gas can be funnelled to the centre of the remnant galaxy that can drive both star bursts \citep{BarnesHernquist91, BarnesHernquist96, MihosHernquist96, DiMatteoEtAl05, HopkinsEtAl06, CoxEtAl08, HopkinsEtAl13}, and also potentially black hole growth and activity \citep{DiMatteoEtAl05, HopkinsEtAl06, HopkinsEtAl07, JohanssonEtAl09, BaraiEtAl14, GaoEtAl20, PierceEtAl23}.
  
While the exact nature of gas delivery to supermassive black holes in the aftermath of a galaxy merger is not yet fully understood, galaxies hosting supermassive black holes \citep[see e.g.,][]{FerrareseMerritt00, HaringRix04, GultekinEtAl09, KormendyHo13, McConnellMa13, SahuEtAl19} are expected to undergo multiple major mergers with similarly sized galaxies over their cosmic lifetime \citep[see e.g.,][and references therein]{O'LearyEtAl21, HuskoEtAl22Mergers}. Hence it is natural to expect that during the merging process, due to the dynamical friction, supermassive black holes present in parent galaxies will sink towards the centre of the merger remnant system and form a $\sim$~parsec scale supermassive black hole binary \citep{BegelmanEtAl80, MayerEtAl07, Mayer13, KhanEtAl16, KhanEtAl18, Colpi14}. Depending on the baryon content of the local environment, the hardening processes of such a supermassive black hole binary prior to gravitational wave (GW) emission may proceed through several different channels, such as three-body encounters with stars \citep{Quinlan96, MilosavljeviMerritt01, MilosavljevicMerritt03, BerczikEtAl06, SesanaEtAl06, RantalaEtAl17} or a third black hole \citep{HoffmanLoeb07, BonettiEtAl18, MannerkoskiEtAl21}, as well as interactions with a gaseous circumbinary disc \citep[CBD, e.g.,][]{CuadraEtAl2009, NixonEtAl11Retrograde, NixonEtAl13, RoedigEtAl2012, D'OrazioEtAl13, FarrisEtAl14, MoodyEtAl19, MunozEtAl19, MunozEtAl20, D'OrazioDuffel21, FranchiniEtAl2021, FranchiniEtAl22}. If both supermassive black holes are accreting then they should appear as dual AGN, with a number of such objects observed on $\gtrsim$~kpc scales \citep[see Fig.~1 in][]{ChenEtAl22}. On scales of $\lesssim 1-10$~pc only a handful of binary candidates have been observed, for example using VLBA \citep{RodriguezEtAl06, BansalEtAl17, KharbEtAl17}, while on even smaller sub-parsec scales candidate binaries have been inferred, for example, from light curves \citep{SillanpaaEtAl88, GrahamEtAl2015, ChenEtAl20, MillonEtal22}. We refer to the review of \citet{DeRosaEtAl2019} for further discussion of dual AGN in the multimessenger era. Notwithstanding the difficulties of these observational efforts and possible systematic biases, the low number of observed binaries tantalisingly suggests that the hardening process may be efficient with the majority of binaries merging into a larger supermassive black hole within the age of the Universe.

The Advanced LiGO and VIRGO collaboration opened the GW window onto our Universe when they made the first detection of merging stellar mass black holes \citep{LigoGW2016}, and has been followed by the observation of an intermediate-mass black hole \citep{LIGO20IMBH}. More recently, the first tentative detections of a nHz GW background, expected to be produced by a cosmic population of merging supermassive black holes, were made by members of the International Pulsar Timing Array \citep[IPTA][]{VerbiestEtAl16, IPTA23}; NANOGrav \citep{NANOGrav23GWB}, the Parkes PTA \citep{ReardonEtAl23} and joint results from the European and Indian PTAs \citep{EPTA23}. An upper limit on the strain amplitude of individual supermassive black hole binaries was also presented \citep{NANOGrav23Individual}. Future space-based missions, such as LISA, will provide a fantastic opportunity to observe $\sim 0.01-100$~mHz frequency GWs emitted by individual supermassive black hole merger events over a wide range of masses (peaking between $\sim 10^4-10^7$~M$_{\odot}$) and out to very high redshifts \citep{Lisa17, LISA23}. It is hence of great importance to provide detailed theoretical predictions on the merger rate of supermassive black holes, binary parameters, and also on the likely black hole spin magnitudes and orientation before merger. The concerted effort of theoretical predictions and future observations will be a unique way not only to probe spin evolution and hence black hole radiative efficiencies but also the amount of gravitational recoil merger remnants will experience \citep{CampanelliEtAl07, GonzalezEtAl2007, LoustoEtAl2011, LoustoEtAl13, LoustoEtAl19, SperhakeEtAl20}. This will shed light on the occupation fraction of supermassive black holes in galaxies, which impacts predicted GW detection rates \citep[e.g.][]{Sesana07, SesanaEtAl09Recoils, BlechaLoeb08}, the likelihood of offset holes and the amount and spatial delivery of feedback from AGN. As well as determining individual GW events, merger rates and binary properties additionally directly impact the expected form of the stochastic GW background \citep[e.g.,][]{RajagopalRomani95, Phinney01, EnokiEtAl04, SesanaEtAl04, SesanaEtAl09PTA, KelleyEtAl17, SiwekEtAl20}. 

A very interesting regime, especially for multi-messenger science, comprises of merging supermassive black hole binaries in gas-rich environments. Here it is expected that both supermassive black hole feeding and feedback processes could be of relevance, which may lead to significant spin evolution both via direct extraction of spin energy \citep{blandford+77, tchekhovskoy+11, TalbotEtAl21} and the impact on inflowing gas dynamics \citep{TalbotEtAl22, TalbotEtAl23}. Simulations of merging gas-rich galaxies suggest that a non-negligible fraction of gas loses angular momentum and is hence efficiently funnelled towards the central region of the merger remnant \citep{BarnesHernquist91, BarnesHernquist96, Barnes2002, MihosHernquist96, DiMatteoEtAl05, HopkinsEtAl06, MayerEtAl07, MayerEtAl2010, CoxEtAl08, JohanssonEtAl09, HopkinsEtAl13, CapeloEtAl15, CapeloDotti2017}. Such a gas-rich environment provides a prime opportunity for the formation of massive CBDs around the supermassive black hole binary, for example via the circularisation of infalling gas clouds \citep{DunhillEtAl14, GoicovicEtAl16, GoicovicEtAl17, GoicovicEtAl18} or the excavation of a cavity within a larger nuclear gas disc by the binary \citep{SouzaLimaEtAl2020}.

In prograde binaries, gravitational torques from the CBD have traditionally been invoked to facilitate the extraction of angular momentum to drive the shrinking of the binary before it enters the GW regime \citep[see e.g.,][]{Pringle91, ArtymowiczLubow94, ArtymowiczLubow96, IvanovEtAl99, GouldRix00, ArmitageNatarajan02, ArmitageNatarajan05, LodatoEtAl09}. However, gas that resides inside the cavity of a CBD, such as streams or mini-discs, can also affect the binary orbital evolution by supplying material for accretion and acting as an additional source of gravitational torques \citep[see e.g.,][]{MacFadyenEtAl08, CuadraEtAl2009, ShiEtAl12, RoedigEtAl2011, RoedigEtAl2012, D'OrazioEtAl13, FarrisEtAl14, MirandaEtAl17, MunozEtAl19, TiedeEtAl20, DittmannRyan22, FranchiniEtAl22, SiwekEtAl23Orbit}. In particular, simulations that resolve the mini-discs and gas flows within the binary orbit ($R<a$) find that net torque from this region is generally positive while the net torque from larger radii is generally negative \citep{RoedigEtAl2012, MunozEtAl19, TangEtAl17, TiedeEtAl20, FranchiniEtAl22, SiwekEtAl23Orbit}. With this in mind, several recent studies have found that the {\it total} net torque acting on a binary can, in some scenarios, be positive and lead to binary expansion \citep[e.g.,][]{MirandaEtAl17, TangEtAl17, MoodyEtAl19, MunozEtAl19, MunozEtAl20, DuffellEtAl20, HeathNixon20, TiedeEtAl20, D'OrazioDuffel21, DittmannRyan22, DittmannRyan23, SiwekEtAl23Orbit, WangEtAl23}. These simulations typically invoke a (locally-)isothermal equation of state and neglect self-gravity. There is only a limited number of simulations that instead consider self-gravitating, massive CBDs \citep{CuadraEtAl2009, RoedigEtAl2011, RoedigEtAl2012, RoedigSesana14, FranchiniEtAl2021}. These simulations, which additionally employ an adiabatic equation of state with $\beta-$cooling (see Section~\ref{sec:num_sims}) to regulate the disc stability, universally find that binaries shrink. A further complication that has only been addressed by a handful of works is the effect of misaligned \citep{NixonEtAl13, DunhillEtAl14, AlyEtAl15, MoodyEtAl19} and retrograde \citep{NixonEtAl11Retrograde, DunhillEtAl14, RoedigSesana14, HeathNixon20, TiedeDOrazio23} CBDs, with the latter behaving quite differently to the prograde case due to lack of resonant torquing by the CBD and providing a promising channel to shrink binaries.

The main aim of this paper is to build on existing literature and model for the first time both the mass and spin evolution of black holes in massive gravito-turbulence CBDs, as well as analyse the general CBD and binary evolution. To achieve this we perform a suite of high-resolution simulations of gas-rich CBDs around live binaries whose properties are expected to place them in the observing range of LISA \citep{Lisa17, LISA23}. We consider different configurations by varying the binary mass ratio, eccentricity and inclination angle (including fully retrograde binaries). In order to resolve streams and mini-discs within the CBD cavity we employ a super-Lagrangian refinement technique \citep{curtis+15} that vastly improves the resolution around each black hole \citep[see also,][]{FranchiniEtAl22}. Additionally, given the important implications of black hole spin on both the GW emission and black hole recoil velocities, we employ the sub-grid accretion disc model of \citet{fiacconi+18} to track both the mass and spin evolution of the black holes. The paper is organised as follows; in Section~\ref{sec:num_sims} we present the numerical code adopted, physical modules included and our setup of the initial conditions for the simulations performed. In Section~\ref{sec:results} we outline the main results, including the evolution of the CBD and binary, an analysis of the expected torques experienced by the binary and the black hole growth and spin evolution. In Section~\ref{sec:discussion} we put these results into the context of the existing literature and discuss their implications while in Section~\ref{sec:conclusion} we summarise our main conclusions.

\section{Numerical simulations}
\label{sec:num_sims}

\subsection{Initial conditions}
\label{sec:ics}

The binary and CBD setup that we model is generally scale-free meaning that it is relevant to a wide range of scenarios and the majority of results we present can be re-scaled with respect to the system mass and length. However, for this work we employ the sub-grid accretion disc model of \citet[][]{fiacconi+18} to track the black hole spin evolution, which requires us to define a physical mass and length scale for our simulations. Therefore, we consider supermassive black hole binaries with initial total mass $M_{\rm bin} = M_{\bullet, 1} + M_{\bullet, 2} = 2 \times 10^{6}$~M$_{\sun}$ and mass ratio $0.1 \leq q \equiv M_{\bullet, 2}/M_{\bullet, 1} \leq 1$. We chose these parameters to be in the binary parameter space that would be of the largest interest for the Laser Interferometer Space Antenna (LISA; \citealt{amaro+13}). The binary is initially on a Keplerian orbit with eccentricity of either $e=0$ or $e=0.5$ and initial semi-major axis $a=2$~pc,  corresponding to a period of $t_{\rm bin} = 0.187$~Myr. The peri- and apo-centre of the binary are aligned along the $x$ axis.

A gaseous CBD surrounds the binary in the $xy$ plane and initially extends between $R_{\rm in}=2a=4$~pc and $R_{\rm out}=7a=14$~pc from its centre of mass. The binary and the disc are not necessarily coplanar. For misaligned systems, the orbital plane of the binary crosses the disc midplane only through the $y$ axis with an inclination angle $i$; for reference, $i=0\degr$ and $i=180\degr$ correspond to prograde and retrograde coplanar orbits, respectively. The disc follows the surface density profile
\begin{equation}
\Sigma(R) = \Sigma \left(\frac{R}{R_{\rm in}} \right)^{-\alpha},
\end{equation}
where $\alpha = 2$. The normalisation $\Sigma$ is chosen to enforce $M_{\rm cbd} / M_{\rm bin} = 0.1$, where $M_{\rm cbd}$ is the total mass of the CBD.

The vertical structure of the CBD is constructed by assuming that the aspect ratio $H/R \approx c_{\rm s}(R)/V_{\rm K}(R)$ is constant, where $c_{\rm s}(R)$ and $V_{\rm K}(R)$ are the local isothermal sound speed and the Keplerian velocity for $M_{\rm bin}$, respectively. Therefore, the temperature profile of the CBD, $c_{\rm s}^2(R) = (H/R)^2 V_{\rm K}^2(R) \propto R^{-1}$, can be normalised after choosing the value of the aspect ratio. Since we include self-gravity in the CBD, we use the Toomre parameter $Q$ to impose the normalisation. Specifically, we choose $Q \approx 3 (H/R) (M_{\rm bin}/M_{\rm cbd}) = 1.5$, which corresponds to an initial aspect ratio $H/R \approx 0.5 M_{\rm cbd}/M_{\rm bin}$. We do not include any additional explicit source of viscosity beyond the intrinsic (low) numerical viscosity, instead relying on the angular momentum transport due to self-gravity \citep[e.g.,][]{CuadraEtAl2009}. The vertical structure of the CBD is locally sampled from a Gaussian distribution with a standard deviation given by $H(R)$. The CBD is initially in nearly Keplerian rotation with azimuthal velocity
\begin{equation}
V_{\phi}^2(R) = \frac{G M_{\rm bin}}{R} \left[1 + \frac{3}{4} \frac{q}{(1+q)^2} \left( \frac{R}{a} \right)^{-2} \right] - 3 c_{\rm s}^2(R),
\end{equation}
where the term in the square brackets represents the correction for the tidal field of the binary and the last term accounts for the gas pressure \citep[see e.g., equation 2 in][]{MunozEtAl20}. We do not include the contribution of the CBD potential in this equation, noting that despite this the CBD remains stable throughout the simulation where the CBD self-gravity is self-consistently taken into account. Note that we do not add any additional velocity dispersion to the gas. The black hole binary and CBD are the only components simulated in this work. While contributions from other components (e.g., stars and dark matter) could be important in dense galactic nucleii for low-mass/large separation binaries, we neglect them here to isolate the effects of the CBD-binary interaction and to ensure that the system remains scale-free.

Table~\ref{tab:runs} summarises the properties of the supermassive black hole binaries explored in this work. We explore 1:1, 1:3 and 1:10 mass ratio binaries denoted by q01, q03, and q10, respectively, and consider circular ($e=0$), eccentric ($e=0.5$), coplanar prograde ($i=0\degr$), misaligned ($i=45\degr$), and retrograde ($i=180\degr$) binaries. For example, the 1:1 mass ratio, circular, coplanar prograde binary simulation is labelled q01e00i00. The binary and CBD, as expected, torque each other from the outset of the simulation, i.e. also during the relaxation phases (see Sections~\ref{sec:code} and \ref{sec:bh_refinement}). To consider the $45\degr$ inclined binary evolution in a pre-relaxed system, we have an additional simulation, q03e00i45mod, in which we take the q03e00i00 run after the two initial relaxation phases, rotate the binary and cavity region about the $y-$axis in order to be inclined with respect to the CBD by $45\degr$ and track its subsequent evolution. 

\begin{table*}
\caption{Summary of the characteristics of the performed runs. Note that: (i) $M_{\rm bin} = 2 \times 10^{6}$~M$_{\sun}$; (ii) $q \equiv M_{\bullet, 2}/M_{\bullet, 1} \leq 1$. From left to right: simulation label, initial mass ratio $q$, initial eccentricity $e$, initial inclination angle $i$ between the binary orbital plane and the CBD midplane, initial Eddington ratio $f_{\rm Edd,1}$ for $M_{\bullet, 1}$, initial Eddington ratio $f_{\rm Edd,2}$ for $M_{\bullet, 2}$, initial $\alpha$ accretion disc and black hole angular momentum ratio for $M_{\bullet, 1}$, initial $\alpha$ accretion disc and black hole angular momentum ratio for $M_{\bullet, 2}$, misalignment angle between $\bmath{J}_{\bullet}$ and $\bmath{J}_{\rm d}$ for $M_{\bullet, 1}$, misalignment angle between $\bmath{J}_{\bullet}$ and $\bmath{J}_{\rm d}$ for $M_{\bullet, 2}$.}
\label{tab:runs}
\begin{tabular}{lccc cccc cc}
\hline
Label & $q$ & $e$ & $i$ & $f_{\rm Edd,1}$ & $f_{\rm Edd,2}$ &  $(J_{\rm d}/J_{\bullet})_1$ & $(J_{\rm d}/J_{\bullet})_2$ & $\theta_1$ & $\theta_2$ \\
 & & & (\degr) & & & & & (\degr) & (\degr) \\
\hline
q01e00i00 & $1$ & 0 & 0 & $3 \times 10^{-3}$ & $3 \times 10^{-3}$ & 0.71 & 0.71 & 76.6 & 64.7 \\
q03e00i00 & $1/3$ & 0 & 0 & $2 \times 10^{-3}$ & $6 \times 10^{-3}$ & 0.65 & 0.81 & 64.7 & 76.6 \\
q10e00i00 & $1/10$ & 0 & 0 & $1.5 \times 10^{-3}$ & $1.5 \times 10^{-2}$ & 0.65 & 1.02 & 76.6 & 64.7 \\
q01e05i00 & $1$ & 0.5 & 0 & $3 \times 10^{-3}$ & $3 \times 10^{-3}$ & 0.71 & 0.71 & 64.7 & 76.6 \\
q03e05i00 & $1/3$ & 0.5 & 0 & $2 \times 10^{-3}$ & $6 \times 10^{-3}$ & 0.65 & 0.81 & 64.7 & 76.6 \\
q10e05i00 & $1/10$ & 0.5 & 0 & $4 \times 10^{-3}$ & $1 \times 10^{-2}$ & 0.49 & 1.15 & 64.7 & 76.6 \\
q01e00i45 & $1$ & 0 & 45 & $3 \times 10^{-3}$ & $3 \times 10^{-3}$ & 0.71 & 0.48 & 86.1 & 96.3 \\
q03e00i45 & $1/3$ & 0 & 45 & $2 \times 10^{-3}$ & $6 \times 10^{-3}$ & 0.65 & 0.55 & 55.0 & 104.4 \\
q10e00i45 & $1/10$ & 0 & 45 & $1.5 \times 10^{-3}$ & $1.5 \times 10^{-2}$ & 0.44 & 1.02 & 113.3 & 14.6 \\
q03e00i45mod & $1/3$ & 0 & 45 & $2 \times 10^{-3} $ & $6 \times 10^{-3}$ & 0.44 & 0.81 & 110.0 & 57.2 \\
q03e00i180 & $1/3$ & 0 & 180 & $1 \times 10^{-3}$ & $3 \times 10^{-3}$ & 0.79 & 0.99 & 64.7 & 76.7 \\
q10e00i180 & $1/10$ & 0 & 180 & $1 \times 10^{-3}$ & $1 \times 10^{-2}$ & 0.72 & 1.15 & 64.7 & 76.6 \\
\hline
\end{tabular}
\flushleft
\end{table*}

\subsection{Numerical code and simulation setup}
\label{sec:code}

The simulations presented in Table~\ref{tab:runs} are performed using the moving-mesh hydrodynamical code {\sc arepo} \citep{springel+10,pakmor+16}, which solves the Euler equations on an unstructured Voronoi tessellation that moves with the flow. The fluid dynamics are solved using a finite-volume, second-order reconstruction scheme coupled with an exact Riemann solver. The unstructured mesh adapts by locally refining and de-refining in a quasi-Lagrangian way to maintain a typical gas mass $m_{\rm g}^{\rm target}$ per cell. The initial conditions are sampled with $10^{6}$ initial mesh elements in the CBD corresponding to a target mass $m_{\rm g}^{\rm target} = 0.2$~M$_{\sun}$. We employ the hierarchical time integration scheme described in \citet{SpringelEtAl2021}, which splits the Hamiltonian into ``fast'' and ``slow'' components, i.e. those on short versus long time-steps, respectively. With this approach, interactions between particles/cells on the smallest timesteps (up to $N_{\rm direct}^{\rm max}=2048$) are solved using direct summation, while a standard octree approach is used for all other interactions. This is primarily done for efficiency, to avoid unnecessary tree constructions for small numbers of particles, but has the added benefit of more accurately calculating the gravitational forces between the black holes and their nearest neighbours, which are on the shortest timesteps, via direct summation. We discuss the implications of using this approach, compared to using solely direct summation or a tree to solve for the gravity in Section~\ref{sec:torques}. The black holes have gravitational softening lengths $\epsilon_{\bullet} = 0.01$~pc, while gas cell softenings adapt with the cell size down to a minimum of $\epsilon_{\rm g} = 0.005$~pc.

We evolve the system assuming an equation of state for an ideal fluid, $P = (\gamma - 1) \rho u$, where $\rho$ is the gas density, $u$ is the specific internal energy, and $\gamma = 5/3$ is the adiabatic index, appropriate for a monoatomic ideal gas. We also allow the gas to cool radiatively through a simple $\beta$-cooling prescription, with the rate of cooling given by
\begin{equation}
\left( \frac{{\rm d}u}{{\rm d}t} \right)_{\rm cool} = - \frac{u}{t_{\rm cool}(R)}\,,
\label{eq:cool}
\end{equation}
where the local cooling timescale $t_{\rm cool}(R)$ is proportional to the local orbital time $\Omega_{K}^{-1}(R)$, namely $t_{\rm cool}(R) = \beta (G M_{\rm bin} R^{-3})^{-1/2}$, where, as stated above, $M_{\rm bin}$ is the total binary mass and $R$ is the distance to the binary centre of mass. We discuss the implications for the chosen cooling prescription and its implementation further in Section \ref{sec:discussion}. We set $\beta = 10$; this choice allows the gas to cool slowly without fragmenting\footnote{We note that while $\beta = 10$ is appropriate in our simulations, it has been shown that in radiation pressure dominated discs much larger vales of $\beta\sim50$ are necessary to prevent fragmentation \citep{ChenEtAl23}}\citep{gammie+01, RiceEtAl05, CuadraEtAl2009, FranchiniEtAl2021}, and eventually makes the CBD settle into a marginally stable configuration with $Q \sim 1$. This choice implicitly assumes a source of heating that prevents rapid cooling and fragmentation of the CBD, which could, for example, come from stellar heating \citep{ThompsonEtAl05, LodatoEtAl09}.

The initial conditions are evolved for 50 binary orbits corresponding to about $1.5t_{\rm cool}$ at the outer edge of the CBD (note this is the first of two relaxation periods, see also Section~\ref{sec:bh_refinement}). During this relaxation period, the binary perturbs the inner edge of the disc, inducing shocks and spiral waves that propagate from the cavity through the CBD. The shocks increase the temperature in the disc, in particular close to the inner rim of the cavity, where the aspect ratio of the CBD, $H/R$, can reach $\approx 0.1$ and the Toomre parameter grows to a few before cooling can efficiently radiate the excess energy. After a few binary orbits, streams of low-density gas from the cavity edge start to develop and bring fresh material close to the two black holes.

\subsection{Black hole refinement}
\label{sec:bh_refinement}
\begin{figure}
\begin{center}
\includegraphics[width=\columnwidth]{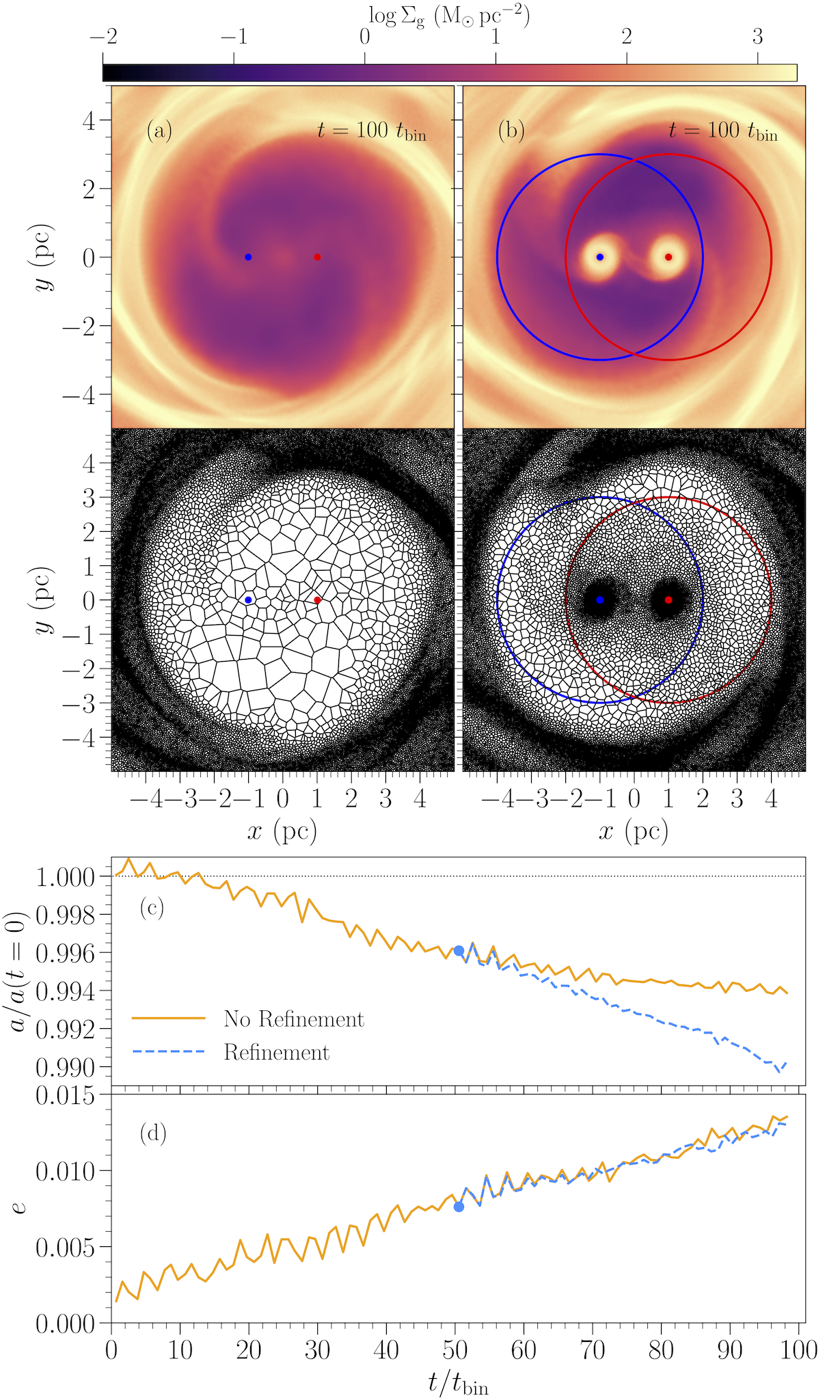}
\caption{Effect of the super-Lagrangian refinement on the evolution of the cavity. Panel (a) and (b) compare the surface density of the gas (top row) and the mesh structure (bottom row) after 100 binary orbits without and with super-Lagrangian refinement switched on during the last 50 binary orbits, respectively. Panel (c) and (d) show the evolution of the binary semi-major axis (normalised to the initial value) and eccentricity, respectively, for the cases without (orange solid curve) and with (light blue dashed curve) this extra refinement, respectively. The light blue dot marks when the super-Lagrangian refinement is activated.}
\label{fig:refinement}
\end{center}
\end{figure}

Inflow from the CBD is set by the interplay between the binary gravitational torque on the inner edge of the cavity and the mechanism of angular momentum transport in the CBD, i.e. gravito-turbulence in our configuration. In addition the binary dynamics are determined by a combination of physical processes including gravitational torques from the CBD, streams and mini-discs within the cavity and accretion. As such, proper modelling of mass transport through the cavity is critically important and relies on the capability of resolving relatively low-density features, such as streams funnelling gas towards the binary. This is intrinsically hard to achieve in (nearly-)Lagrangian approaches that would naturally maintain low resolution in under-dense regions. We can overcome this limitation in {\sc arepo} by using the super-Lagrangian refinement scheme presented by \citet{curtis+15,curtis+16} \citep[see also the recent simulations of][for a similar approach]{FranchiniEtAl22}. Super-Lagrangian refinement allows us to focus both mass and spatial resolution locally around each black hole particle in the simulation. Specifically, the scheme splits and merges cells within $R_{\rm ref}$ around each black hole to force the maximum cell size of the mesh elements to decrease linearly from $R_{\rm max}^{\rm cell}$ at $R_{\rm ref}$ to $R_{\rm min}^{\rm cell}$ at the centre. Cell radii are allowed to remain within a factor 2 of the local maximum. We set $R_{\rm max}^{\rm cell} = 0.2$~pc and $R_{\rm min}^{\rm cell} = 0.01$~pc; $R_{\rm max}^{\rm cell}$ is chosen to match the typical size of the mesh cell just inside the CBD cavity, while $R_{\rm min}^{\rm cell}$ is equal to the gravitational softening of the black holes. We have chosen an $e$-dependent value for $R_{\rm ref}$, namely $R_{\rm ref} = 3$~pc and $R_{\rm ref} = 5$~pc for $e=0$ and $e=0.5$, respectively; moreover, we set $R_{\rm ref} = 2$~pc for runs with $i=180\degr$ because the cavity edge forms at closer distances from the binary (see Section~\ref{sec:cbd}). These choices guarantee that the refinement region slightly crosses the edge of the cavity into the CBD, enforcing continuity in the mesh structure. The refinement may lead to the creation of very low mass cells; such occurrences are prevented by the lower limit $m_{\rm min} = 10^{-5} m_{\rm g}^{\rm target}$ for cells within the refinement region.

Fig.~\ref{fig:refinement} shows the effect of the super-Lagrangian refinement on the evolution of the cavity for run q01e00i00 as a representative example. We compare the evolution after $100$ binary orbits without any refinement and with the refinement activated during the last $50$ orbits to allow for the system to relax. Gaseous streams connecting the cavity edge to the black holes that can be barely seen without refinement are significantly better resolved thanks to the super-Lagrangian refinement. The streams effectively bring material close to the black holes and lead to the formation of gaseous ``mini-discs'', otherwise absent without refinement. The development of these features also has an impact on the dynamical evolution of the binary. As already shown by previous work \citep{RoedigEtAl2011,RoedigEtAl2012, TangEtAl17, MunozEtAl19, TiedeEtAl20, SiwekEtAl23Orbit, FranchiniEtAl22} and further discussed in Section~\ref{sec:torques}, the gravitational torques caused by the gas streams inside the cavity fundamentally contribute to the binary evolution. As the super-Lagrangian refinement allows for the streams penetrating the cavity to be better resolved, the gravitational torques are more prominent. While in the run without refinement the shrinking of the binary semi-major axis appears to slow after $\sim50~t_{\rm bin}$, the binary in the run with refinement continues to shrink after this time, though we do not see appreciable differences in the evolution of the eccentricity. The lack of difference suggests that the eccentricity evolution is primarily due to the binary being torqued by the CBD, which remains unchanged between runs with and without refinement. This is consistent with the findings of \citet{RoedigEtAl2011} and \citet{RoedigEtAl2012}, who claim that eccentricity evolution arises due to gravitational interactions between over-densities at the inner edge of the CBD and the binary. After the $\beta$-cooling relaxation, all simulated systems are furthermore relaxed with the super-Lagrangian refinement for $50$ additional binary orbits. Thus, the overall relaxation phase lasts $\approx 18.7$~Myr, corresponding to $\approx 3 t_{\rm cool}(R_{\rm out})$, after which the CBD attains a nearly steady state and we use the resulting snapshot as initial conditions to be further evolved for $500~t_{\rm bin}$ including the accretion and spin evolution model originally presented by \citet{fiacconi+18} and discussed in the next section.

\subsection{Black hole spin model}
\label{sec:bh_spin_model}

The accretion disc model of \citet{fiacconi+18} assumes that each black hole is surrounded by a sub-grid accretion disc following the analytical thin $\alpha$-disc solution\footnote{In the following, we will refer to the large-scale, hydrodynamically-resolved gaseous disc surrounding the binary as the ``CBD'', to the compact, hydrodynamically-resolved gaseous discs around each black hole as the ``mini-discs'', and to the sub-grid, analytic $\alpha$-discs as ``accretion discs''.} \citep{shakura+73}. The black holes and accretion discs are described by their masses and angular momenta, self-consistently coupled through mass accretion and the Bardeen-Petterson effect \citep{bardeen+75,scheuer+96,king+05,martin+07,perego+09, DottiEtAl13}. Moreover, the system composed of the black hole and the accretion disc is subjected to an inflow of matter and angular momentum as dictated by the hydrodynamical simulations. We refer the reader to \citet{fiacconi+18} for a thorough description of the model.

The accretion model requires the initial black hole mass and spin, the initial mass and accretion rate of the accretion disc, and the directions of the angular momenta to be specified (see Table~\ref{tab:runs}). The spin of a black hole depends on its merger and accretion history, with the properties of the accretion flow (i.e. how chaotic or coherent it is) determining whether or not a black hole is effectively spun up \citep{VolonteriEtAl05, VolonteriEtAl07, KingPringle06, KingPringle07, DottiEtAl13, SesanaEtAl14}. We assume that both black holes are initially highly spinning as a result of previous evolution and we assign black hole spins of $a_{\bullet, 1,2} = 0.9$, consistent with current observational constraints for $\sim 10^{6}$~M$_{\odot}$ black holes \citep{Reynolds19}. The timescale, $\tau_{\rm align}$, on which black hole spin aligns with respect to the accretion disc angular momentum scales as $a_{\bullet}^{5/7}$ (see Equation~(\ref{eq:alignment})) and so in choosing high spin values we are able to apply something of an upper limit on values of $\tau_{\rm align}$. The accretion disc is initiated with a mass of $M_{\rm d} = 10^{-3}M_{\bullet, 1,2}$, and angular momentum that is aligned with the angular momentum of the surrounding mini-disc. On the other hand, the black hole spins have been given arbitrary initial orientations with respect to the accretion disc angular momentum (shown in Table \ref{tab:runs}, noting that some of the accretion disc angular momenta are counter-rotating with respect to the black hole spin in some of the misaligned set-ups). Given that the initial spins of individual black holes will be determined by their accretion and merger histories prior to forming the binary, they do not necessarily need to be aligned, neither with each other nor their initial mini-discs, at the outset of our simulations. The initial accretion rate through the sub-grid accretion disc is set to be about the same for both black holes and close to the initial mass inflow rate measured inside the mini-discs. Finally, the initial accretion disc - black hole angular momentum ratio, $J_{\rm d}/J_{\bullet}$, is set to ensure consistency between the initial $f_{\rm Edd}$ and other black hole and accretion disc properties (see Equation~(\ref{eq:fedd})). 

Finally, we note an important modification we made to the original accretion model. In \citet{fiacconi+18}, the mass inflow rate $\dot{M}_{\rm inflow}$ and the specific angular momentum of the inflowing material $\bmath{L}_{\rm inflow}$ were calculated with a kernel-weighted interpolation from a volume encompassing a mass of $32 m_{\rm g}^{\rm target}$, typically slightly larger than the size of the mini-discs. While we employ the same approach to estimate $\dot{M}_{\rm inflow}$, we have adjusted how we measure $\bmath{L}_{\rm inflow}$ as despite the central weighting this approach tends to overestimate its magnitude compared to that of the gas resolved on scales of the sub-grid accretion disc (typically by a factor of a few). Instead, we only take the direction of $\bmath{L}_{\rm inflow}$ and we assign the magnitude of the inflowing gas specific angular momentum to $L_{\rm inflow} = L_{\rm d}$, where $L_{\rm d}$ is the instantaneous specific angular momentum of the accretion disc. This means that the condition originally imposed in \citet{fiacconi+18} requiring $L_{\rm inflow}\leq L_{\rm d}$ is automatically satisfied and we assume that all inflowing gas would be able to circularise onto the sub-grid accretion disc. In practice we typically find reasonable agreement between $L_{\rm d}$ and the specific angular momentum of gas on scales of the sub-grid accretion disc.

\section{Results}
\label{sec:results}
\begin{figure*}
\begin{center}
\includegraphics[width=1.78\columnwidth]{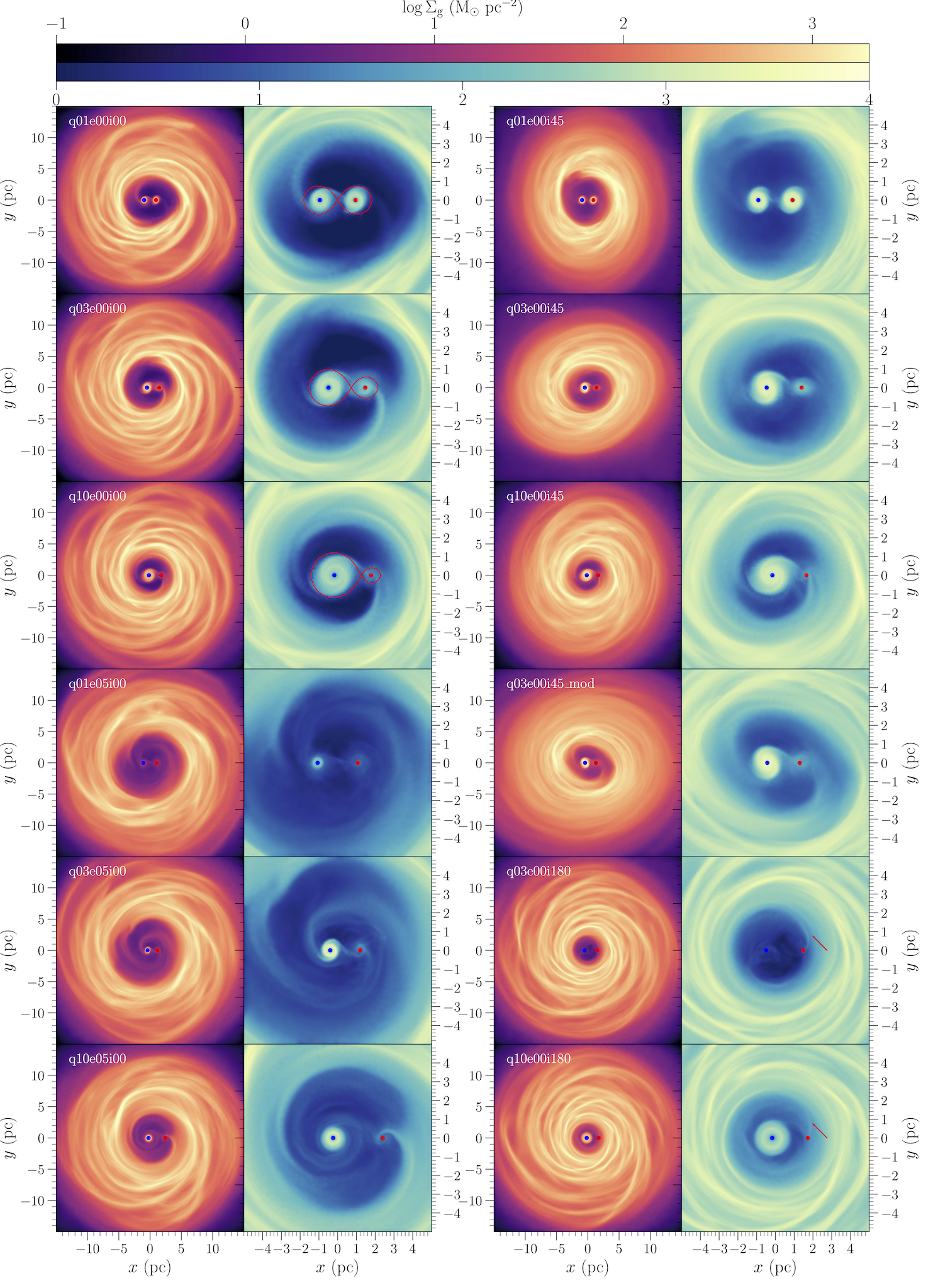}
\caption{Gas surface density maps projected on the $xy$ plane along the $z$ axis of all the simulations in Table~\ref{tab:runs} at time $t = 100~t_{\rm bin}$. For each pair of images, the one on left-hand side shows the CBD, while the one on the right-hand side shows a zoom in the inner cavity region. The blue and red filled circles mark the position of the primary and secondary black hole, respectively. The red curves for the first three simulations show the Roche lobes projected on the $xy$ plane. The red arrows in the last two simulations highlight the location of the dynamical friction wake. Both the appearance and properties of the CBD and the mini-discs are affected by the binary mass ratio and orbital parameters.}
\label{fig:cbd_maps}
\end{center}
\end{figure*}

\begin{figure*}
\begin{center}
\includegraphics[width=2\columnwidth]{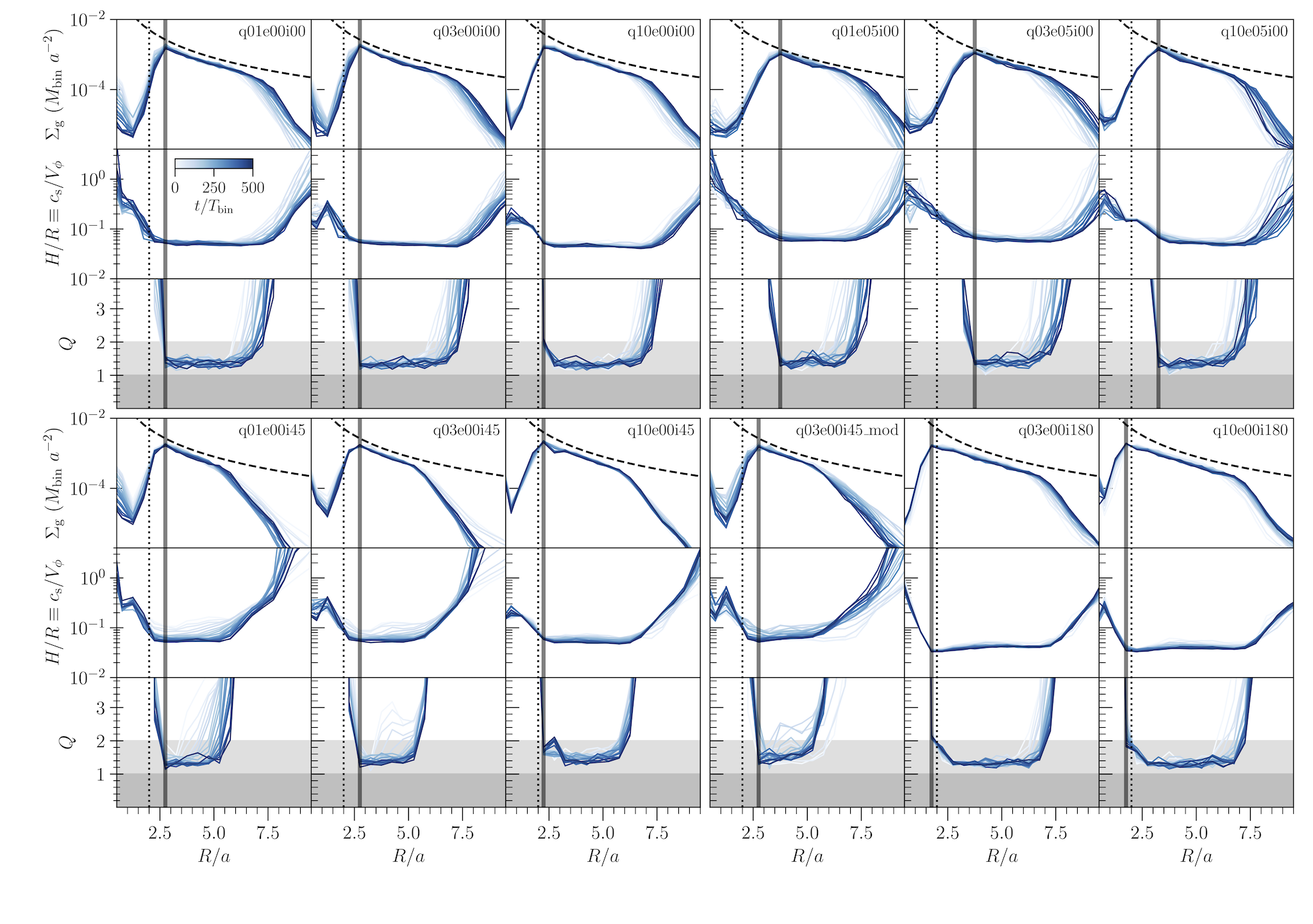}
\caption{Time-evolution of radial profiles of key CBD properties for our simulation suite. For each row, the top panels show the gas surface density, with dashed curves highlighting the initial $\propto R^{-2}$ profile; the middle panels show the aspect ratio $H/R$ estimated as $c_{\rm s} / V_{\phi}$; and the bottom panels show the Toomre parameter $Q$. Light-grey and dark-grey shaded areas highlight the marginal stability and instability regions for $Q$, respectively, in the limit of a razor-thin disc.
The vertical dotted lines in all panels indicate the original inner edge of the CBD at $2a=4$~pc, while for comparison the vertical grey lines indicate the location of the CBD density peak, illustrating that for prograde binaries the CBD moves outwards (especially in the case of eccentric binaries), while for retrograde binaries it moves inwards.
}
\label{fig:cbd_profiles}
\end{center}
\end{figure*}

\subsection{Properties of the circumbinary discs}
\label{sec:cbd}
After the relaxation phases, we run the system for an additional $500~t_{\rm bin}$ during which the CBD outside the cavity evolves in a nearly steady state. Fig.~\ref{fig:cbd_maps} compares density projections for all runs at time $t=100~t_{\rm bin}$. Each run is presented by two panels, one showing the whole disc (orange/purple colourmap) and the other zoomed in to the cavity region (blue/green colourmap). As expected, all simulations show rather similar CBDs at large distances from the cavity edge, except for the cases of misaligned binaries with $i=45\degr$, where the CBD becomes warped owing to the gravitational torque of the binary (see further discussion in  Section~\ref{sec:misaligned}). Since we have aligned the image in the binary orbital plane, the CBD appears distorted because of projection effects (i.e. it does not lie in the binary orbital plane), while otherwise, it resembles the companion runs with $i=0\degr$. Prominent spiral arms across the CBD piling up close to the cavity edge represent a common feature across different simulations \citep[including previous works that model self-gravitating CBDs, e.g.,][]{CuadraEtAl2009, RoedigEtAl2012, FranchiniEtAl2021}. The spiral arms are maintained by the interplay of cooling and local shock heating induced by self-gravity in the CBD, preventing fragmentation into self-bound clumps and maintaining a gravito-turbulent state in the disc. Gravito-turbulence is the main process responsible for mass and angular momentum transport in the CBD. According to the analysis of \citet{gammie+01}, gravito-turbulence may be described through the effective $\alpha$-parameter $\alpha_{\rm SG} = (4/9) \left[ \gamma ( \gamma - 1) \beta \right]^{-1} \approx 0.04$, where the last approximate equality holds for our parameter choices. For typical values of gas surface density and sound speed as measured from the simulations close to the cavity edge, we expect mass transport rates $\dot{M} = 3 \pi \nu \Sigma \approx 5 \times 10^{-3}$~M$_{\sun}$~yr$^{-1}$. However, the complex nature of the system, driven by binary torques and gas hydrodynamics in full 3D geometry, means that the mass flows through the disc vary with time and while instantaneous mass inflow rates similar to this estimate are achieved, gas can also move outwards leading to time-averaged net inflow rates $\gtrsim 10^{-4}$~M$_{\sun}$~yr$^{-1}$ measured directly from the simulations, with fluctuations of factor $\sim$a few between different runs (see Fig.~\ref{fig:mdot_time}). This suggests that at least qualitatively and on average over time, the CBD behaves like an $\alpha$-disc, albeit with a lower net inflow rate than predicted from idealised analytic estimates.

It is interesting to note that the strength and spatial distribution of spiral arms and the overall shape of the CBD is affected by the binary mass ratio and our initial choice of orbital parameters. However, most of the morphological diversity between various runs appears close to and inside the cavity. The cavity edge is not circular as its shape is continuously stirred by the gravitational perturbation from the central binary. This in turn triggers the formation of streams of gas penetrating the inner region of the cavity and providing fresh gas to the mini-discs \citep[see also,][noting, however, that not all of these simulations capture the mini-discs themselves.]{RoedigEtAl2012, ShiEtAl12, D'OrazioEtAl13, FarrisEtAl14, RagusaEtAl16, MirandaEtAl17, MunozEtAl19, TiedeEtAl20, FranchiniEtAl22}.

Except for the $q=1$ case, where the whole system is symmetric and both black holes alternatively induce the formation of the streams depending on surface density fluctuations close to the cavity edge, the secondary black hole is typically associated with the most prominent stream. The size of the cavity decreases for lower values of the mass ratio $q$, with a similar trend being seen in Fig.~3 of \citet{SiwekEtAl23Acc}. This indicates that the torques exerted by the binaries that move the inner edge of the CBD outwards become less effective at doing so for low q values. Indeed, we find that the cavity becomes larger for all prograde binaries, except for the q10e00i00 case, which stays roughly constant albeit with a less well-defined edge. 

On the other hand, the retrograde ($i=180\degr$) cases all exhibit shrinking cavities as the CBD inner edge moves inwards. This is expected due to differences between the binary-disc interaction in the prograde vs. retrograde cases. In prograde discs, gravitational torques act to reduce the binary separation while transporting angular momentum and the inner edge of the CBD outwards \citep[e.g.,][]{Pringle91, CuadraEtAl2009, LodatoEtAl09}. On the other hand, in the retrograde CBDs the net gravitational torques are negative and due to a lack of resonances are significantly weaker, which promotes continued inflow of material and a smaller cavity \citep{NixonEtAl11Retrograde, RoedigSesana14, BankertEtAl15}. 

Typically, the sharp inner edge at $2a$ is lost rapidly during relaxation. For the eccentric cases in particular the inner edges of the CBDs are disturbed and move outwards. In the high eccentricity runs, the cavity edge is less sharp than in the cases with nearly circular orbits, with cavities also being more extended and containing more diffuse material. The diffuse material is produced by the hydrodynamical interaction between the streams and the central binary. At the apocentre, the cavity edge is maximally perturbed and the streams start to develop; by the time the black holes reach the pericentre, their relative velocity with respect to the streams increases and the faster motion of the black holes makes streams wind up. The streams are eventually mixed and dispersed just after the pericentric passage when the separation between the two black holes increases again. 

The binary parameters also have a large impact on the sizes of the mini-discs. As already pointed out by \citet{ArtymowiczLubow94}, the radial extent of the mini-discs changes with the mass ratio of the binary and the eccentricity of the orbit. For circular orbits, the mini-discs surrounding the primary and the secondary black holes grow and decrease, respectively, as the binary mass ratio decreases, approximately following the shape of the Roche lobes as shown in Fig.~\ref{fig:cbd_maps}. When highly eccentric orbits are considered, the sizes of the mini-discs are reduced by the tidal truncation induced at the pericentre. In all cases, the mini-discs show spiral arm patterns induced by the interaction with and the circularisation of the infalling streams, as well as bridges of material between the two mini-discs typically following the primary. 

Barring the mini-disc around the primary black hole in the q10e00i180 case, mini-discs appear significantly less developed in retrograde binaries compared to the prograde cases. In the $q=1/3$ retrograde run, a tiny circum-primary mini-disc forms, while the secondary black hole does not develop a clear mini-disc and accretion mostly proceeds through episodic and very thin streams of gas plunging toward the black hole. Similarly, no mini-disc forms around the secondary black hole in the retrograde $q=1/10$ run either. The retrograde binary orbits in a cavity that is less depleted of material when compared with the corresponding prograde cases. No obvious gas streams seem to penetrate the cavity. The binaries develop tiny perturbations in the background material that resemble the wakes caused by dynamical friction, in particular for the $q=1/10$ simulations, where more material diffuses in from the CBD inner edge. While the whole cavity resembles a Type II gap between the mini-disc around the primary and the CBD, as observed in protoplanetary disc calculations \citep{Rafikov02, NelsonEtAl03, CridaMorbidelli07}, we note that the processes controlling gap formation differ for prograde and retrograde systems \citep[see e.g.,][]{IvanovEtAl2015}. 

\begin{figure*}
\begin{center}
\includegraphics[width=2\columnwidth]{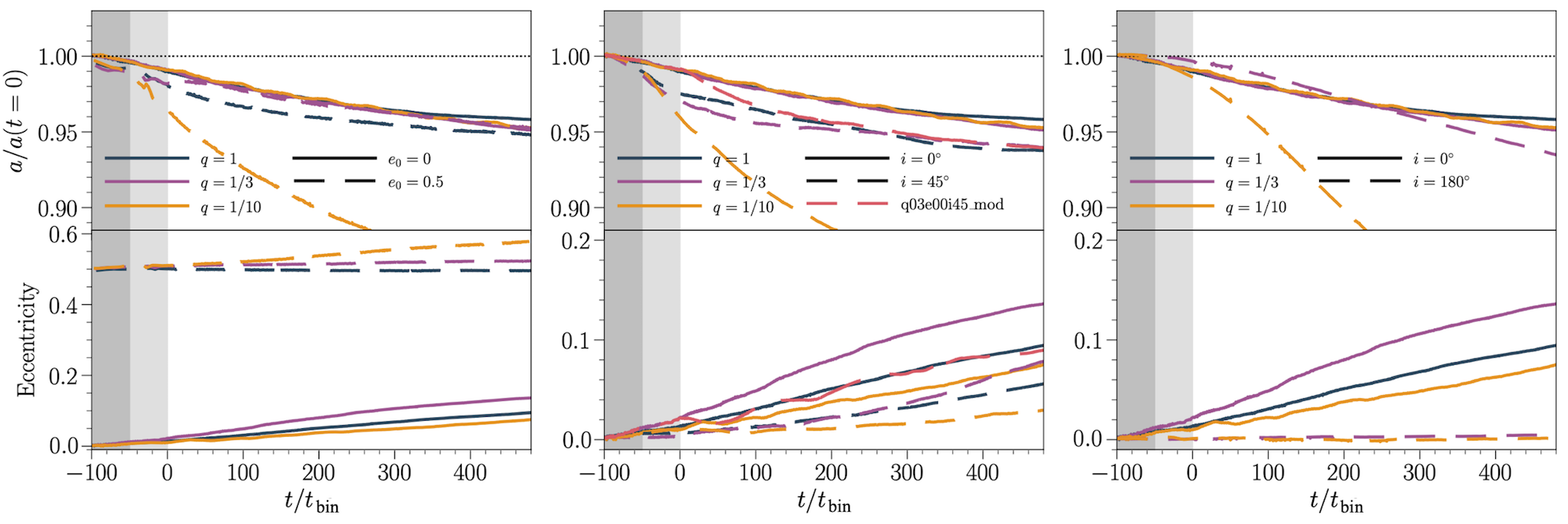}
\caption{Time evolution of the semi-major axis (top panels), and eccentricity (bottom panels), of all the simulated binaries. From left to right: comparison between circular and eccentric binaries; comparison between co-rotating coplanar and co-rotating misaligned binaries; comparison between coplanar and counter-rotating binaries. Blue, purple and yellow curves refer to mass ratios $q=1$, $q=1/3$, and $q=1/10$, respectively. The dark and light grey shaded areas indicate the initial relaxation phases without and with the super-Lagrangian refinement, respectively. The middle panels also include the q03e00i45mod binary evolution. Overall we find that the interaction of the binaries with their CBD leads to shrinking of the semi-major axis and eccentricity growth, and expect the binaries to be driven towards merger.}
\label{fig:bbh_dyn}
\end{center}
\end{figure*}

Fig.~\ref{fig:cbd_profiles} shows the time evolution of the CBD surface density, aspect ratio, and Toomre parameter $Q$ profiles for our different simulations. Despite fluctuations due to the formation of spiral arms, the CBD roughly maintains the original $\propto R^{-2}$ profile over time. The surface density profiles peak near the cavity edge, which we indicate by the vertical grey lines in each panel. This can be compared to the vertical black dotted line, which indicates the original location of the CBD inner edge and illustrates whether or not the CBD moves outwards (as for prograde binaries) or inwards (as for retrograde binaries). The surface density filling the cavity (i.e. to the left of the grey vertical line) of binaries with mass ratios $q \geq 1/3$ and $i=0\degr$ tends to decrease over time, particularly for circular binaries. Similar but weaker behaviour is seen for $q=1/10$ binaries (i.e. run q10e00i00 and q10e05i00); however, the cavity surface density is typically larger, therefore confirming the tendency of the $q=1/10$ cases to have more diffuse material inside the cavity. 

The evolution of the cavity gas content is determined by the balance between the inflow/outflow of material from/to the CBD driven by the interaction with the binary and accretion onto the black holes that slowly depletes the mini-discs that are built up during the second relaxation phase of the simulation. On the other hand, we see an opposite trend for counter-rotating binaries, where, particularly in the case of the q10e00i180 binary, the surface density inside the cavity slightly grows over time, as the torque exerted by the binary is less effective in maintaining the cavity than in the prograde cases. For the eccentric cases, the CBD edge is significantly further out and is not so sharp due to interaction between the secondary black hole and the CBD.

For systems in which the binary and CBD are initially coplanar ($i=0\degr$ and $i=180\degr$), the aspect ratio and Toomre parameter are generally constant within the CBD for $R \lesssim 6-7 a$. For these systems, the aspect ratio typically ranges between $\approx 0.03$ and  $\approx 0.06$ and  the Toomre parameter is $\approx1.3-1.5$, with little dependence on the binary properties. Such values are indeed suggestive of a gravito-turbulent steady state in the CBDs, where spiral arms continuously form and disperse with no fragmentation into self-gravitating bound clumps. However, deviations from these values can be seen at early times for binaries that are not initially coplanar with their CBD ($i=45\degr$), which undergo significant evolution as the binary and warped CBD come into alignment (see Section~\ref{sec:misaligned}). For all binaries, inside the cavity, the aspect ratio grows to $\gtrsim 0.1$ because of the presence of the mini-disc and the Toomre parameter $Q \gg 1$. The mini-discs are  indeed generally less dense and hence thicker than the CBD , however, the  exact values of these quantities within about $1.5 a$ should be taken with a pinch of salt  due to the mini-discs and streams resulting in the geometry of the system  departing significantly from axisymmetry as assumed by the profiles. On the other hand, outside $\sim 7 a$, i.e. beyond the original extent of the CBD in the initial conditions, all the profiles evolve significantly. We typically see a progressive reduction of both the aspect ratio and the Toomre parameter over time, corresponding to an increase in the surface density. This happens because the very outer and under-dense regions of the CBD, produced by the initial readjustment of the disc during the relaxation phase, are still experiencing the slow action of radiative cooling. As the temperature decreases, the outer regions of the discs become progressively thinner and denser with time, explaining the evolution of the $\Sigma$, $H/R$ and $Q$ profiles.

\subsection{Binary evolution and torques}
\label{sec:binary_evo}
\subsubsection{General behaviour}

While some studies of CBDs include analytic binaries, with the binary orbit pre-determined and fixed \citep[e.g.,][]{MunozEtAl19, MunozEtAl20, SiwekEtAl23Orbit} in the simulations presented here we include a live binary \citep[see e.g.,][]{CuadraEtAl2009, RoedigEtAl2011, RoedigEtAl2012, FranchiniEtAl2021} whereby its evolution arises naturally through its interaction with the gas via gravitational torques and accretion. Fig.~\ref{fig:bbh_dyn} shows the time evolution of every binary's semi-major axis and eccentricity. The figure is further divided into three columns that compare circular and eccentric binaries (left), co-rotating coplanar and co-rotating misaligned binaries (middle), and coplanar and counter-rotating binaries (right). As already seen in previous studies that consider the evolution of a live binary in a self-gravitating CBD \citep{CuadraEtAl2009, RoedigEtAl2012, FranchiniEtAl2021}, the general behaviour is that the semi-major axis of the binary decreases and the eccentricity increases over time. As a consequence, the angular momentum of the binaries, defined as:
\begin{equation}
J_{\rm bin} = \mu \sqrt{G M a \left( 1 - e^2 \right)},
\label{eq:Jbin}
\end{equation}
where $\mu = M_{1} M_{2} / (M_{1} + M_{2})$ and $M = M_{1} + M_{2}$ are the binary reduced and total mass, respectively\footnote{$M_{i} = M_{\bullet, i} + M_{{\rm d}, i}$ is the sum of the mass of the black hole and the accretion disc.}, decreases over time (by $\sim$a~few$-10\%$), even considering the mass evolution of each component due to external accretion that may lead to an increase in the binary angular momentum \citep{RoedigEtAl2012}. We should also caveat that, as we discuss further in Section~\ref{sec:torques}, the gravitational torques experienced by the binary can be sensitive to numerical effects, in particular the choice of gravity solver. 

We do not observe any obvious trend in the evolution of $a$ as a function of the mass ratio $q$; however, we notice that for eccentric, inclined and retrograde $q = 1/10$ binaries, their semi-major axis decreases most significantly. As we discuss in Section~\ref{sec:torques}, the evolution of the binary separation and hence angular momentum is dominated by the gas gravitational torques acting on the black holes. Different regions provide positive and negative torques on the binaries and it is the balance of torques from gas in the vicinity of the binary at $R<a$ (typically positive), and gas at $R>a$ that largely comprises of streams and the CBD (typically negative) that are important in determining the {\it net torque} experienced by the binary. In all cases where the $q=1/10$ binaries shrink more rapidly we find that they experience strong negative torques (see Fig.~\ref{fig:ave_torques}), while the q10e00i00 binary experiences a relatively weaker simulated torque. However, upon accurately calculating torques in post-processing we find that the simulated torque is likely an underestimate in this case and that the q10e00i00 binary should likely shrink more rapidly (see Section~\ref{sec:torques} for further discussion).

Some recent studies that employ an exact analytic binary potential have been used to explore binary eccentricity evolution, with the expectation being that binaries with zero initial eccentricity should remain circular, while non-zero initial eccentricities should either shrink or grow towards some critical value \citep[e.g.,][]{D'OrazioDuffel21, ZrakeEtAl21, SiwekEtAl23Orbit}. This being said studies that include live binaries find that eccentricity can grow even when the binary is initially circular \citep[e.g.,][]{CuadraEtAl2009, FranchiniEtAl2021}. We find mild eccentricity growth in all of our binaries except for in the retrograde cases \citep[see also,][for retrograde eccentricity evolution]{TiedeDOrazio23}. The eccentricity generally tends to grow slightly faster for the $q=1/3$ cases than for equal-mass binaries, regardless of the other parameters. This tendency with the binary mass ratio is inverted for the $q=1/10$ cases, where the eccentricity growth is the slowest, except for the binaries starting with high initial eccentricity $e=0.5$. \citet{CuadraEtAl2009} found that eccentricity growth of an initially circular binary only starts once the CBD develops asymmetries. However, while our CBDs develop asymmetries fairly quickly, the eccentricity for our initially circular binaries starts to grow from $t=0$. This being said, recent work of \citet{FranchiniEtAl2021}, also finds mild early eccentricity growth when simulating massive gravito-turbulent CBDs around circular binaries. 

\begin{figure*}
\begin{center}
\includegraphics[width=2\columnwidth]{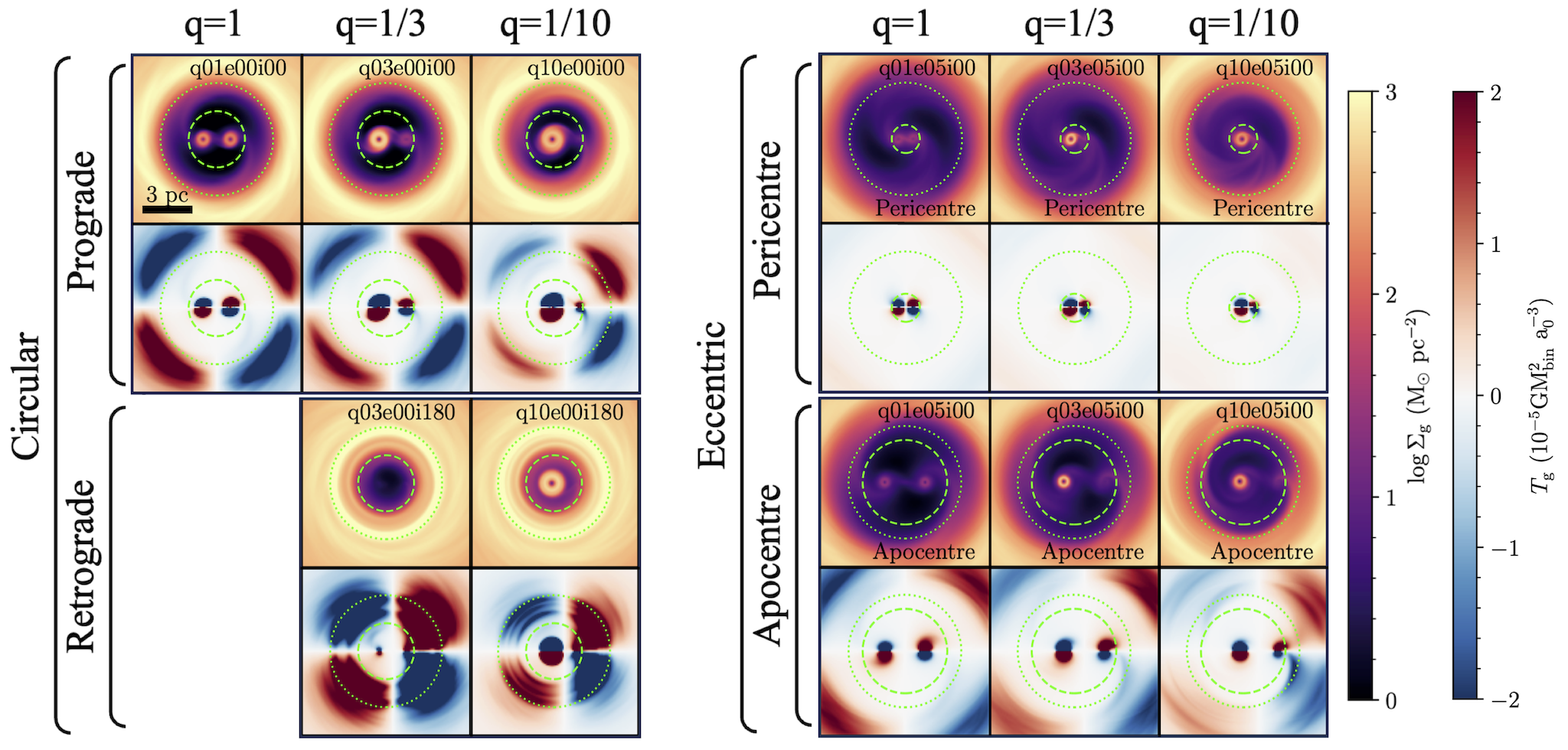}
\caption{Time-averaged surface density and torque maps for all the coplanar runs between $200~t_{\rm bin}$ and $300~t_{\rm bin}$.
Each simulation is identified by its label from Table~\ref{tab:runs}.
For eccentric runs ($e=0.5$), both the averages at the apocentre and at the pericentre are shown. The inner dashed circle indicates the radius corresponding to the semi-major axis for runs with $e=0$ and to the apo- and pericentre for the runs with $e=0.5$; the outer dashed circle at $R=2a=4$~pc indicates the initial cavity rim location. It can be seen that there are both positive and negative torques acting on the binary, in the mini-discs, the central cavity, including the gas streams and further out in the CBD. The spatial location and strength of these torques depends on the binary mass ratio and orbital parameters.}
\label{fig:combined_ave_maps}
\end{center}
\end{figure*}

A very different behaviour is shown by the counter-rotating binaries. Their eccentricity does not significantly grow over time, as has been seen for circular retrograde binaries in the recent work of \citet{TiedeDOrazio23}. However, \citet{NixonEtAl11Retrograde} discuss that the eccentricity growth depends on accretion onto (or gas captured by) the secondary, showing that if the secondary is able to accrete efficiently at apocentre (but not pericentre) then an initially circular orbit can gain eccentricity. Unlike in \citet{NixonEtAl11Retrograde}, the secondary black holes in our retrograde simulations accrete a negligible amount of mass (see Section~\ref{sec:bh_accretion}), as such, a similar channel of eccentricity growth due to accretion does not occur in our simulations. The overdense wake lagging behind the counter-rotating binary and the low eccentricity are features reminiscent of the action of dynamical friction on a massive body embedded in a rotating background \citep{dotti+07,dotti+09}. Additionally, as the cavity is more permeated by diffuse material, the counter-rotating black holes experience a headwind gas motion. As we show in Section~\ref{sec:torques}, the retrograde binaries experience strong negative torques, with the continuous refilling of the cavity leading to the steep and continuous shrinking. This is in contrast to the prograde cases in which the cavity is slowly cleared of gas, the CBDs expand and the shrinking of the binary appears to slow down.

\subsubsection{Torque Analysis}
\label{sec:torques}

Gravitational torques from the CBD and gas within the cavity, including the mini-discs and streams, are critical in determining the binary evolution \citep[e.g.,][]{CuadraEtAl2009, RoedigEtAl2012, MunozEtAl19}. In this section, we calculate the gravitational torques exerted by the gas on the binaries, and decompose them by their spatial contributions. This means we are able to analyse the dominant origins of such torques and their net influence on the binary separation. As well as allowing us to predict whether or not the binaries should grow or shrink, comparisons between the simulated binary evolution and post-processed torques from both direct-summation and tree based force calculations let us assess the appropriateness of the numerical techniques we employ.

Fig.~\ref{fig:combined_ave_maps} shows time-averaged maps of the gas surface density and torque surface density for every coplanar run. Panels on the left-hand side show initially circular binaries and are split into prograde (top) and retrograde (bottom) runs, while panels on the right-hand side are for eccentric binaries and show maps at both pericentre (top) and apocentre (bottom). For each run, the top panel shows gas surface density, while the bottom panel shows the surface density of the torque component directed along the binary angular momentum vector. The maps are produced by averaging over $\sim100$ snapshots between $t=200~t_{\rm bin}$ and $300~t_{\rm bin}$. Torque surface density maps show both positive (red) and negative (blue) torques with respect to the binary angular momentum vector, which act to change the binary angular momentum. The surface density maps further illustrate many of the features discussed in Section~\ref{sec:cbd} and provide a useful reference when considering which gas primarily torques the binaries. 

 If we first consider the prograde, circular binaries, we find that all the binaries experience torques from gas in the dense inner edge of the CBD and just inside the cavity. In the case of $q=1/10$ binary, there are clear torques from the gas that has moved into the cavity, with the stream feeding the secondary clearly visible. We also note that there are significant torque contributions from the gas in and around the black holes. For the retrograde binaries, the inner edge of the CBD has moved inwards and there is more diffuse gas within the cavity that torques the binary. For the $q=1/3$ case only small, tenuous mini-discs form around the black holes, although gas around the black holes and in the streams feeding them still contributes to the torques. In the $q=1/10$ case a clear mini-disc forms around the primary and contributions to the torques from gas around the primary and streams feeding the secondary are seen. Finally, in the case of the eccentric binaries, the torquing of the binary varies depending on the stage of the orbit -- at pericentre, when the black holes are furthest from the CBD inner edge, the dominant contribution to the torques comes from gas around the black holes, while at apocentre there are additional important contributions from material in the CBD and from the streams and diffuse gas that exists in the cavity.

\begin{figure*}
\begin{center}
\includegraphics[width=2\columnwidth]{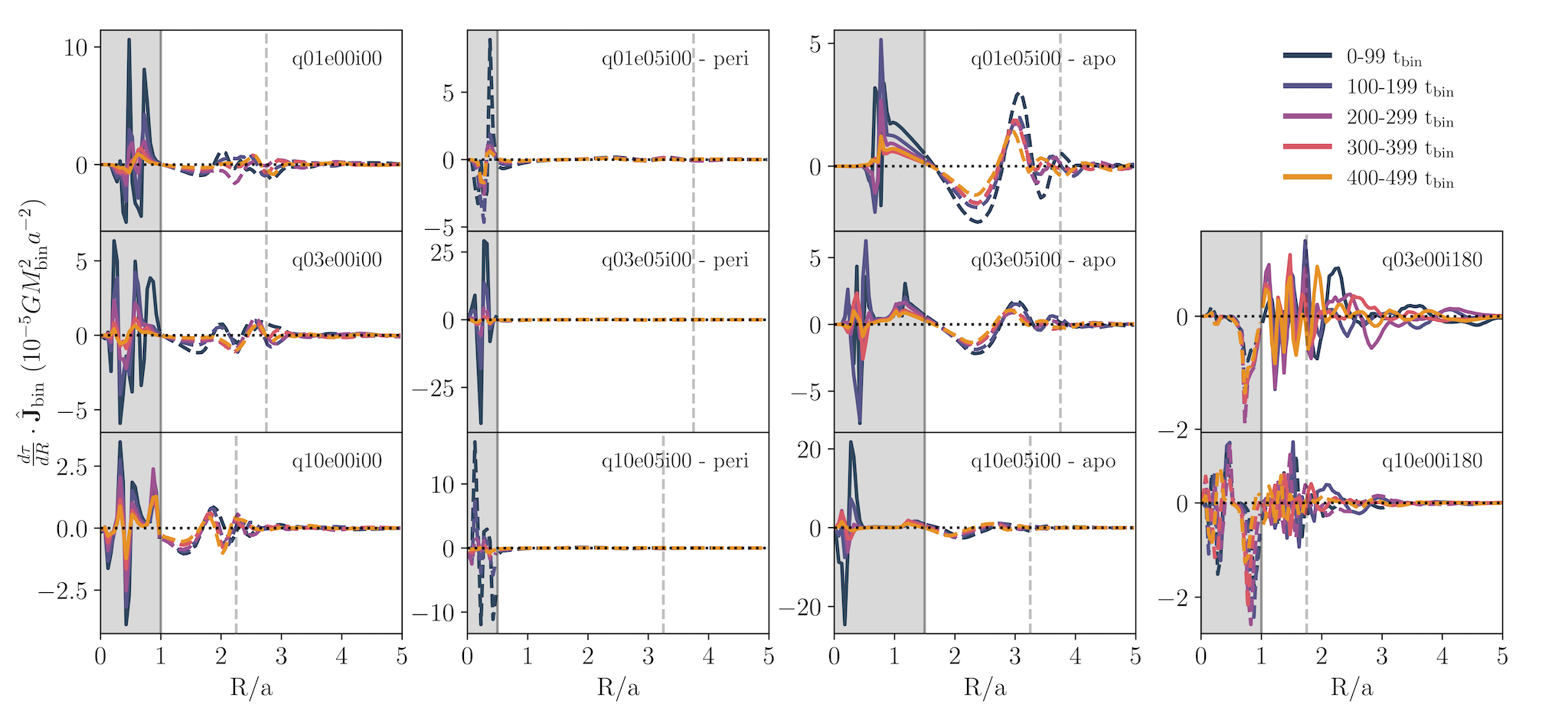}
\caption{Average radial profiles of the gas torque density, ${\rm d}{\boldsymbol \tau} / {\rm d}R$, acting along $\hat{\bf J}_{\rm bin}$, for all in-plane binaries. The left and right most columns show the circular pro-grade and retrograde binaries, respectively, while the central two columns show the initially eccentric binaries at pericentre (left) and apocentre (right). Each row shows binaries with mass ratios of $q=1$, $1/3$ and $1/10$, from top to bottom, respectively. Each profile is produced by averaging over $100$ orbits$^{\ref{foot:orbit_ave}}$, while the line colours indicate the time range as shown in the legend. Profiles are split into two radial regions representing gas inside and outside the binary region ($R=a=2$~pc for circular orbits or $R=a(1\pm e)=2\pm1$~pc for eccentric runs) with solid/dashed lines indicating whether the integrated net torque in each region is positive/negative. Vertical dashed lines indicate the peak CBD density from Fig.~\ref{fig:cbd_profiles}. In general we find that material close to the binary, i.e. comprising the mini-discs, acts to torque up the binary, while gas at larger radii provides a net negative torque. Overall, the net torque acting on the binaries tends to be negative, driving them to shrink.}
\label{fig:torque_profiles}
\end{center}
\end{figure*}

Taking a more quantitative approach, similar to \citet{RoedigEtAl2012}, we calculate average profiles of the gas torque density, ${\rm d}{\boldsymbol \tau} / {\rm d}R$, acting on the binary, along the direction of the binary angular momentum. From left to right, the columns in Fig.~\ref{fig:torque_profiles} show circular prograde binaries, eccentric binaries at pericentre, eccentric binaries at apocentre and circular retrograde binaries, respectively. From top to bottom, the panels are for binary mass ratios of $q=1$, $1/3$ and $1/10$, respectively (misaligned binaries are discussed in Section~\ref{sec:misaligned}). For each binary, we create five profiles, each of which is generated by averaging over $100$ orbits\footnote{\label{foot:orbit_ave}To ensure consistency with the eccentric binary profiles, for which the apocentre/pericentre snapshot from each orbit is used, for the non-eccentric binaries we randomly select one snapshot per orbit when calculating the average profiles. The same approach is additionally taken for the profiles in Fig.~\ref{fig:inc_torque_profiles}.}. We additionally split each profile into the gas torques from two radial regions; namely close to the binary (within $R=a=2$~pc for circular orbits or $R=a(1\pm e)=2\pm1$~pc for eccentric runs) indicated by the grey shaded region, and at larger radii without shading. For each region, we integrate the profile to find the net torque and use solid lines if  the net torque in the region is positive (i.e. would increase the binary angular momentum) and dashed lines if the net torque in the region is negative (i.e. would decrease the binary angular momentum). The vertical dashed line in each panel corresponds to the CBD density peak from Fig.~\ref{fig:cbd_profiles} and indicates the location at which the CBD transitions to the cavity region.

\begin{figure}
\begin{center}
\includegraphics[width=1\columnwidth]{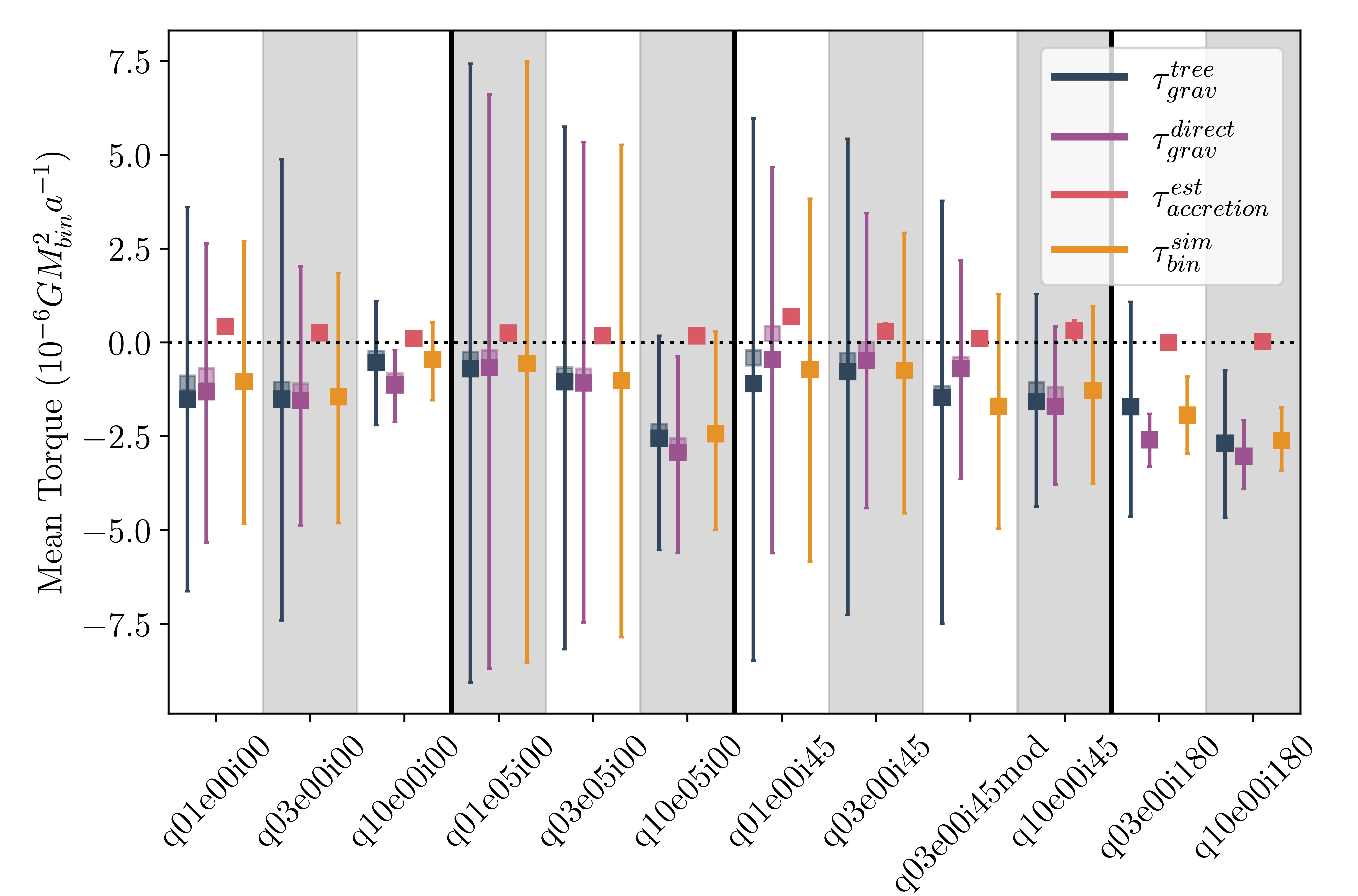}
\caption{Average values for the torque component along the binary angular momentum vector experienced by each binary compared to the estimated gravitational and accretion contributions. Coloured squares show the average torques over $500$ binary orbits while error bars show the one-$\sigma$ variation on an orbit by orbit basis. Specifically we show the actual torque experienced by the binary based the evolution of its angular momentum (yellow, $\tau_{\rm bin}^{\rm sim}$) as well as two estimates for the gravitational torque using either a pure tree approach (blue, $\tau_{\rm grav}^{\rm tree}$) or pure direct summation (purple, $\tau_{\rm grav}^{\rm direct}$) and an estimated accretion torque (pink, $\tau_{\rm accretion}^{\rm est}$). Shaded squares show the estimated total torque experienced by the binary when adding $\tau_{\rm accretion}^{\rm est}$ to either $\tau_{\rm grav}^{\rm tree}$ or $\tau_{\rm grav}^{\rm direct}$. The average gravitational torques are always negative, while accretion torques are always positive. All total average torque experienced by binaries in the simulation is negative, with most values being generally consistent with those found by post processing methods. Some binaries, particularly misaligned binaries can show larger differences between actual and estimated torques, however, only the $q01e00i45$ binary exhibits a qualitative disagreement between the method in terms of the sign of the overall torque.}
\label{fig:ave_torques}
\end{center}
\end{figure}

All of the runs show spikes in the torque profiles at small radii (corresponding to gas in and around the mini-discs), followed by smoother variations in the cavity and CBD. Specifically, there is typically a negative torque within the cavity followed by varying positive and negative peaks at the outer edge of the cavity/inner edge of the CBD. First considering the circular, prograde binaries, as highlighted by \citet{RoedigEtAl2012}, while peaks in the torque profiles at large radii (such that they are within the main body of the CBD) could be associated with outer Lindblad resonances \citep[OLRs, see also,][]{ArtymowiczLubow94}, one should be careful applying such an interpretation to peaks that appear within the cavity region (but beyond the mini-discs). These are associated with gas streams and are expected to have a kinetic origin. Gas within $R<a$ provides net positive torques on the binary, while outside of this region, the gas provides a net negative torque, dominated by the gas in the cavity and inner edge of the CBD. As we discuss later (see Fig.~\ref{fig:ave_torques} and Table~\ref{tab:torques}), on average the net torque acting on the binary is negative and acts to shrink it. This is due to the magnitude of the net negative torque at $R\gtrsim a$ generally being larger than the magnitude of the net positive torque at $R\lesssim a$, however, as we discuss in Section~\ref{sec:discussion} the exact balance is likely sensitive to the gas thermodynamics \citep[see e.g.,][]{RoedigEtAl2012, FranchiniEtAl22, WangEtAl23}. 

In the case of the eccentric binaries, we see that, as discussed in relation to Fig.~\ref{fig:combined_ave_maps}, at pericentre the torques are dominated by gas close to the black holes, with its net torque being either positive or negative. At apocentre, the profile features look somewhat more akin to the circular case, with net torques from gas close to the black hole being positive but those from gas in the cavity and CBD being negative. Again, on average the overall net torque acting on the binary is negative. 

While the rough location of peaks and troughs in the prograde cases remain the same, their magnitude reduces over time, particularly in the central region. This is likely a reflection of the declining average surface density in the cavity (see Fig.~\ref{fig:cbd_profiles}), which arises due to a combination of black hole accretion and the CBD moving outwards due to being torqued by the binary (see discussion in Section~\ref{sec:bh_accretion}). 

In Table~\ref{tab:torques} we provide the average torques experienced by each binary (directed along its own angular momentum vector) split into separate time bins of $\sim 100$ orbits each and decompose the contributions into being from gas at $R<a_{\rm sep}$, $R>a_{\rm sep}$, and all gas. Specifically, $a_{\rm sep}$ is the measured binary separation at the time torques are calculated, i.e.  this would be $a$ for circular binaries and vary between $a(1-e)$ and $a(1+e)$ for eccentric binaries moving between apo- and pericenter, taking into account the evolution of $a$ and $e$ over the course of the simulation (see Fig.~\ref{fig:bbh_dyn}). The colour coding scales linearly from $0$ to $\pm 5$, i.e., brighter colours represent stronger torques, with red and blue representing positive and negative torques, respectively. We find that in the majority of cases, even though the absolute average torque exerted from inner and outer regions typically decrease over time, the net gravitational torque from all gas only shows relatively mild evolution (although short-term fluctuations do occur, see Fig.~\ref{fig:ave_torques}). 

Finally, the retrograde circular binary exhibits many more sharp variations in the torque profiles, indicative of the more complex distribution of gas that fills the cavity region. As discussed in \citet{NixonEtAl11Retrograde}, a retrograde circular binary does not excite resonances within the CBD, and so peaks and troughs seen here should not be interpreted as such. The fact that the radial location of peaks and troughs within the CBD region itself vary further supports this by showing that the torques are not produced from consistent well-defined resonant radii. Overall, while throughout the simulation some regions switch from providing net positive and net negative torques to the binary, on average the binary experiences a strong negative torque, as shown quantitatively in Table~\ref{tab:torques}. Specifically, we find that the torques within the cavity and CBD roughly cancel each other out in the retrograde run and it is the gas within the vicinity of the black holes that dominates the net torquing of the binary. 

\begin{table*}
\caption{Averages of the gravitational torque component along the binary angular momentum vector (in units of $10^{-6}GM_{\rm bin}^{2}a^{-1}$)  experienced by the different binary systems. We show the total torque exerted by all gas, as well as individual contributions from gas at $R<a_{\rm sep}$ and $R>a_{\rm sep}$, where $a_{\rm sep}$ is the measured binary separation at each time torques are calculated ($a$ for circular binaries and varying between $a(1-e)$ and $a(1+e)$ for eccentric binaries). Averages are calculated in five time bins of $100$ orbits each. From left to right the columns show: simulation label, region over which torque is calculated, and five columns for separate time bins. The colour coding scales linearly from $0$ to $\pm 5$, i.e., whereby brighter colours represent stronger torques, with red and blue representing positive and negative torques, respectively. In the majority (although not all) of cases the absolute torques exerted by the inner and outer regions tend to decrease, although only mild evolution is seen in the overall average gravitational torque acting on the binary.}
\label{tab:torques}
\begin{tabular}{lcccccc}
\hline
Label & Region & $0-100\times t_{\rm bin}$ & $100-200\times t_{\rm bin}$ & $200-300\times t_{\rm bin}$ & $300-400\times t_{\rm bin}$ & $400-500\times t_{\rm bin}$ \\
\hline\hline
\multicolumn{7}{c}{\bf Circular prograde binaries} \\
\hline
{q01e00i00} & all gas & \cellcolor[rgb]{0.7808627450980392, 0.8785882352941177, 0.9312941176470588}$-1.4$ & \cellcolor[rgb]{0.8154509803921569, 0.8974705882352941, 0.9417058823529412}$-1.173$ & \cellcolor[rgb]{0.7946980392156863, 0.8861411764705882, 0.9354588235294118}$-1.312$ & \cellcolor[rgb]{0.7908549019607843, 0.884043137254902, 0.9343019607843137}$-1.335$ & \cellcolor[rgb]{0.7939294117647059, 0.885721568627451, 0.9352274509803922}$-1.314$\\
 & $R<a_{\rm sep}$ & \cellcolor[rgb]{0.9381137254901961, 0.7597607843137255, 0.731521568627451}$1.893$ & \cellcolor[rgb]{0.924807843137255, 0.7055137254901961, 0.6707921568627451}$2.331$ & \cellcolor[rgb]{0.9475960784313726, 0.7984196078431374, 0.7747999999999999}$1.586$ & \cellcolor[rgb]{0.9538666666666666, 0.8239843137254902, 0.8034196078431373}$1.378$ & \cellcolor[rgb]{0.9639607843137254, 0.8651372549019608, 0.8494901960784313}$1.051$\\
 & $R>a_{\rm sep}$ & \cellcolor[rgb]{0.48955294117647064, 0.719556862745098, 0.8436039215686274}$-3.293$ & \cellcolor[rgb]{0.4572705882352941, 0.7019333333333333, 0.8338862745098039}$-3.504$ & \cellcolor[rgb]{0.5502745098039216, 0.7527058823529411, 0.8618823529411764}$-2.898$ & \cellcolor[rgb]{0.5794823529411764, 0.7686509803921568, 0.8706745098039216}$-2.712$ & \cellcolor[rgb]{0.6325176470588235, 0.7976039215686275, 0.8866392156862745}$-2.364$\\
\hline
{q03e00i00} & all gas & \cellcolor[rgb]{0.8461960784313726, 0.9142549019607844, 0.9509607843137255}$-0.975$ & \cellcolor[rgb]{0.8054588235294118, 0.8920156862745098, 0.9386980392156863}$-1.238$ & \cellcolor[rgb]{0.7162980392156864, 0.8433411764705883, 0.9118588235294117}$-1.821$ & \cellcolor[rgb]{0.7201411764705883, 0.8454392156862744, 0.9130156862745098}$-1.797$ & \cellcolor[rgb]{0.7009254901960784, 0.8349490196078431, 0.9072313725490195}$-1.922$\\
 & $R<a_{\rm sep}$ & \cellcolor[rgb]{0.8957490196078431, 0.587043137254902, 0.5381647058823529}$3.279$ & \cellcolor[rgb]{0.9174666666666667, 0.6755843137254902, 0.6372862745098039}$2.569$ & \cellcolor[rgb]{0.9381137254901961, 0.7597607843137255, 0.731521568627451}$1.895$ & \cellcolor[rgb]{0.9599843137254902, 0.8489254901960784, 0.8313411764705883}$1.179$ & \cellcolor[rgb]{0.9697725490196079, 0.8888313725490196, 0.8760156862745097}$0.858$\\
 & $R>a_{\rm sep}$ & \cellcolor[rgb]{0.3419764705882353, 0.638992156862745, 0.7991803921568627}$-4.254$ & \cellcolor[rgb]{0.4111529411764706, 0.6767568627450979, 0.8200039215686274}$-3.806$ & \cellcolor[rgb]{0.42498823529411767, 0.6843098039215686, 0.8241686274509804}$-3.715$ & \cellcolor[rgb]{0.5387450980392158, 0.7464117647058823, 0.8584117647058823}$-2.976$ & \cellcolor[rgb]{0.568721568627451, 0.7627764705882353, 0.8674352941176471}$-2.779$\\
\hline
{q10e00i00} & all gas & \cellcolor[rgb]{0.7993098039215686, 0.8886588235294117, 0.9368470588235294}$-1.279$ & \cellcolor[rgb]{0.8077647058823529, 0.8932745098039216, 0.9393921568627451}$-1.224$ & \cellcolor[rgb]{0.825443137254902, 0.9029254901960784, 0.944713725490196}$-1.108$ & \cellcolor[rgb]{0.825443137254902, 0.9029254901960784, 0.944713725490196}$-1.109$ & \cellcolor[rgb]{0.8515764705882354, 0.9171921568627451, 0.9525803921568627}$-0.941$\\
 & $R<a_{\rm sep}$ & \cellcolor[rgb]{0.9092078431372549, 0.6419137254901961, 0.5995921568627451}$2.839$ & \cellcolor[rgb]{0.918078431372549, 0.678078431372549, 0.6400784313725489}$2.55$ & \cellcolor[rgb]{0.9315372549019608, 0.7329490196078432, 0.7015058823529412}$2.108$ & \cellcolor[rgb]{0.9359725490196078, 0.7510313725490196, 0.7217490196078431}$1.965$ & \cellcolor[rgb]{0.9364313725490196, 0.7529019607843137, 0.7238431372549019}$1.949$\\
 & $R>a_{\rm sep}$ & \cellcolor[rgb]{0.36272941176470586, 0.6503215686274509, 0.8054274509803921}$-4.118$ & \cellcolor[rgb]{0.4157647058823529, 0.6792745098039216, 0.821392156862745}$-3.774$ & \cellcolor[rgb]{0.5018509803921568, 0.7262705882352941, 0.8473058823529411}$-3.216$ & \cellcolor[rgb]{0.5233725490196078, 0.7380196078431372, 0.8537843137254901}$-3.074$ & \cellcolor[rgb]{0.5518117647058823, 0.7535450980392157, 0.8623450980392157}$-2.89$\\
\hline\hline
\multicolumn{7}{c}{\bf Eccentric binaries} \\
\hline
{q01e05i00} & all gas & \cellcolor[rgb]{0.9760941176470588, 0.9851686274509803, 0.9900627450980393}$-0.13$ & \cellcolor[rgb]{0.9107607843137255, 0.9495019607843138, 0.9703960784313725}$-0.556$ & \cellcolor[rgb]{0.8308235294117647, 0.9058627450980392, 0.9463333333333334}$-1.077$ & \cellcolor[rgb]{0.8169882352941177, 0.8983098039215687, 0.9421686274509804}$-1.164$ & \cellcolor[rgb]{0.9322823529411766, 0.961250980392157, 0.9768745098039215}$-0.414$\\
 & $R<a_{\rm sep}$ & \cellcolor[rgb]{0.8491019607843138, 0.3968666666666667, 0.3252627450980392}$4.805$ & \cellcolor[rgb]{0.9105843137254902, 0.6475254901960785, 0.6058745098039215}$2.797$ & \cellcolor[rgb]{0.9443843137254903, 0.7853254901960784, 0.7601411764705882}$1.69$ & \cellcolor[rgb]{0.9546313725490196, 0.8271019607843137, 0.8069098039215685}$1.357$ & \cellcolor[rgb]{0.9457607843137255, 0.7909372549019609, 0.7664235294117647}$1.644$\\
 & $R>a_{\rm sep}$ & \cellcolor[rgb]{0.23821176470588235, 0.5823450980392156, 0.7679450980392156}$-4.929$ & \cellcolor[rgb]{0.4810980392156863, 0.7149411764705882, 0.8410588235294117}$-3.348$ & \cellcolor[rgb]{0.5710274509803922, 0.764035294117647, 0.8681294117647058}$-2.763$ & \cellcolor[rgb]{0.6086901960784313, 0.7845960784313726, 0.8794666666666666}$-2.519$ & \cellcolor[rgb]{0.6801725490196079, 0.8236196078431373, 0.9009843137254901}$-2.055$\\
\hline
{q03e05i00} & all gas & \cellcolor[rgb]{0.8715607843137255, 0.9281019607843137, 0.9585960784313725}$-0.809$ & \cellcolor[rgb]{0.8231372549019608, 0.9016666666666666, 0.9440196078431373}$-1.127$ & \cellcolor[rgb]{0.8362039215686274, 0.9088, 0.9479529411764707}$-1.041$ & \cellcolor[rgb]{0.8062274509803922, 0.892435294117647, 0.9389294117647059}$-1.235$ & \cellcolor[rgb]{0.8077647058823529, 0.8932745098039216, 0.9393921568627451}$-1.226$\\
 & $R<a_{\rm sep}$ & \cellcolor[rgb]{0.8897843137254902, 0.5627254901960784, 0.5109411764705882}$3.475$ & \cellcolor[rgb]{0.913643137254902, 0.6599960784313725, 0.619835294117647}$2.697$ & \cellcolor[rgb]{0.9304666666666667, 0.7285843137254902, 0.6966196078431373}$2.144$ & \cellcolor[rgb]{0.9451490196078431, 0.788443137254902, 0.7636313725490196}$1.665$ & \cellcolor[rgb]{0.9547843137254902, 0.8277254901960784, 0.8076078431372549}$1.35$\\
 & $R>a_{\rm sep}$ & \cellcolor[rgb]{0.3404392156862745, 0.6381529411764706, 0.7987176470588235}$-4.266$ & \cellcolor[rgb]{0.40961568627450984, 0.6759176470588235, 0.8195411764705882}$-3.814$ & \cellcolor[rgb]{0.5072313725490196, 0.7292078431372548, 0.8489254901960784}$-3.181$ & \cellcolor[rgb]{0.5502745098039216, 0.7527058823529411, 0.8618823529411764}$-2.898$ & \cellcolor[rgb]{0.6002352941176471, 0.7799803921568627, 0.876921568627451}$-2.574$\\
\hline
{q10e05i00} & all gas & \cellcolor[rgb]{0.500313725490196, 0.7254313725490196, 0.8468431372549019}$-3.225$ & \cellcolor[rgb]{0.5018509803921568, 0.7262705882352941, 0.8473058823529411}$-3.216$ & \cellcolor[rgb]{0.5764078431372549, 0.7669725490196078, 0.8697490196078431}$-2.731$ & \cellcolor[rgb]{0.5956235294117647, 0.7774627450980391, 0.8755333333333333}$-2.604$ & \cellcolor[rgb]{0.5564235294117648, 0.7560627450980392, 0.8637333333333334}$-2.858$\\
 & $R<a_{\rm sep}$ & \cellcolor[rgb]{0.959678431372549, 0.8476784313725491, 0.8299450980392156}$1.191$ & \cellcolor[rgb]{0.9631960784313726, 0.8620196078431372, 0.8460000000000001}$1.075$ & \cellcolor[rgb]{0.9688549019607844, 0.8850901960784313, 0.871827450980392}$0.89$ & \cellcolor[rgb]{0.9709960784313726, 0.8938196078431373, 0.8815999999999999}$0.822$ & \cellcolor[rgb]{0.9777254901960785, 0.9212549019607843, 0.9123137254901961}$0.602$\\
 & $R>a_{\rm sep}$ & \cellcolor[rgb]{0.31814901960784314, 0.6259843137254901, 0.7920078431372548}$-4.41$ & \cellcolor[rgb]{0.3365960784313725, 0.6360549019607843, 0.7975607843137255}$-4.29$ & \cellcolor[rgb]{0.4395921568627451, 0.6922823529411765, 0.8285647058823529}$-3.62$ & \cellcolor[rgb]{0.4695686274509804, 0.7086470588235294, 0.8375882352941176}$-3.427$ & \cellcolor[rgb]{0.4641882352941177, 0.7057098039215686, 0.8359686274509803}$-3.46$\\
\hline\hline
\multicolumn{7}{c}{\bf Misaligned binaries} \\
\hline
{q01e00i45} & all gas & \cellcolor[rgb]{0.8554196078431372, 0.9192901960784314, 0.9537372549019607}$-0.913$ & \cellcolor[rgb]{0.9852196078431372, 0.951807843137255, 0.9465176470588235}$0.355$ & \cellcolor[rgb]{0.9730196078431372, 0.9834901960784314, 0.9891372549019608}$-0.149$ & \cellcolor[rgb]{0.8869333333333334, 0.9364941176470588, 0.9632235294117647}$-0.708$ & \cellcolor[rgb]{0.8692549019607843, 0.926843137254902, 0.9579019607843138}$-0.826$\\
 & $R<a_{\rm sep}$ & \cellcolor[rgb]{0.8495607843137255, 0.39873725490196077, 0.327356862745098}$4.792$ & \cellcolor[rgb]{0.8706666666666667, 0.4847843137254902, 0.4236862745098039}$4.102$ & \cellcolor[rgb]{0.912878431372549, 0.6568784313725491, 0.6163450980392157}$2.719$ & \cellcolor[rgb]{0.9410196078431372, 0.771607843137255, 0.7447843137254901}$1.798$ & \cellcolor[rgb]{0.9411725490196079, 0.7722313725490196, 0.7454823529411765}$1.797$\\
 & $R>a_{\rm sep}$ & \cellcolor[rgb]{0.22745098039215686, 0.5764705882352941, 0.7647058823529411}$-5.706$ & \cellcolor[rgb]{0.4203764705882353, 0.6817921568627451, 0.8227803921568627}$-3.747$ & \cellcolor[rgb]{0.5556549019607844, 0.755643137254902, 0.8635019607843137}$-2.867$ & \cellcolor[rgb]{0.6109960784313726, 0.7858549019607843, 0.8801607843137255}$-2.506$ & \cellcolor[rgb]{0.5925490196078431, 0.7757843137254902, 0.8746078431372548}$-2.623$\\
\hline
{q03e00i45} & all gas & \cellcolor[rgb]{0.835435294117647, 0.9083803921568627, 0.9477215686274509}$-1.045$ & \cellcolor[rgb]{0.943043137254902, 0.9671254901960784, 0.9801137254901962}$-0.347$ & \cellcolor[rgb]{0.9753254901960784, 0.9847490196078431, 0.9898313725490197}$-0.135$ & \cellcolor[rgb]{0.9899294117647058, 0.992721568627451, 0.9942274509803921}$-0.041$ & \cellcolor[rgb]{0.8677176470588235, 0.9260039215686274, 0.9574392156862745}$-0.834$\\
 & $R<a_{\rm sep}$ & \cellcolor[rgb]{0.8521607843137254, 0.40933725490196077, 0.3392235294117647}$4.705$ & \cellcolor[rgb]{0.8900901960784314, 0.5639725490196079, 0.5123372549019608}$3.463$ & \cellcolor[rgb]{0.9053843137254902, 0.6263254901960784, 0.5821411764705882}$2.963$ & \cellcolor[rgb]{0.8931490196078431, 0.5764431372549019, 0.5262980392156863}$3.367$ & \cellcolor[rgb]{0.9150196078431373, 0.665607843137255, 0.6261176470588234}$2.65$\\
 & $R>a_{\rm sep}$ & \cellcolor[rgb]{0.22745098039215686, 0.5764705882352941, 0.7647058823529411}$-5.753$ & \cellcolor[rgb]{0.4103843137254902, 0.6763372549019607, 0.8197725490196078}$-3.809$ & \cellcolor[rgb]{0.5202980392156863, 0.7363411764705883, 0.8528588235294118}$-3.097$ & \cellcolor[rgb]{0.4718745098039216, 0.7099058823529412, 0.8382823529411765}$-3.409$ & \cellcolor[rgb]{0.4603450980392157, 0.7036117647058824, 0.8348117647058824}$-3.484$\\
\hline
{q03e00i45mod} & all gas & \cellcolor[rgb]{0.9307450980392157, 0.9604117647058823, 0.9764117647058823}$-0.423$ & \cellcolor[rgb]{0.9230588235294118, 0.9562156862745098, 0.9740980392156863}$-0.477$ & \cellcolor[rgb]{0.8554196078431372, 0.9192901960784314, 0.9537372549019607}$-0.913$ & \cellcolor[rgb]{0.8208313725490196, 0.9004078431372549, 0.9433254901960785}$-1.14$ & \cellcolor[rgb]{0.917678431372549, 0.9532784313725491, 0.972478431372549}$-0.512$\\
 & $R<a_{\rm sep}$ & \cellcolor[rgb]{0.8723490196078432, 0.49164313725490194, 0.43136470588235293}$4.045$ & \cellcolor[rgb]{0.8463490196078431, 0.38564313725490196, 0.3126980392156863}$4.893$ & \cellcolor[rgb]{0.9342901960784313, 0.7441725490196078, 0.714070588235294}$2.018$ & \cellcolor[rgb]{0.9772666666666667, 0.9193843137254902, 0.9102196078431373}$0.615$ & \cellcolor[rgb]{0.9599843137254902, 0.8489254901960784, 0.8313411764705883}$1.181$\\
 & $R>a_{\rm sep}$ & \cellcolor[rgb]{0.30969411764705884, 0.6213686274509803, 0.7894627450980392}$-4.467$ & \cellcolor[rgb]{0.22745098039215686, 0.5764705882352941, 0.7647058823529411}$-5.37$ & \cellcolor[rgb]{0.5456627450980392, 0.7501882352941176, 0.8604941176470589}$-2.931$ & \cellcolor[rgb]{0.7262901960784314, 0.8487960784313726, 0.9148666666666667}$-1.755$ & \cellcolor[rgb]{0.7362823529411765, 0.8542509803921569, 0.9178745098039216}$-1.692$\\
\hline
{q10e00i45} & all gas & \cellcolor[rgb]{0.6778666666666666, 0.8223607843137255, 0.9002901960784313}$-2.069$ & \cellcolor[rgb]{0.7040000000000001, 0.8366274509803922, 0.908156862745098}$-1.902$ & \cellcolor[rgb]{0.6694117647058824, 0.8177450980392157, 0.8977450980392156}$-2.124$ & \cellcolor[rgb]{0.7654901960784314, 0.8701960784313725, 0.9266666666666666}$-1.499$ & \cellcolor[rgb]{0.8485019607843137, 0.9155137254901962, 0.9516549019607843}$-0.961$\\
 & $R<a_{\rm sep}$ & \cellcolor[rgb]{0.9186901960784314, 0.6805725490196078, 0.6428705882352941}$2.531$ & \cellcolor[rgb]{0.8997254901960784, 0.6032549019607844, 0.556313725490196}$3.149$ & \cellcolor[rgb]{0.8989607843137255, 0.6001372549019608, 0.5528235294117647}$3.177$ & \cellcolor[rgb]{0.8521607843137254, 0.40933725490196077, 0.3392235294117647}$4.706$ & \cellcolor[rgb]{0.8431372549019608, 0.37254901960784315, 0.2980392156862745}$5.182$\\
 & $R>a_{\rm sep}$ & \cellcolor[rgb]{0.28970980392156864, 0.6104588235294117, 0.7834470588235294}$-4.597$ & \cellcolor[rgb]{0.22745098039215686, 0.5764705882352941, 0.7647058823529411}$-5.05$ & \cellcolor[rgb]{0.22745098039215686, 0.5764705882352941, 0.7647058823529411}$-5.3$ & \cellcolor[rgb]{0.22745098039215686, 0.5764705882352941, 0.7647058823529411}$-6.204$ & \cellcolor[rgb]{0.22745098039215686, 0.5764705882352941, 0.7647058823529411}$-6.143$\\
\hline\hline
\multicolumn{7}{c}{\bf Retrograde binaries} \\
\hline
{q03e00i180} & all gas & \cellcolor[rgb]{0.6663372549019607, 0.8160666666666666, 0.8968196078431372}$-2.146$ & \cellcolor[rgb]{0.5748705882352941, 0.7661333333333333, 0.8692862745098039}$-2.74$ & \cellcolor[rgb]{0.5564235294117648, 0.7560627450980392, 0.8637333333333334}$-2.861$ & \cellcolor[rgb]{0.5825568627450981, 0.7703294117647059, 0.8715999999999999}$-2.688$ & \cellcolor[rgb]{0.6063843137254902, 0.7833372549019608, 0.8787725490196078}$-2.534$\\
 & $R<a_{\rm sep}$ & \cellcolor[rgb]{0.6724862745098039, 0.8194235294117647, 0.8986705882352941}$-2.107$ & \cellcolor[rgb]{0.5594980392156863, 0.7577411764705883, 0.8646588235294117}$-2.842$ & \cellcolor[rgb]{0.5402823529411764, 0.7472509803921568, 0.8588745098039215}$-2.963$ & \cellcolor[rgb]{0.5694901960784314, 0.7631960784313725, 0.8676666666666666}$-2.775$ & \cellcolor[rgb]{0.596392156862745, 0.7778823529411765, 0.8757647058823529}$-2.6$\\
 & $R>a_{\rm sep}$ & \cellcolor[rgb]{0.9906980392156863, 0.9931411764705882, 0.9944588235294118}$-0.033$ & \cellcolor[rgb]{0.9927137254901961, 0.9823607843137255, 0.9807215686274509}$0.108$ & \cellcolor[rgb]{0.9927137254901961, 0.9823607843137255, 0.9807215686274509}$0.108$ & \cellcolor[rgb]{0.9933254901960784, 0.9848549019607843, 0.983513725490196}$0.092$ & \cellcolor[rgb]{0.9939372549019608, 0.9873490196078432, 0.9863058823529411}$0.071$\\
\hline
{q10e00i180} & all gas & \cellcolor[rgb]{0.5825568627450981, 0.7703294117647059, 0.8715999999999999}$-2.689$ & \cellcolor[rgb]{0.4111529411764706, 0.6767568627450979, 0.8200039215686274}$-3.804$ & \cellcolor[rgb]{0.4311372549019608, 0.6876666666666666, 0.8260196078431372}$-3.676$ & \cellcolor[rgb]{0.5341333333333333, 0.7438941176470588, 0.8570235294117647}$-3.004$ & \cellcolor[rgb]{0.6878588235294117, 0.8278156862745099, 0.9032980392156862}$-2.007$\\
 & $R<a_{\rm sep}$ & \cellcolor[rgb]{0.6601882352941176, 0.8127098039215686, 0.8949686274509804}$-2.187$ & \cellcolor[rgb]{0.5226039215686274, 0.7376, 0.8535529411764706}$-3.08$ & \cellcolor[rgb]{0.5364392156862745, 0.7451529411764706, 0.8577176470588235}$-2.99$ & \cellcolor[rgb]{0.6217568627450981, 0.7917294117647059, 0.8834}$-2.433$ & \cellcolor[rgb]{0.7424313725490196, 0.8576078431372549, 0.9197254901960784}$-1.652$\\
 & $R>a_{\rm sep}$ & \cellcolor[rgb]{0.9192156862745098, 0.9541176470588235, 0.9729411764705882}$-0.502$ & \cellcolor[rgb]{0.8846274509803922, 0.935235294117647, 0.9625294117647059}$-0.724$ & \cellcolor[rgb]{0.8907764705882353, 0.9385921568627451, 0.9643803921568628}$-0.685$ & \cellcolor[rgb]{0.9084549019607844, 0.948243137254902, 0.9697019607843137}$-0.57$ & \cellcolor[rgb]{0.9415058823529412, 0.966286274509804, 0.9796509803921569}$-0.355$\\
\hline\hline
\end{tabular}
\flushleft
\end{table*}

To get a sense of the overall importance of the gravitational torques on the evolution of the binary we can compare the total gravitational torque exerted on the binary by the gas to the accretion torque (the effective torque experienced by the binary due to linear momentum conservation when accreting gas) and to the simulated rate of change of the binary angular momentum. The total gravitational torque acting on the binary is given by 
\begin{equation}
    {\boldsymbol\tau}_{\rm grav}=M_{1}({\bf r}_1\times{\bf a}_1)+M_{2}({\bf r}_2\times{\bf a}_2)\,,
\label{eq:total_grav_torque}
\end{equation}
where $M_i$, ${\bf r}_i$ and ${\bf a}_i$ are the mass, position and gravitational acceleration\footnote{In practice, ${\bf a}_i$ can be either the total gravitational acceleration or that only due to the gas, because for each black hole the acceleration due to the other black hole provides no contribution to the total torque in Equation~(\ref{eq:total_grav_torque}).} of each black hole particle (which includes the sub-grid $\alpha$ accretion disc) and $i=1$ or $2$ refer to the primary and secondary black hole, respectively. Additionally, the linear momentum of accreted material is added to each black hole during each accretion event, which results in a change in the $i^{th}$ black hole's angular momentum of
\begin{equation}
    \Delta{\bf J}_{{\rm acc}, i}=\Sigma_j\left({\bf r}_i\times m_j{\bf v}_j\right)\,,
\end{equation}
where $m_j$ and ${\bf v}_j$ are, respectively, the mass accreted from and velocity of the the $j^{th}$ cell and the sum is performed over all cells from which gas is accreted. In practise we do not log every accretion event explicitly and for the purpose of analysing the torques experienced by the binary we estimate it from the simulation output. To do this, we calculate the expected specific angular momentum contribution of accreted gas for each black hole, i.e., ${{\bf L}}_{{\rm gas}, i}=\Delta{\bf J}_{{\rm acc}, i}/\Sigma_j m_j$ from each snapshot and estimate the total change in angular momentum due to accretion as
\begin{equation}
    \Delta{\bf J}_{\rm acc}^{\rm est}=\Delta M_{1}\bar{{\bf L}}_{{\rm gas}, 1}+\Delta M_{2}\bar{{\bf L}}_{{\rm gas}, 2}\,,
\label{eq:torque}
\end{equation}
 where $\Delta M_i$ and ${\bar{\bf L}}_{\rm gas, i}$ are, respectively, the change in mass of the $i^{th}$ black hole particle and the specific angular momentum of accreted gas averaged over consecutive snapshots. We then define an effective accretion torque as ${\boldsymbol\tau}_{\rm accretion}^{\rm est}=\Delta{\bf J}_{\rm acc}^{\rm est}/\Delta t$, where $\Delta t$ is the time between snapshots. 

 Torques can act to change both the magnitude and direction of ${\bf J}_{\rm bin}$. To isolate the torque that acts only to change the magnitude of the binary angular momentum, which determines whether the binary shrinks or expands, we take only the component along the binary angular momentum vector. For in-plane binaries this is essentially the total torque, however, this is not the case for misaligned binaries (see Section~\ref{sec:misaligned}). We calculate averages for the relevant torque component over 500 binary orbits and show values found for each simulation by the coloured squares in Fig.~\ref{fig:ave_torques}. For the gravitational torques, we make two estimates, one in which the black hole accelerations are calculated via a pure tree method ($\tau_{\rm grav}^{\rm tree}$, shown in blue) and one in which the accelerations are calculated via direct summation over all gas cells ($\tau_{\rm grav}^{\rm direct}$, purple). This allows us to compare these methods with each other and to the simulated binary angular momentum evolution -- which in general shows reasonably consistent results. Estimated accretion torques, $\tau_{\rm accretion}^{\rm est}$, are shown by the pink squares, which we additionally add to the gravitational torque estimates and show as faded squares to illustrate the total expected torque in each case. Finally, by directly measuring the change in $|{\bf J}_{\rm bin}|$ from the simulations we can calculate the average total torque, $\tau_{\rm bin}^{\rm sim}$, experienced by the binary along its angular momentum vector, which is shown by the yellow squares. The vertical bars represent the one sigma scatter in orbit averaged values, i.e. how much the torques vary on an orbit-by-orbit basis. 

In the majority of cases, the gravitational torques measured using the tree versus direct summation are in reasonable agreement with each other and, at least for coplanar binaries, with torques derived from the binary evolution when taking estimated accretion torques into account. However, non-negligible differences between the tree and direct summation method are evident in some runs, particularly for misaligned binaries. In all cases, the mean torque experienced by the binary is closer to the value calculated using the tree approach. This suggests that for these runs the hybrid force calculation we employ does not perform so well at recovering the torque one would expect from a pure direct summation force calculation and would indicate that we may require a larger value of $N_{\rm direct}^{\rm max}$. This being said, it does result in less variation in the torques experienced by the binary that are in general more consistent with variations expected from direct summation calculations. For all binaries, we find that the gravitational torque estimates result in the same qualitative behaviour in providing negative torquing of the binary. 

Referring back to Fig.~\ref{fig:bbh_dyn} and the discussion in Section~\ref{sec:binary_evo}, apart from in the circular prograde systems, the $q=1/10$ mass ratio binaries shrink more rapidly compared to their higher mass ratio counterparts. From Fig.~\ref{fig:ave_torques} we can now see that this is because the $q=1/10$ binaries typically experience strong negative net torques leading to rapid extraction of their angular momentum. In the case of the q10e00i00 binary, the average simulated torque is likely under-estimated by a factor of $\sim 2$, and so like the other $q=1/10$ binaries, we would expect it to show more rapid shrinking than its higher mass ratio counter-parts (although still likely less rapidly than other $q=1/10$ systems). While we reserve a discussion of torques in the misaligned binaries to Section~\ref{sec:misaligned}, we note here that the faster shrinking of inclined binaries compared to their in-plane counterparts exhibited in Fig.~\ref{fig:bbh_dyn} is potentially due to over-estimating the gravitational torques acting on the binary, with the predicted behaviour of the q01e00i45 binary from this torque analysis to be that it gains angular momentum opposed to losing it. These small inconsistencies highlight the need to accurately calculate gravity forces on the fly for full self-consistent evolution of a live binary. We also note, however, that our estimates for the accretion torques may be over-smoothed as we only estimate their values between snapshots and may be missing episodes of low or high angular momentum gas accretion at intermediate times. Finally, for retrograde binaries, we find a negligibly small accretion torque. This is because the accretion prescription we employ, which requires net gas inflow onto the black hole, results in very little accretion onto the secondary (see Section~\ref{sec:bh_accretion}). If we were to use less conservative accretion criteria for which more accretion occurs onto the secondary black hole, we would expect it to accrete material with angular momentum opposite to that of the binary and hence lead to more rapid shrinking of the binary \citep[e.g.,][]{NixonEtAl11Retrograde}.

\subsubsection{Misaligned binary evolution}
\label{sec:misaligned}
\begin{figure*}
\begin{center}
\includegraphics[width=2\columnwidth]{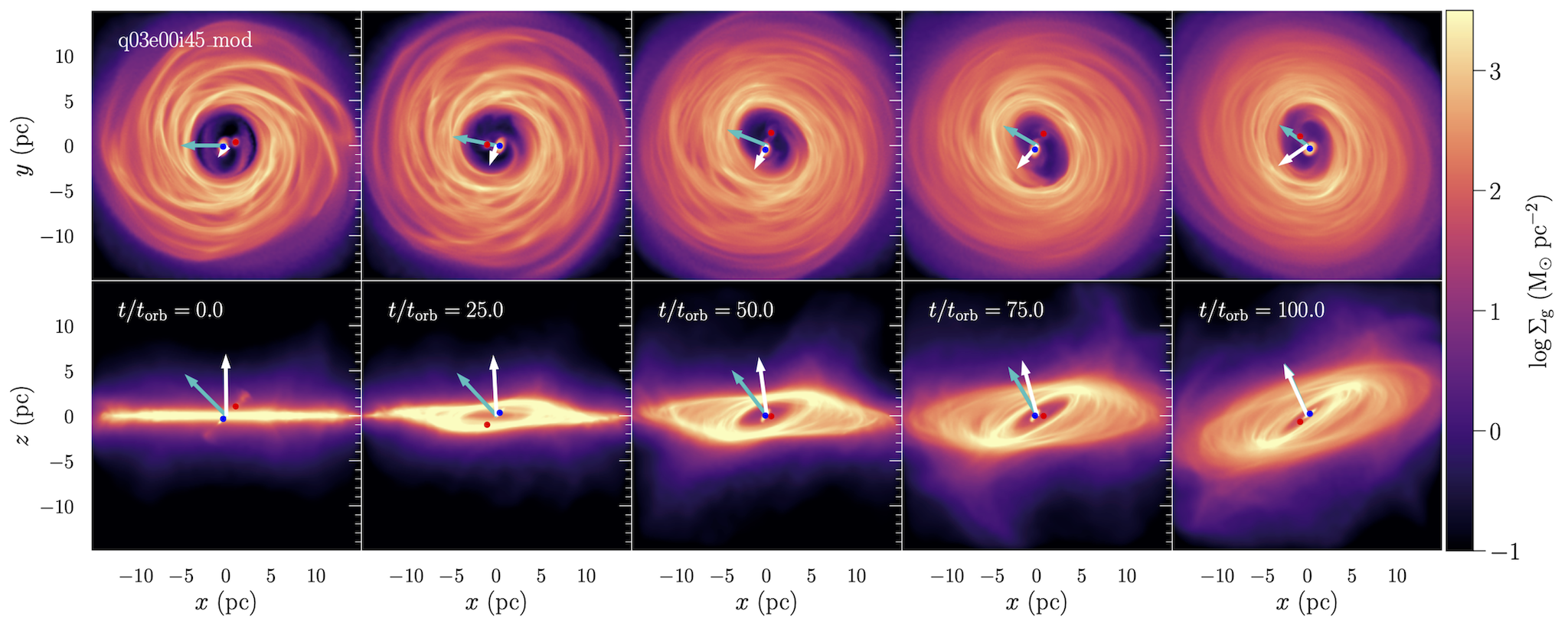}
\caption{Surface density maps for the q03e00i45mod binary illustrating the evolution of the CBD and black hole binary. The $x-y$ and $x-z$ projections are shown in the top and bottom rows,  respectively with the time of the image shown in the top left of each panel in the bottom row. The primary and secondary black holes are shown by the blue and red points, while the green and white arrows show the angular momentum vectors of the binary and CBD, respectively. As the system evolves, the binary and CBD torque each other such that their angular momenta move to align, resulting in the production of a warp in the CBD that propagates outwards with time.}
\label{fig:incl_time_evo}
\end{center}
\end{figure*}
So far our analysis has focused on binaries for which the angular momentum vector is aligned or anti-aligned with that of the CBD. In this section, we consider binaries that are initially inclined by $45\degr$ with respect to the CBD. Fig.~\ref{fig:incl_time_evo} shows evolving density projections for the q03e00i45mod run in the $x-y$ (top) and $x-z$ (bottom) planes. As is most clearly seen in the bottom row, the binary and gas torque each other in such a way as to align the angular momentum of the CBD and the binary, as shown by the white and green arrows, respectively. This proceeds by driving a warp through the CBD such that gas at the inner edge of the disc aligns first and gas at progressively larger radii aligns as the system evolves. This process has been explored in several previous works \citep[e.g.,][]{MillerKrolik13, NixonEtAl13, DunhillEtAl14, AlyEtAl15, MoodyEtAl19}, and is analogous to the Bardeen-Petterson effect that acts to align black holes and their accretion discs \citep{bardeen+75, king+05, NixonEtAl11Alignment}.

\begin{figure}
\begin{center}
\includegraphics[width=1\columnwidth]{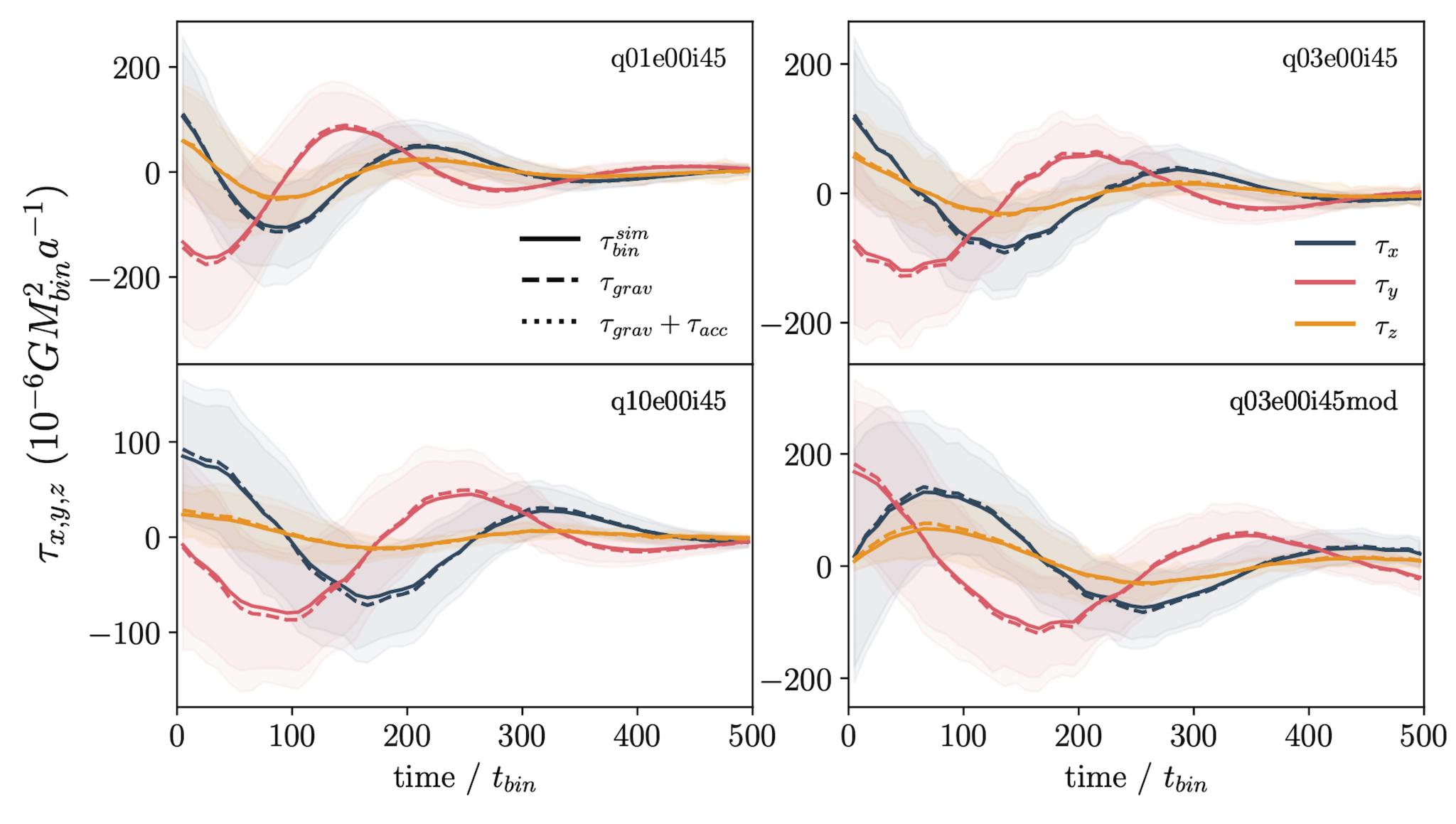}
\caption{Evolution of the $x$ (blue), $y$ (pink) and $z$ (yellow) components of the torque experienced by the binary in misaligned cases, as shown by the solid line. For comparison the post processed gravitational torque experienced by the binary estimated with a direct summation method is shown by the dash lines, while the estimated gravitational plus accretion torque is shown by the dotted lines. The actual and post-processed torques match very well and show that the torquing is dominated by the gravitational component. Initially large torques oscillate and damp as the binary and CBD angular momentum realign. The q03e00i45mod run shows a $\sim 100$~$t_{\rm bin}$ offset compared to the other misaligned binaries due to their misalignment being initialised after the relaxation.}
\label{fig:inc_torque_evo}
\end{center}
\end{figure}

The evolution of the torques acting on misaligned binaries is explored in Fig.~\ref{fig:inc_torque_evo}, which shows the evolution of the $x$, $y$ and $z$ components of the gas torque acting on the binary. We calculate both the torque experienced by the binary from the rate of change of the binary angular momentum ($\tau_{\rm bin}^{\rm sim}$, solid) and that calculated in post-processing using a direct summation method ($\tau_{\rm grav}$, dashed), as well as adding the estimated accretion torque ($\tau_{\rm grav} + \tau_{\rm acc}$, dotted, although this is essentially indistinguishable from the $\tau_{\rm grav}$ curve). The lines represent the torque averaged over $10$ orbits while the shaded regions indicate the one-$\sigma$ scatter. Each panel represents a different simulation, with all four initially misaligned binaries shown, as labelled. There are only very small differences in $\tau_{\rm bin}^{\rm sim}$ and $\tau_{\rm grav}$, showing that the instantaneous evolution of the binary angular momentum is well captured. Additionally, there is a negligible impact from accretion torques, showing that the global binary angular momentum evolution is dominated by the gravitational interaction with the gas. However, as we discuss below this is not necessarily the case for the magnitude of the angular momentum. The binaries initially experience large torques, with the amplitude of each component oscillating and damping as the binary and CBD align. As discussed in Section~\ref{sec:ics}, the runs in which the binary is initially misaligned undergo some realignment during the relaxation phase and to verify this has a limited impact on our results we perform an additional run, q03e00i45mod, in which we artificially misalign the cavity region of the q03e00i00 run to be at $45\degr$ to the CBD after the initial relaxation phase. Comparing this run to the q03e00i45 run, we see that they show similar behaviour albeit with a time offset. The offset roughly corresponds to $100$~orbits, i.e., the relaxation time, although this is not exact. 

\begin{figure}
\begin{center}
\includegraphics[width=1\columnwidth]{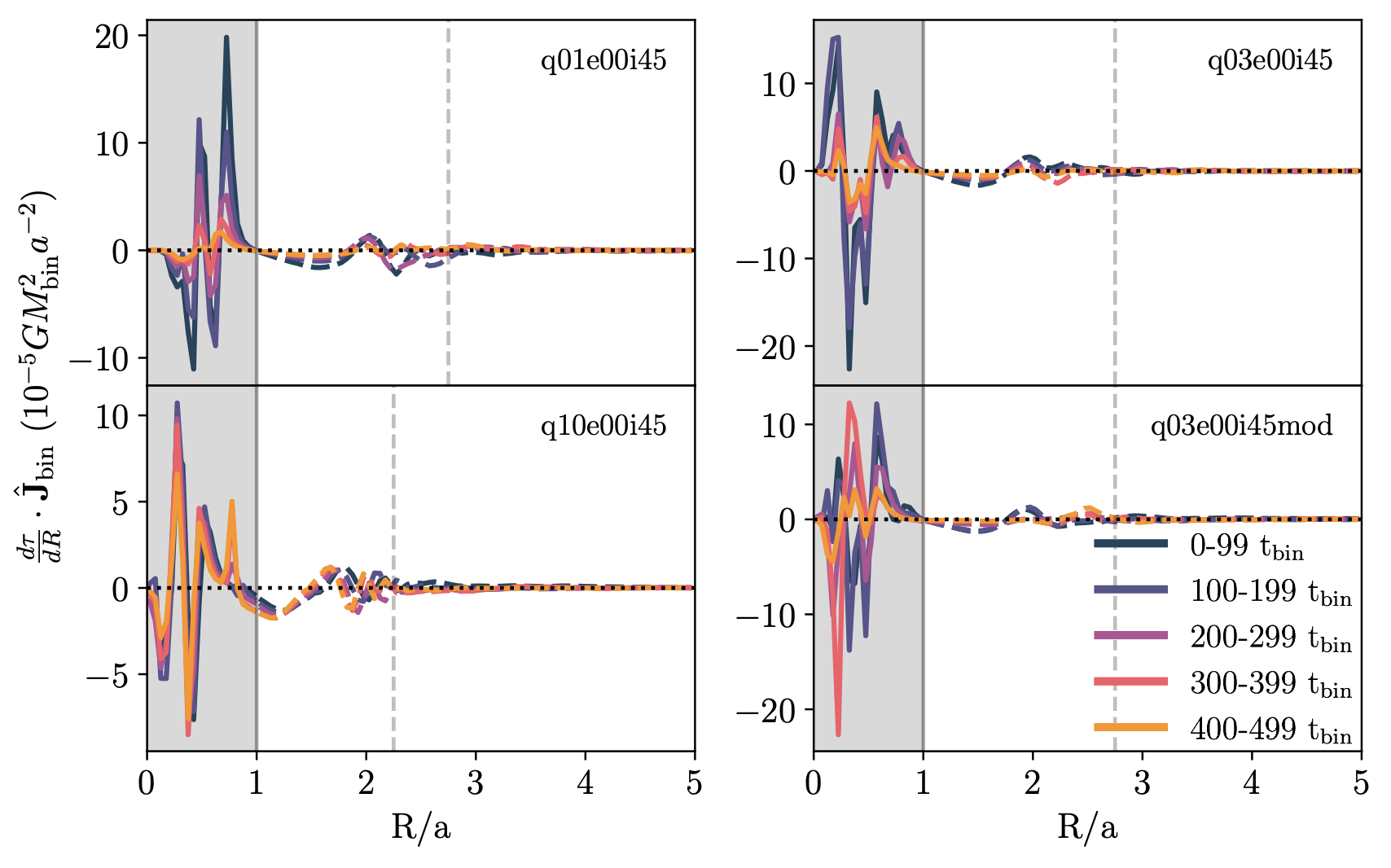}
\caption{As in Fig.~\ref{fig:torque_profiles} but showing the average radial profiles of the gas torque density, ${\rm d}{\boldsymbol \tau} / {\rm d}R$, acting along $\hat{\bf J}_{\rm bin}$ for all initially misaligned binaries. Each panel is labelled with the binary run name, the time period over which each profile is calculated is indicated by the colours given in the legend and as in Fig.~\ref{fig:torque_profiles}, solid and dashed lines represent regions that exert positive and negative net torques on the binary, respectively. Vertical dashed lines indicate the peak CBD density from Fig.~\ref{fig:cbd_profiles}. The profiles generally show similar features to the in-plane binaries, with material close to the binary typically exerting net positive torques, while material at larger radii exert net negative torques.}
\label{fig:inc_torque_profiles}
\end{center}
\end{figure}

Isolating the torques that solely change the magnitude of the binary angular momentum, Fig.~\ref{fig:inc_torque_profiles} shows, as in Fig.~\ref{fig:torque_profiles}, average radial profiles of the gas torque density, ${\rm d}{\boldsymbol \tau} / {\rm d}R$, acting along the direction of the binary angular momentum vector for the inclined runs. Each panel represents a different binary and colours illustrate different time ranges with each profile being produced by averaging over $100$ orbits$^{\ref{foot:orbit_ave}}$. The profiles show broadly similar trends to their in-plane circular counterparts with several sharp features from gas within the binary orbit ($R<a=2$~pc) and smoother peaks and troughs at larger radii. Additionally, as in the in-plane binaries, the gas at smaller radii provides a net positive torque to the binary, while gas at larger radii provides a net negative torque, which is also shown quantitatively in Table~\ref{tab:torques}. Comparing to the equivalent coplanar binaries, we expect from the direct summation calculations, that the misaligned $q=1$ and $q=1/3$ systems should experience stronger positive torques in the $R<a_{\rm sep}$ region and generally less negative gravitational torques overall. The misaligned binaries typically also experience mildly stronger positive accretion torques. For the q01e00i45 system in particular, combining the estimated accretion torque with the torque estimated from direct summation would lead to a net positive torque on the binary. This suggests that accretion torques could play a particularly important role in determining the binary evolution for inclined systems. However, as we outline in Section~\ref{sec:discussion}, even though the total torque acting on the q01e00i45 binary that we estimate in post-processing is positive, we still expect the binary to shrink.

We finally note that comparing the q03e00i45 and q03e00i45mod runs we find that the `modified' run experiences a stronger negative torque leading to its more rapid shrinking at later times as seen in Fig.~\ref{fig:bbh_dyn}. However, it is worth noting that the post-processed torques estimated from direct summation are much more similar between these two runs, and differences in the accretion torques that may lead to different binary evolution are likely an artefact of how the modified binary was created, as at least initially, it continues to experience somewhat coherent accretion from the mini-discs that were artificially realigned with the binary after relaxation, while the q03e00i45 run experiences a less coherent flow throughout its evolution (see Section~\ref{sec:bh_spin_evo} and Fig.~\ref{fig:spin_proj}). 

\begin{figure*}
\begin{center}
\includegraphics[width=2\columnwidth]{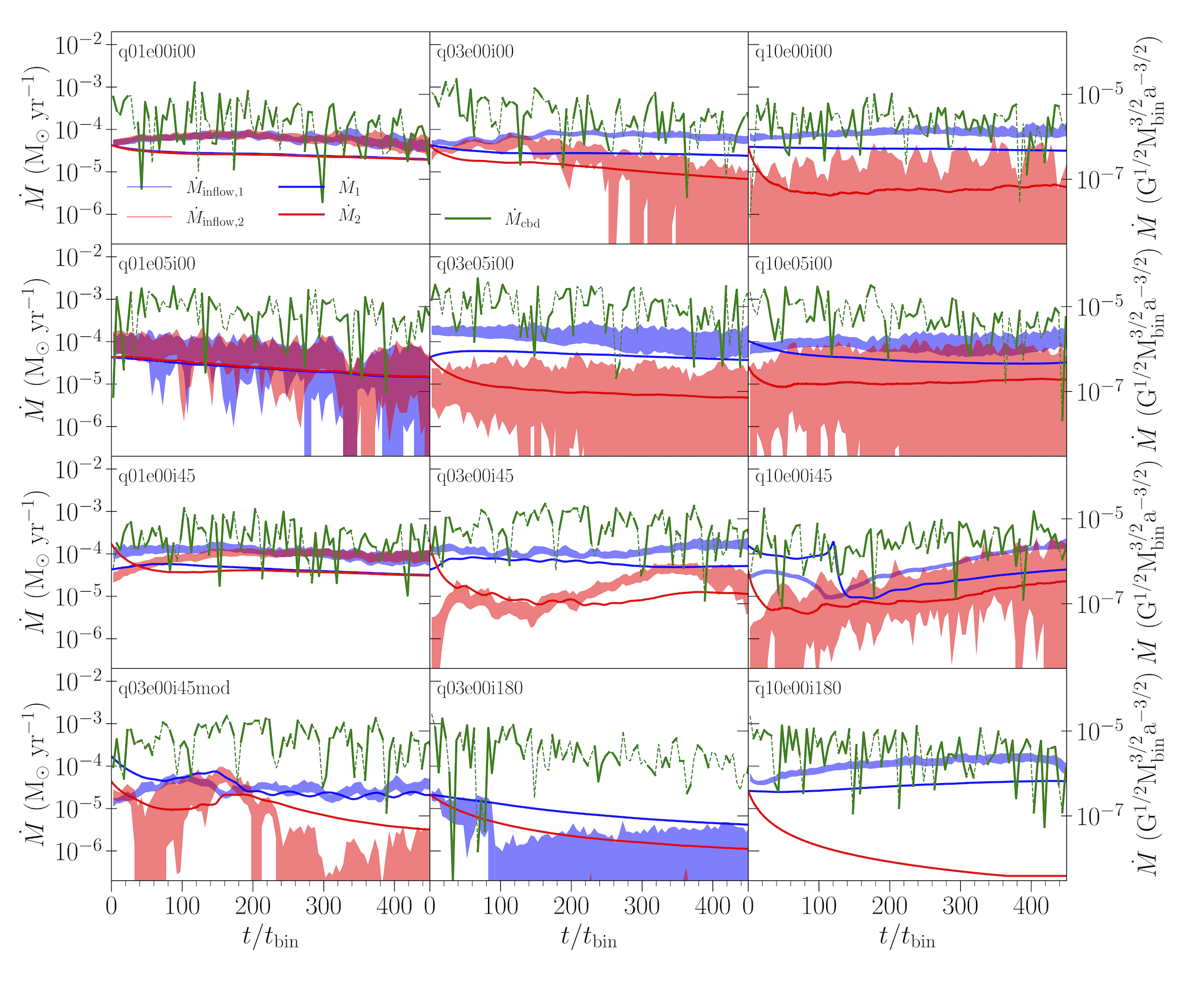}
\caption{Evolution of inflow and accretion rates onto each black hole. Specifically the primary and secondary black holes are shown in blue and red, respectively, with shaded regions indicating the range of inflow rates calculated onto the $\alpha$ accretion discs, while the solid (red and blue) lines show the accretion rate onto each black hole from the sub-grid $\alpha$ accretion disc. Additionally, inflow/outflow rates through the cavity are shown by the green solid/dashed lines, respectively. In general, primary black holes experience higher inflow rates and accretion rates. Inflow rates onto secondary black holes, as well as onto black holes in high eccentricity binaries, exhibit more fluctuations than inflows onto primary black holes and for circular binaries. Note that scale-free units provided on the right-hand side of the figure should only be applied to quantities within the simulation domain, i.e. inflow/outflow rates, and not the black hole accretion rates from the sub-grid accretion disc model.}
\label{fig:mdot_time}
\end{center}
\end{figure*}

\begin{figure*}
\begin{center}
\includegraphics[width=2\columnwidth]{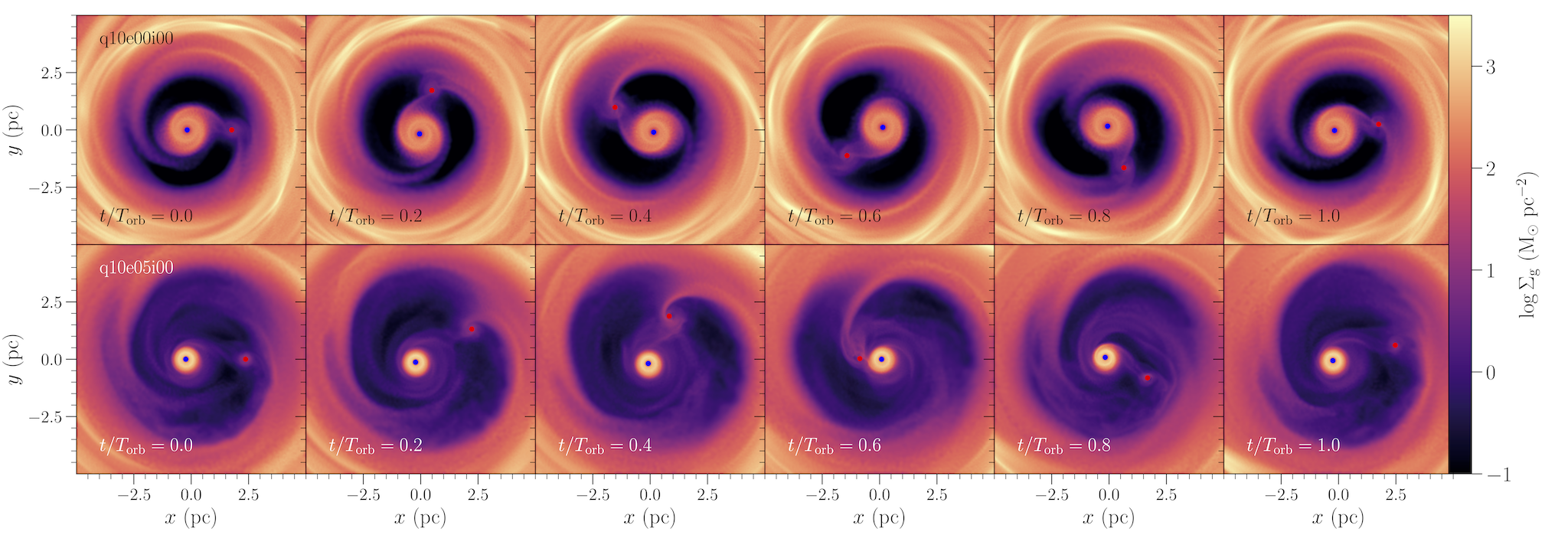}
\caption{Time sequence of gas surface density maps focusing on the cavity region for one representative orbit of an initially circular (q10e00i00, top row) and eccentric  (q10e05i00, bottom row) binary. The sequences show more erratic dynamics in the eccentric case, with triggering, development and destruction of gas streams reaching the secondary black hole first and then partially feeding the mini-disc surrounding the primary black hole.}
\label{fig:bbh_q10_ecc}
\end{center}
\end{figure*}

\subsection{Black hole and accretion disc evolution}
\label{sec:bh_evolution}

Having considered the evolution of the binary and its interaction with the CBD, in this section, we delve deeper into the evolution of the black holes and their sub-grid accretion discs. Specifically, we look at gas accretion onto the black holes and hence their mass as well as spin evolution. This is important as both the binary mass ratio, the magnitude and the orientation of the black hole spins have important consequences for gravitational wave emission and the remnant black hole properties, including its recoil velocity. 

\subsubsection{Accretion}
\label{sec:bh_accretion}

 Considering accretion onto the black holes, Fig.~\ref{fig:mdot_time} shows mass inflow rates onto each black hole particle by the blue (primary) and red (secondary) shaded regions. Specifically, these regions show the $5^{\rm th}$ to $95^{\rm th}$ percentile range of the inflow rates within time bins of length $5~t_{\rm bin}$. The corresponding solid blue and red lines show the accretion rate through the sub-grid accretion disc onto each black hole. While there are fluctuations in the inflow rates on to the accretion disc-black hole system, the rate at which material ultimately reaches a black hole is mediated by the sub-grid accretion disc (see Equation~(\ref{eq:fedd}) below for analytic form of Eddington rate through the accretion disc). This results in the black hole accretion rates being smoother, and at times lower, than the instantaneous inflow rates. The green line shows mass flow rates through the cavity, calculated at $R=2a=4$~pc for circular binaries and $R=3a=6$~pc for eccentric binaries. Solid lines show net inflow and dashed lines show net outflow, both averaged over  $\sim 5$ binary orbits. Mass flows through the inner edge of the CBD and cavity fluctuate between being inflowing and outflowing, with the net rate being roughly consistent with, although generally slightly lower than, the binary accretion rate. This is likely an effect of the relaxation period we impose during which accretion is prohibited and the mini-discs build up reservoirs from which the black holes subsequently accrete. While in our simulations the CBD provides a roughly constant supply of material to the cavity, leading to a somewhat steady-state, in reality black hole and/or stellar feedback, or a more dynamic environment, may change this scenario. As mentioned in Section~\ref{sec:ics}, quantities on resolved scales of the simulation can generally be presented in a scale-free manner, however, the sub-grid accretion disc model requires us to set a physical scale. As such, physical values applicable to our fiducial model parameters are shown on the left-hand axis of Fig.~\ref{fig:mdot_time} and can be applied to all quantities presented. To illustrate how quantities might change for different system parameters we additionally show scale-free quantities on the right-hand axis, however, strictly speaking these are only applicable to CBD inflow/outflow rates and the simulated inflow rates onto the black hole particles, but not the mass flow rates through the sub-grid accretion disc.
 
As discussed in Section~\ref{sec:bh_spin_model}, the sub-grid accretion disc model requires black hole accretion rates to be initialised and are set to the values provided in Table~\ref{tab:runs} such that  $\dot{M}_{{\rm acc}, i}=f_{\rm Edd,i}M_{\bullet,i}/\eta_{i}\tau_{\rm Salp}$, where $\tau_{\rm Salp}=450$~Myr is the Salpeter time, and the growth rate of the black hole is given as $\dot{M}_{\bullet,i}=(1-\eta)\dot{M}_{{\rm acc}, i}$. Subsequent evolution then depends on the mass and angular momentum evolution of the sub-grid accretion disc, which itself depends on the resolved gas inflows feeding the disc. In all cases, we find inflow rates onto the primary black hole to exceed those onto the secondary. Inflow rates onto the secondary black hole in a system often exhibit much larger fluctuations when compared to inflow rates onto the primary, which are typically smoother. Counter to our findings, a number of previous works have found that due to the secondary black hole typically inducing a stronger gas stream from the cavity it experiences preferential accretion over the primary \citep[e.g.,][]{RoedigEtAl2012, FarrisEtAl14, DuffellEtAl20, MunozEtAl20, DittmannRyan21, SiwekEtAl23Acc}. We expect that differences in our findings are related to the modelling of the gas viscosity and thermodynamics within the cavity. While we employ an adiabatic equation of state with $\beta-$cooling many of the  previous works  mentioned above employ a (locally-)isothermal equation of state instead. In addition the exact modelling of the black hole accretion may have an impact, with previous works making use of sink particle prescriptions opposed to the inflow rate calculation that we employ. We discuss these points and their implications for our results further in Section~\ref{sec:discussion_bhevo}, noting that several other previous studies also find preferential accretion onto the primary black hole \citep[e.g.,][]{OchiEtAl05, HanawaEtAl10, YoungEtAl2015}. In the most extreme case, inflow rates onto the retrograde binaries, particularly onto the secondaries can be greatly inhibited. This leads to declining black hole accretion rates as the sub-grid accretion disc mass is depleted. This is likely a result of our conservative accretion criterion of requiring net inflow onto the black hole and may have important implications for the evolution of retrograde binaries \citep[see e.g.,][]{NixonEtAl11Retrograde}.

Comparing the top two rows, inflow rates onto eccentric binaries show increased scatter due to the more complicated gas dynamics within the cavity compared to the circular binaries. This is illustrated in Fig.~\ref{fig:bbh_q10_ecc}, which shows a face-on gas density projection for the q10e00i00 and q10e05i00 binaries in the top and bottom rows, respectively. In the circular orbit case, the streams feeding both black holes and the primary mini-disc persist throughout the orbit. However, for the eccentric binary, the streams form and disperse, and although a mini-disc remains around the primary it is much less extended and receives new gas from streams more sporadically than in the circular binary case. The misaligned binaries also give rise to more complex gas dynamics in the cavity, however, as shown in the third row of Fig.~\ref{fig:mdot_time}, instead of leading to a general increase the scatter of inflow rates (compared to the in-plane circular binaries), exhibit more long time variations, particularly onto the secondary black holes. One feature to note is for the q10e00i45 run, in which the sub-grid accretion rate onto the primary experiences a slight spike followed by a sudden drop at $\sim120$~orbits. This is simply down to the nature of the equation determining the black hole accretion rate, $\dot{M}_{\rm acc}=f_{\rm Edd}M_{\bullet}/\eta\tau_{\rm Salp}$, where from \citet{fiacconi+18} the dimensionless Eddington rate is given as
\begin{multline}
\label{eq:fedd}
f_{\rm Edd}\approx0.76 \\
\times\left(\frac{\eta}{0.1}\right)\left(\frac{M_{\rm d}}{10^4 M_{\odot}}\right)^5\left(\frac{M_{\bullet}}{10^6 M_{\odot}}\right)^{-47/7}\left(\frac{a_{\bullet}J_{\rm d}/J_{\bullet}}{3}\right)^{-25/7}\,,
\end{multline}
where $\eta$ is the spin-dependant radiative efficiency and $\tau_{\rm Salp}$ is the Salpeter time. Variations in $f_{\rm Edd}$ are dominated by changes in the disc mass, $M_{\rm d}$ and angular momentum, $J_{\rm d}$, which both decrease during the first $\sim 120$~orbits as $\dot{M}_{\rm acc}>\dot{M}_{\rm in}$ and almost balance with each other. However, after a brief increase in the accretion rate the $M_{\rm d}$ decreases sufficiently to result in a sharp decline in the black hole accretion rate. After this $\dot{M}_{\rm acc}<\dot{M}_{\rm in}$ and the accretion rate stabilises for the remainder of the simulation. A similar although less dramatic process is seen after $\sim 140$~orbits in the q03e00i45mod run.

\subsubsection{Spin evolution}
\label{sec:bh_spin_evo}
\begin{figure}
\begin{center}
\includegraphics[width=0.92\columnwidth]{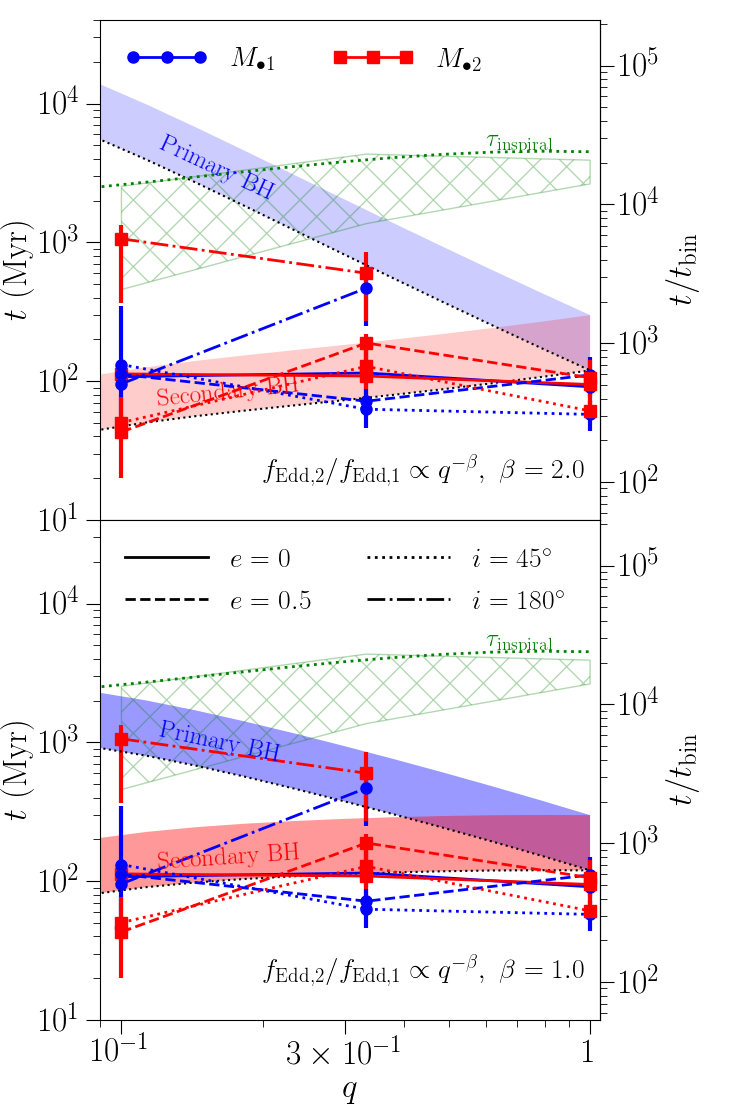}
\caption{Spin alignment and binary inspiral timescales extracted from the simulations and based on the analytical model of \citet{GerosaEtAl15}. Analytic predictions are illustrated by the shaded blue and red regions, which show spin-alignment timescales as a function of binary mass ratio for the primary and secondary black hole, respectively, and the dotted green line, which shows the binary inspiral time. Simulated black hole alignment time-scales (with associated scatter) are shown by the blue and red points for the primary and secondary black hole, respectively, with different runs indicated by the lines styles. The range of binary inspiral times found in the simulations are shown by the green hatched region. The top and bottom panels differ by the assumed accretion rate ratio between the primary and secondary black hole, $f_{\rm Edd,2}/f_{\rm Edd,1}=q^{-\beta}$, with $\beta=2.0$ and $1.0$, respectively. In general we find that black hole spins align with their accretion discs on timescales of $\lesssim$ the binary inspiral time and in the majority of cases much more quickly. The lack of preferential accretion onto secondary black holes means that spin alignment can also efficiently occur for the primary black hole in low mass ratio systems. Note that the scale-free units provided on the right-hand side of the figure are exact for the binary inspiral time and at best provide only an approximate guide for spin alignment timescales (see the main text for the necessary assumptions).}
\label{fig:spin_tal_q}
\end{center}
\end{figure}

As well as modelling black hole mass growth, the sub-grid black hole model allows us to study the evolution of the black hole spin and angular momentum of the sub-grid accretion disc. In particular, an understanding of black hole spin alignment is important to determine the recoil velocities of black hole merger remnants \citep{CampanelliEtAl07, GonzalezEtAl2007, LoustoEtAl2011, LoustoEtAl13, LoustoEtAl19, SperhakeEtAl20}. Specifically, relativistic numerical simulations predict that very high recoil velocities, up to $\sim 5000$~km~s$^{-1}$, can be achieved if the black hole spins are large and misaligned \citep{LoustoEtAl2011, LoustoEtAl13, LoustoEtAl19}. Such velocities would be able to eject black holes even from very massive galaxies, remove them from the pool of future black hole mergers and hence impact the GW background \citep{RajagopalRomani95, Phinney01, EnokiEtAl04, SesanaEtAl04}. On the other hand, if the black hole spins are both aligned with the binary angular momentum, much smaller recoil velocities are expected. Furthermore, black hole spin magnitude and orientation is fundamental if we want to study black hole-driven jet (and associated wind) feedback \citep[see e.g.,][]{tchekhovskoy+11, YuanNarayan14, liska+17, TalbotEtAl21, TalbotEtAl22}. 

\citet{GerosaEtAl15} developed an analytic model of black hole binary evolution in a CBD assuming the system is coplanar. Based on this model, the timescale on which the spin of each black hole would align with its accretion disc's angular momentum is given as
\begin{multline}
\tau_{\rm align}=3.4\frac{\tau_{\rm Salp}}{f_{\rm CBD}}\frac{\alpha_{\rm AD}}{F(q)}\left(\frac{1}{\alpha_{2}}\frac{H}{R}\right)^{2/3}\\
\times \begin{cases}
a_{\bullet, 1}^{2/3}\left(1+q^{\beta-1}\right)/\left(q^{\beta-1}+q^{\beta}\right) &\text{[Primary]}\\
a_{\bullet, 2}^{2/3}\left(q+q^{\beta}\right)/\left(1+q\right) &\text{[Secondary]}\,,
\end{cases}
\label{eq:align_gerosa}
\end{multline}
where $f_{\rm CBD}=\dot{M}_{\rm CBD}\tau_{\rm Salp}/M_{\rm bin}$ is the dimensionless Eddington rate of material feeding the cavity from the CBD, $\alpha_{\rm AD}=0.1$ is the assumed $\alpha-$viscocity of the accretion disc, $\alpha_{2}$ is the vertical $\alpha-$viscosity parameter given as \citep{Pringle92, Ogilvie99}
\begin{equation}
\alpha_2 = \frac{1}{\alpha}\frac{2\left(1+7\alpha^2\right)}{4+\alpha^2},
\end{equation}
$H/R$ is the accretion disc aspect ratio, which for this analysis we set to $0.01$, and $F(q)$ is the fraction of material flowing into the cavity that is accreted by the binary components and takes the form $F(q)=0.8054+0.984\log(q)+0.3818\log^2(q)$, where the numerical coefficients were found by \citet{GerosaEtAl15} as best fits to the simulation results of \citet{D'OrazioEtAl13}. Finally, $\beta$ determines how material accreted by the binary is distributed between the primary and secondary black holes such that $f_{\rm Edd,2}/f_{\rm Edd,1}=q^{-\beta}$, whereby values of $\beta>1$ imply preferential accretion onto the secondary black hole. \citet{GerosaEtAl15} compared the alignment timescales to the predicted binary inspiral time  due to interaction with a CBD which is given by
\begin{equation}
\tau_{\rm in}^{\rm CBD}=\frac{3}{4}\frac{\left(1+q\right)M_{\rm cbd}/M_{\rm bin} + q}{\left(1+q\right)^2}\frac{\tau_{\rm Salp}}{f_{\rm CBD}}.
\label{eq:t_in}
\end{equation}
The full derivation of this equation is given in \citet{GerosaEtAl15}, but in essence is an interpolation between the analytical predictions for binary inspiral times in three different mass regimes, either $M_{\bullet, 2}<<M_{\rm cbd}<<M_{\bullet, 1}$ \citep{LinPaploizou79, ArtymowiczLubow94, ArmitageNatarajan02}, $M_{\bullet, 2}\sim M_{\rm cbd}<<M_{\bullet, 1}$ \citep{SyerClarke95, IvanovEtAl99, LodatoEtAl09, BaruteauEtAl13} and $M_{\bullet, 2}\lesssim M_{\bullet, 1}$ \citep{Rafikov13}.

Applying the \citet{GerosaEtAl15} model to our setup parameters, in Fig.~\ref{fig:spin_tal_q}, we plot the predicted spin alignment timescales as a function of binary mass ratio, $q$, for the primary and secondary black hole in the blue and red shaded regions, respectively, while the green dotted line shows the predicted inspiral time. The alignment timescales are represented by shaded regions to illustrate the $\sim 2.5$ uncertainty in warp propagation theory. Similarly to Fig.~\ref{fig:mdot_time}, we show physical values applicable to our fiducial setup on the left-hand axis, and scale-free values on the right-hand axis. The latter can be freely used for the inspiral timescales, but only approximately apply to the alignment timescales under certain simplifying assumptions, which are discussed later in this section. The top panel shows the results of the analytic modelling assuming preferential accretion onto the seconday black hole, i.e., $\beta=2$, as originally assumed by \citet{GerosaEtAl15}. This model leads to inhibited accretion onto the primary for low $q$ binaries and results in spin alignment timescales for the primary black hole that exceed the inspiral time and hence spin alignment is not expected for these systems. This issue can be alleviated if a larger fraction of material feeding the binary ends up accreting on to the primary black hole. To illustrate this in the bottom panel we set $\beta=1$, implying equal accretion rates onto each black hole. In this case, alignment timescales reduce for the primary and increase for the secondary black hole, with all times being less than the inspiral time down to $q=1/10$.

We stress that behaviour discussed in the above paragraph arises from the analytic modelling of \citet{GerosaEtAl15}. We additionally plot alignment times taken directly from our sub-grid accretion disc model to illustrate expected behaviour from our simulated systems.  The accretion disc model of \citet{fiacconi+18} assumes that spin-alignment occurs on a timescale of
\begin{equation}
    \label{eq:alignment}
    \tau_{\rm align}\approx\frac{0.17}{\cos\left(\pi/7\right)}\left(\frac{M_{\bullet}}{10^{6}M_{\odot}}\right)^{-2/35}\left(\frac{f_{\rm Edd}}{\eta_{0.1}}\right)^{-32/35}a_{\bullet}^{5/7}{\rm Myr}\,.
\end{equation}
 Using this definition, median alignment timescales for primary and secondary black holes are shown by the blue circles and red squares, respectively, while different line styles indicate different binary eccentricities and inclination angles, as shown on the legend. Vertical bars on each point illustrate the $5^{th}$ to $95^{th}$ percentile range. We only include snapshots after the first $100$~orbits to limit any possible influence from the initial sub-grid parameter choices. The evolution of $\tau_{\rm align}$ given by Equation~(\ref{eq:alignment}) as well as the black hole spin--black hole spin, $\tau_{\rm bh}^{\rm bh}$, and black hole spin -- binary angular momentum, $\tau_{\rm bh}^{\rm bin}$, alignment timescales measured directly from the simulations are shown in Fig.~\ref{fig:spin_alignment_evo}. We note here that in the case of initially coplanar binaries both the analytic timescale and the timescales measured from the simulations generally match very well. However, in initially misaligned systems, non-negligable differences can occur between these different alignment timescale estimates, which we discuss further below and in more detail in Appendix~\ref{apndx:spin_alignment}. Finally, the green hatched regions indicates the range of inspiral times calculated from the simulation data over the same time period. The upper bound of the green hatched region follows the analytical expectation for inspiral times from Equation~(\ref{eq:t_in}), but in general, the simulations exhibit faster rates of binary shrinking notwithstanding that our CBD is in a quasi steady state. 

Several interesting features are clear from our simulations, firstly, while the alignment timescales for secondary black holes are close to the region predicted by the analytical model, the majority of primary black holes also have alignment timescales in this region and do not follow the blue track predicated by \citet{GerosaEtAl15}. This behaviour largely stems from differences in the black hole accretion rates in our simulated binaries and those assumed by \citet{GerosaEtAl15}, with the fact that our simulations do not exhibit preferential accretion onto the secondary black hole playing a key factor. We find that $\beta$ is always less than one and often close to zero, and a larger fraction of material entering the cavity accretes onto the primary black hole (and onto the binary in general) in low $q$ systems. These conditions result in more efficient spin-alignment for primary black holes even at low $q$ values. The only outliers from this general behaviour occur for the retrograde binaries, which, apart from the $q=1/10$ primary black hole, experience much less accretion than in other binary setups. As such these points are almost entirely determined by the initial parameters of the sub-grid accretion disc and their points could be seen as something of an upper bound. Had these black holes experienced similar levels of accretion as that observed in the other binaries then they would show much shorter alignment timescales. Barring these outliers, spin alignment occurs on timescales shorter than the inspiral timescale, often at rates $\lesssim 0.2\times\tau_{\rm in}$, meaning that for the parameter space studied here, and for our assumed accretion disc model (see Section~\ref{sec:discussion_future} for a discussion of caveats) we would expect spins to be aligned by the time black holes reach merger.

In this work, we have assumed that the binary lives in an environment where hardening due to stellar interactions can be neglected. However, if the binary were to reside in a massive stellar bulge or central nuclear star cluster, stellar scattering may provide an additional channel to shrink the binary \citep{Quinlan96, MilosavljeviMerritt01, MilosavljevicMerritt03, BerczikEtAl06, SesanaEtAl06, RantalaEtAl17}. Combining results from \citet{MilosavljevicMerritt03} with the $M-\sigma$ relation of \citet{FerrareseEtAl06}, \citet{CuadraEtAl2009} provide an approximate estimate for the timescale of binary shrinking due to stellar interactions as a function of binary mass and separation. It suggests that for our binary mass ($M_{\rm bin} = 2 \times 10^{6}$~M$_{\sun}$), on $\sim$~parsec scales stellar scattering could be more effective than a CBD at shrinking the binary, and would proceed on timescales shorter than spin alignment. However, this should be seen as an approximate guide, given the dependence on the chosen $M-\sigma$ relation and more generally on the exact stellar properties of the system. In any case, for sub-parsec separations as well as scenarios in which the binary does not reside in a stellar bulge or nuclear star cluster we expect that the CBD would dominate binary shrinking. In addition, given that $\tau_{\rm align}$ used in our sub-grid model is approximately inversely proportional to $f_{\rm Edd}$ and only weakly depends on M$_{\bullet}$ (see Equation~(\ref{eq:alignment})), one might expect the ratio of $\tau_{\rm align}/\tau_{\rm in}^{\rm CBD}$ to be roughly constant upon re-scaling of the system, i.e., for spin alignment to always occur on timescales shorter than binary shrinking due to the CBD. However, this behaviour should be seen as a qualitative guide and only approximately applies under the assumption that the mass flow rate through the sub-grid accretion disc follows the mass inflow rate from larger scales. As evident from Equation~(\ref{eq:fedd}), $f_{\rm Edd}$ depends on a range of factors that may change upon re-scaling the large scale simulation properties, not withstanding the fact that changes in accretion rates that would arise from such re-scaling could lead to different accretion disc structures that would alter the alignment timescales.

\begin{figure*}
\begin{center}
\includegraphics[width=2\columnwidth]{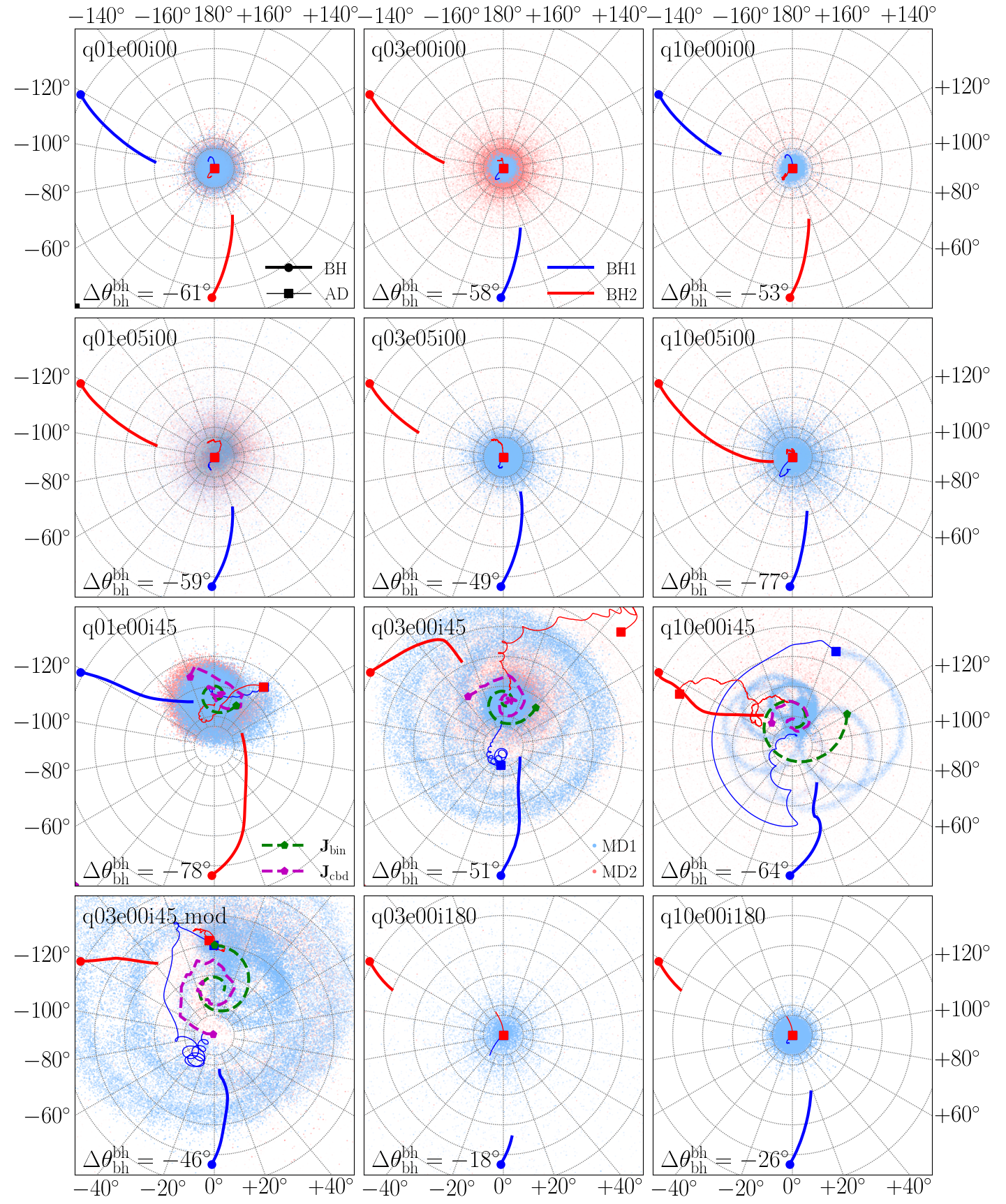}
\caption{Polar projections of the evolution of the black hole spins (thick solid lines initiated by circles) and sub-grid accretion disc angular momenta (thin solid lines initiated by squares), with quantities associated with the primary and secondary black holes are shown by blue and red, respectively. The angular momenta of mini-disc material is shown by the blue and pink points (around the primary and secondary black hole, respectively), with their transparency level scaling with the gas inflow rate onto the corresponding black hole. For misaligned runs, the binary and CBD angular momenta are additionally shown by thick dashed green and purple lines initiated by diamonds, respectively. The total change in the angle between the black hole spins, $\Delta\theta_{\rm bh}^{\rm bh}$, is shown in the bottom left of each panel. In general, coplanar runs clearly tend towards alignment between black hole spins and the global angular momentum vector, however, mini-discs in low $q$ misaligned systems do not align rapidly which may initially impact the global alignment in such systems.}
\label{fig:spin_proj}
\end{center}
\end{figure*}

In addition to the potential role of stellar scattering shrinking the binary, there is another caveat to the \citet{GerosaEtAl15} model and analysis we perform here -- namely that for this line of argument to work one has to assume that the whole system is coplanar. However, if this is not the case, while black hole spins may align with their accretion flow, there is no requirement, at least initially, for the accretion flows and hence the black hole spins to be aligned with each other. To explore this further, in Fig.~\ref{fig:spin_proj} we show the angular momentum evolution of different components of the system as polar projections. Quantities associated with the primary and secondary black holes are shown by blue and red, respectively, with spins shown by thick lines initiated by circles and accretion disc angular momenta by thin lines initiated by squares. We also show the angular momenta of accreted material, which we refer to as the mini-disc angular momenta, although note that in some instances a well defined mini-disc may not exist and instead this is simply representative of the gas local to the black hole. We only show mini-disc angular momenta at times when there is inflow onto the black hole as illustrated by faded blue and pink points for the primary and secondary black hole, respectively. The points have transparency levels that scale with the instantaneous inflow rate. The binary and CBD angular momenta are shown for misaligned binary runs by thick dashed lines initiated by diamonds. Each projection represents a different binary setup as labelled, and the total change in the angle between the black hole spins, $\Delta\theta_{\rm bh}^{\rm bh}$, is shown in the bottom left of each panel.

First, for all in-plane binary cases the accretion disc angular momenta generally remain aligned with the $z-$axis (i.e. at the pole), albeit with some small deviation and possible precession. The black hole spins, which are initially not aligned with this axis, progressively re-orientate towards the $z-$axis, and hence each other. Concerning the mini-disc angular momentum, there are a few behaviours to note. In the circular binary case, the mini-disc angular momenta are on average aligned with the $z-$axis, although the distributions of orientations in the case of the secondary black holes become broader for low mass ratios.  We find that larger deviations from $z-$axis alignment generally correspond to times in which inflow rates onto the sub-grid accretion disc are weaker, which fits with a picture in which the lower mass mini-discs around secondary black holes in low $q$ systems are more susceptible to disruption. Somewhat larger variations in the mini-disc angular momenta are seen in the eccentric binaries, for both the primary and secondary black holes, reflecting the dynamics of the cavity gas and episodic nature of the mini-discs as the binary moves between apo- and pericentre. In the retrograde binaries, there are clear inflows onto the primary black holes that are on average aligned with the $z-$axis, with more scatter in the $q=1/3$ case compared to the $q=1/10$ case, reflecting that the latter exhibits a long-lived coherent mini-disc. On the other hand, the $q=1/10$ binary shows no inflows onto the secondary black hole, and while in the $q=1/3$ binary there is some inflow, its angular momentum can vary over a wide range of angles, with no coherent or long-lived mini-disc or accretion flow forming. Barring the primary black hole in the $q=1/10$ case, the lack of coherent long-lived inflows in the retrograde binaries results in limited evolution of the  black hole spin orientations. 

In all of the coplanar cases, the nature of the systems means that the assumptions used for the analysis presented in Fig.~\ref{fig:spin_tal_q} are valid. Next, we consider cases in which the binary and CBD are initially misaligned. In these cases, we additionally show the binary and CBD angular momenta by the green and purple dashed lines, which evolve towards aligning with each other. In the $q=1$ case, the mini-disc angular momenta also tend to reorient towards a common axis with little scatter, with the accretion disc angular momentum approaching the same direction. The black hole spins also appear to move towards alignment, with any periods of growing misalignment between the black hole spins and binary angular momentum being short and occuring early in the evolution (see Fig.~\ref{fig:spin_alignment_evo}). This suggests that in the $q=1$ case, the analysis presented in Fig.~\ref{fig:spin_tal_q} should still hold if the system remains coplanar in the long term. 

For lower mass ratio binaries, while the binary and CBD angular momentum move towards alignment along a common axis, the mini-disc angular momentum varies significantly \citep[see also,][]{MoodyEtAl19}, even after the misalignment between the binary and CBD has reduced significantly. This is especially evident for mini-disc material around primary black holes, and leads to the accretion disc angular momenta also varying; sometimes exhibiting a precession-like motion. Considering the evolution of the black hole spin, we find that the misaligned accretion can result in deviations with respect to the initially coplanar systems. This is particularly evident, for example, for the secondary black holes in the q03e00i45 and q10e00i45 runs, as well as the primary black hole in the latter run. This being said, the overall trend is for the relative angle between the black hole spins to decrease in all binaries considered (see $\Delta\theta_{\rm bh}^{\rm bh}$ values on each panel of Fig.~\ref{fig:spin_proj}, as well as its evolution in Fig.~\ref{fig:spin_spin_theta}). We do, however, note that when considering the alignment evolution between black hole spin and binary angular momentum (see Appendix~\ref{apndx:spin_alignment} and Fig.~\ref{fig:spin_alignment_evo}), their relative motion leads to periods during which the spins become more misaligned, suggesting that, at least in the early stages, alignment between a black hole's spin and the binary can be inhibited. In the long term, if the system is able to effectively globally align to become coplanar and subsequently evolve unperturbed, spin alignment should be expected \citep{MillerKrolik13, GerosaEtAl15}. We stress, however, that we have studied only a small parameter space in initial black hole spins, CBD inclination angles and eccentricities here. It is also important to highlight that we have not considered regimes in which disc breaking/tearing occurs, and that the sub-grid accretion disc angular momentum only couples to the resolvable scales via accretion and is not impacted by the external tidal field of the gas and companion black hole, which could also effect the evolution of both the accretion disc and black hole spin \citep{GerosaEtAl20, NealonEtAl22}. We discuss these points further in Sections~\ref{sec:discussion_bhevo} and \ref{sec:discussion_gw} and note that we suggest that longer term simulations in a range of more realistic environments would be critical to fully understand the late time evolution of the black hole spin orientation in low-$q$, misaligned systems and its implications for GW emission and black hole recoil velocities. We additionally note that for two such systems (q10e00i45 and q03e00i45mod), the black hole spin and accretion disc are initially highly misaligned and in a retrograde configuration. The extent of any initial misalignment of the black hole spin could be an important factor in the alignment process that requires further exploration.

\section{Discussion}
\label{sec:discussion}

The simulation of binaries and their interaction with surrounding CBDs is a rich area of research \citep[see e.g.,][for a recent review]{LaiMunoz23}, however, the vast parameter space in terms of both the binary and CBD properties make developing a comprehensive understanding of this topic a formidable task. Many simulations of CBDs \citep[e.g.,][]{D'OrazioEtAl13, MirandaEtAl17, MoodyEtAl19, MunozEtAl19, MunozEtAl20, D'OrazioDuffel21, SiwekEtAl23Orbit} make two key assumptions: {\it i)} the gas can be treated as isothermal (or locally isothermal), meaning that heating due to compression or shocks is assumed to rapidly radiate away, and {\it ii)} the CBD self-gravity can be neglected. We instead consider the self-gravitating regime, which has only been studied in a handful of previous works \citep[e.g.,][]{CuadraEtAl2009, RoedigEtAl2011, RoedigEtAl2012, FranchiniEtAl2021}, and similarly employ an adiabatic equation of state with $\beta$~-cooling such that the CBD remains marginally Toomre stable. The CBD, which is prevented from fragmenting, forms spiral structures that act to transport angular momentum \citep[e.g.,][]{LodatoRice04, LodatoRice05, FranchiniEtAl2021}. We have performed a parameter study with respect to the binary mass ratio, $q$, eccentricity, $e$ and inclination angle, $i$, while fixing the initial total binary mass ($M_{\rm bin} = 2 \times 10^{6}$~M$_{\sun}$) and properties of the CBD. In this section, we discuss our results further and put them into the context of the wider literature. 

\subsection{Evolution of prograde binaries}
\label{sec:discussion_prograde_binaries}

One of the key questions to address is how the interaction between the binary and CBD alters the separation of the binary and whether or not CBDs can drive binaries from pc to mpc scales on which GW emission can bring the binary to coalescence. Specifically, both gravitational torques (Section~\ref{sec:torques}) and accretion onto the black holes (Section~\ref{sec:bh_accretion}) can modify the binary angular momentum. Gravitational torquing by a CBD has traditionally been invoked as a mechanism to extract angular momentum from a binary and drive it towards merger \citep[e.g.][]{Pringle91, ArtymowiczLubow94, ArtymowiczLubow96, IvanovEtAl99, GouldRix00, ArmitageNatarajan02, ArmitageNatarajan05, LodatoEtAl09}. In particular, OLRs within prograde CBDs provide a mechanism through which the binary and CBD can exchange angular momentum \citep{LyndenBellKalnajs72, GoldreichTremaine79, GoldreichTremaine80, LinPapaloizou86}. However, as we have shown, gas within the cavity, including streams and mini-discs, also torque the binary and provide material for accretion, which offers additional channels for binary evolution \citep[see also,][for previous examples of such processes]{MacFadyenEtAl08, CuadraEtAl2009, ShiEtAl12, RoedigEtAl2011, RoedigEtAl2012, D'OrazioEtAl13, FarrisEtAl14, MirandaEtAl17, MunozEtAl19, TiedeEtAl20, FranchiniEtAl22, DittmannRyan22, DittmannRyan23, SiwekEtAl23Orbit}.

\citet{RoedigEtAl2012} utilised high-resolution simulations to analyse the gravitational torques acting on the binary. Similar to our results, they found that gas within the cavity makes a telling contribution; gas at $R<a$ provides a net positive torque and gas at larger radii provides net negative torques. Similar results have also been found in multiple simulations that resolve mini-discs \citep[e.g.][]{MunozEtAl19, TangEtAl17, TiedeEtAl20, FranchiniEtAl22, SiwekEtAl23Orbit}. \citet{RoedigEtAl2012} attribute positive torque within $R<a$ to super-Keplerian inflows that bend in front of the black holes, while at $R>a$, there is a combination of torques facilitated by OLRs in the CBD and torques within the cavity that are purely kinematic in origin. This picture is consistent with the torques identified in our prograde binary systems, for example in Fig.~\ref{fig:combined_ave_maps} and Fig.~\ref{fig:torque_profiles}.

Accretion onto the binary can additionally provide effective torquing due to both changing the black hole masses and velocities \citep[e.g.,][]{RoedigEtAl2012, FranchiniEtAl2021, FranchiniEtAl22, ShiEtAl12, MunozEtAl19, TiedeEtAl20, SiwekEtAl23Orbit}. The anisotropic component of the accretion torque (arising from differences in velocity between the black hole and accreting material), is generally found to be small compared to the gravitational torque \citep{TangEtAl17, TiedeEtAl20, DuffellEtAl20, D'OrazioDuffel21}, although not necessarily negligible \citep{MunozEtAl19, SiwekEtAl23Orbit}. Accretion torques presented in this paper refer to the {\it total} change in binary angular momentum due to accretion, which is always positive and in the majority of cases subdominant compared to the gravitational torques given the low accretion rates. The overall evolution of the semi-major axis is given as \citep{RoedigEtAl2012}
\begin{equation}
\frac{\dot{a}}{a}=\frac{2\tau_{\rm bin}}{J_{\rm bin}}+\frac{2e}{1-e^2}\dot{e}-\frac{\dot{M}_{\rm bin}}{M_{\rm bin}}-\frac{2\dot{\mu}}{\mu}\,,
\label{eq:adot}
\end{equation}
which is found by differentiating Equation~(\ref{eq:Jbin}), rearranging and setting $\dot{J}_{\rm bin}=\tau_{\rm bin}$. Neglecting the eccentricity term and rearranging Equation~(\ref{eq:adot}), we can define the maximum torque under which an equal mass, circular binary shrinks,
\begin{multline}
\tau_{\rm bin} < \frac{3}{8}\dot{M}_{\rm bin}\sqrt{GM_{\rm bin}a},\\
\simeq 1.12\times\left(\frac{\dot{M}_{\rm bin}}{2\times 10^{-4}~{\rm M_{\odot}yr}^{-1}}\right)10^{-6}GM_{\rm bin}^{2}a^{-1}\,,
\label{eq:torque_limit}
\end{multline}
where we use $M_{1}=M_{2}=M_{\rm bin}/2$ and assume $\dot{M}_1 = \dot{M}_2 = \dot{M}_{\rm bin}/2$. As such, binaries experiencing a positive torque can still shrink provided $\dot{M}_{\rm bin}$ is sufficiently large \citep[see e.g.,][]{RoedigEtAl2012}. From our post-processed direct summation torque estimates, the q01e00i45 binary should experience a positive torque (see Section~\ref{sec:torques} and Fig.~\ref{fig:ave_torques}), that would increase $J_{\rm bin}$. However, given the binary's average accretion rate of $\sim 2\times 10^{-4}$~M$_{\odot}$yr$^{-1}$, Equation~(\ref{eq:torque_limit}) suggests that the limiting torque is roughly an order of magnitude greater than the average torque experienced by the q01e00i45 binary. Therefore we expect this binary to shrink. 

In all bar one case (the q01e00i45 binary) we expect the average {\it total} torque (i.e. gravitational plus accretion) experienced by binaries in our simulations to be negative, and that every binary should shrink. Previous works simulating high-mass, self-gravitating CBDs have similarly found binary shrinking \citep[e.g.,][]{CuadraEtAl2009, RoedigEtAl2012, RoedigSesana14, FranchiniEtAl2021}. However, this behaviour is not universal. Recent works, which typically neglect self-gravity and assume a (locally-)isothermal equation of state, have found that net torques acting on a binary are sufficiently positive to drive expansion \citep[e.g.,][]{MirandaEtAl17, TangEtAl17, MoodyEtAl19, MunozEtAl19, MunozEtAl20, DuffellEtAl20, HeathNixon20, TiedeEtAl20, D'OrazioDuffel21, SiwekEtAl23Orbit, WangEtAl23}. As highlighted in the review of \citet{LaiMunoz23}, circular binaries above a mass ratio of $q\sim 0.1-0.3$ expand, while for mass ratios below $q\sim 0.05-0.1$ they shrink \citep[e.g.,][]{DuffellEtAl20, MunozEtAl20, SiwekEtAl23Orbit}, although recent parameter studies suggest that such behaviours are sensitive to both the disc thickness and viscosity \citep{DittmannRyan22, DittmannRyan23}. Eccentric binaries have also been observed to expand for particular combinations of $e$ and $q$ \citep[e.g.][]{MunozEtAl19, D'OrazioDuffel21, SiwekEtAl23Orbit}, although \citet{D'OrazioDuffel21} notes that initially eccentric binaries ($e\gtrsim 0.1$) tend towards an equilibrium eccentricity ($e\simeq 0.4$) where the binaries are likely to shrink. Clearly, to elucidate this issue in future works it will be of crucial importance to explore both the roles of gas self-gravity and realistic thermodynamic modelling both within CBDs and critically the mini-discs.

 Simulations of massive CBDs have traditionally employed Lagrangian simulations leading to restricted resolution in the cavity. We address this issue using a super-Lagrangian refinement scheme. Similarly, \citet{FranchiniEtAl22} compared traditional ``Lagrangian'' simulations of a massive ($M_{\rm cbd}/M_{\rm bin}=0.1$) CBD to simulations employing ``hyper-Lagrangian'' refinement. They find that while positive torques from gas at $R\lesssim a$ can lead to mild binary expansion, this is sensitive to choices of disc temperature and thickness. Like many other simulations that find binary expansion, \citet{FranchiniEtAl22} assumed an isothermal equation of state and did not account for self-gravity. Varying the level of $\beta-$cooling, from $\beta=0$ (locally-isothermal) to $\beta=4$, \citet{WangEtAl23} found binary expansion for the majority of values explored except for $0.7\lesssim\beta\lesssim 0.9$. The simulations presented in our work, as well as others that regularly find binary shrinking \citep[e.g.][]{CuadraEtAl2009, RoedigEtAl2012, RoedigSesana14, FranchiniEtAl2021} instead employ higher values of $\beta\simeq 10$. More massive CBDs, such as those considered by these simulations, potentially drive more rapid initial shrinking of the binary due to stronger gravitational interactions \citep{FranchiniEtAl2021}. We also note that \citet{RoedigEtAl2012} found that using an isothermal equation of state in the cavity leads to significantly weaker gravitational torques compared to the adiabatic case, leading to slower evolution of the binary semi-major axis.

The viscous torque, which is proportional to $\alpha\left(H/R\right)^{2}$, determines how rapidly material is transported to OLR locations, with the expectation that thicker and more viscous discs promote resonant torquing and hence more effective shrinking of the binary \citep[see e.g. discussions in][]{HeathNixon20, FranchiniEtAl2021}. Despite our simulations having low intrinsic viscosity and equilibrium thickness ($H/R\sim 0.05$), we find that binaries shrink. Other recent works have found that thicker discs and/or higher viscosities may lead to binary expansion. For example, \citet{HeathNixon20} found that for isothermal CBDs, with fixed $\alpha-$viscosity, thick discs led to binary expansion due to accretion of high angular momentum material \citep[see also,][]{FranchiniEtAl22}. While by varying $H/R$ (or Mach number) at fixed kinematic viscosity, $\nu$, \citet{TiedeEtAl20} found expansion of binaries in thick discs and shrinking in thin discs driven by gravitational torques, \citet{DittmannRyan22} found that binary expansion is insensitive to the value of $\nu$ in thick discs; however, in thin discs binaries continue to expand for high values of $\nu$, but shrink ever more rapidly for decreasing values of $\nu$. Other works have instead varied $\alpha$ for a fixed $H/R$, finding only negligible differences in the accretion rate normalised torques \citep{MunozEtAl20, DuffellEtAl20}, although \citet{FranchiniEtAl22} found that the level of viscosity determined whether or not their binary entered a phase of expansion. 

\subsection{Retrograde and misaligned binaries}
\label{sec:discussion_retro_misaligned}

While many simulations model coplanar prograde systems, the  initial relative alignment between a CBD and binary could be arbitrary. Black hole growth via randomly orientated accretion events, such as chaotic accretion via many small clouds \citep{KingPringle06, KingPringle07}, means that the binary may interact with many misaligned accretion events and CBDs before eventually merging \citep{NixonEtAl13}. For misaligned binaries ($i=45\degr$), we find, as expected from analytical theory \citep{NixonEtAl11Alignment}, that the binary and CBD torque each other and eventually align via the outward propagation of a warp (see Fig.~\ref{fig:incl_time_evo}). The radial gravitational torque profiles (Fig.~\ref{fig:inc_torque_profiles}) are qualitatively similar to those of in-plane counterparts, although the $q=1$ and $q=1/3$ misaligned binaries experience overall weaker negative gravitational torques, which is a consequence of stronger positive net torques from gas at $R<a$ (see Table~\ref{tab:torques}), and in general experience stronger positive accretion torques. Simulations of a misaligned $q=1$ binary by \citet{MoodyEtAl19} exhibit similar general behaviours to ours in terms of the warping and binary-CBD alignment, although for both misaligned and in-plane systems they find binary separation growth. Similar to our work, they also find that the mini-disc orientations are variable in misaligned systems. We discuss the implications of such variability in the context of our simulations on spin alignment, recoil velocities and GW detection rates in Sections~\ref{sec:bh_spin_evo} and~\ref{sec:discussion_gw}.

While in our simulations, and those of \citet{MoodyEtAl19}, the disc warps, under certain conditions the disc may instead break or tear \citep{NixonEtAl13, AlyEtAl15}. \citet{NixonEtAl13} found that discs can break at a wide range of inclination angles. They provide analytical limits for the radius within which disc breaking is expected for both viscous and inviscid CBDs. Based on their viscous limit we might naively expect to see disc breaking in our simulations. However, there are several differences between our work and that of \citet{NixonEtAl13}, such as the gas equation of state (isothermal vs adiabatic with $\beta-$cooling), the inclusion of self-gravity that induces stabilising spiral arms, and the treatment of viscosity. The inviscid regime whereby the warp propagates as a wave \citep{papaloizou+83}, and may be more applicable to our low viscosity simulations, suggests a breaking radius within the cavity and hence we would not expect disc breaking. While all CBDs in our simulations end up aligning with the binary, \citet{NixonEtAl11Alignment} show that counter-alignment is also possible if $\cos i<-J_{\rm cbd}/2J_{\rm bin}$. \citet{AlyEtAl15} further found that alignment to stable polar orbits is possible around highly eccentric binaries. Both \citet{NixonEtAl13} and \citet{AlyEtAl15} find that disc tearing can lead to angular momentum cancellation and accretion of low angular momentum gas by the binary that could aid binary shrinking. 

\citet{DunhillEtAl14} studied the formation of CBDs from infalling turbulent gas clouds on both prograde and retrograde orbits, finding that misaligned discs can align or counter-align with the binary, with the retrograde CBDs driving more significant binary evolution. Retrograde CBDs, therefore, provide a promising avenue to drive binaries towards coalescence. Resonant torques, facilitated by OLRs in prograde CBDs, do not occur in retrograde systems \citep{NixonEtAl11Retrograde}. Therefore, while prograde CBDs are expected to move outwards, retrograde CBDs are not, with gas more readily filling the cavities \citep[e.g.,][]{NixonEtAl11Retrograde, RoedigSesana14, BankertEtAl15}. This behaviour is seen in our simulations (e.g., in Figs.~\ref{fig:cbd_maps}, \ref{fig:cbd_profiles} and \ref{fig:combined_ave_maps}), and while shrinking of $a$ appears to slow for prograde binaries, it continues to decrease quite rapidly for retrograde cases (see Fig.~\ref{fig:bbh_dyn}). We find that the evolution of retrograde binaries is dominated by gravitational torques originating from gas at $R<a$ (see Table~\ref{tab:torques}), which, unlike the prograde case, provide net negative torques with respect to the binary angular momentum. Similar results were found by \citet{TiedeDOrazio23}, who additionally found that while initially, circular binaries remain circular (as we also find, see Fig.~\ref{fig:bbh_dyn}), eccentric binaries become ever more eccentric. \citet{TiedeDOrazio23} additionally find that for $e\gtrsim 0.55$, retrograde Lindblad resonances can be driven in the CBD \citep[see also,][]{NixonLubow15}. An alternative mechanism to shrinking retrograde binaries via gravitational torques was shown by \citet{NixonEtAl11Retrograde} who instead find that accretion can drive binaries to coalescence. If accretion is efficient at apocentre but not pericentre, eccentricity growth can occur even for an initially circular binary. Negligible growth of the secondary black holes in our simulated retrograde binaries is likely due to a combination of differences in the gas thermodynamic modelling, the lack of a robust mini-disc around these black holes, and the requirement for net inflow in order to accrete gas. With respect to the latter, a less restrictive accretion criteria may lead to eccentricity growth similar to that seen in \citet{NixonEtAl11Retrograde}.

\subsection{Evolution of black hole properties}
\label{sec:discussion_bhevo}

The varied population of supermassive black holes that reside in the centres of galaxies \citep{FerrareseMerritt00, HaringRix04, GultekinEtAl09, KormendyHo13, McConnellMa13, SahuEtAl19}, including dwarfs \citep{ReinesEtAl13, ReinesEtAl20, MezcuaEtAl16, KavirajEtAl19, BirchallEtAl20}, gain their mass through a combination of accretion and merger events over cosmic time. Numerous simulations have been performed to study accretion onto binaries \citep[e.g.,][]{ShiEtAl12, MacFadyenEtAl08, D'OrazioEtAl13, FarrisEtAl14, MirandaEtAl17, DuffellEtAl20, MunozEtAl20, WangEtAl23, SiwekEtAl23Acc}. As highlighted by \citet{RoedigEtAl2012}, simulations that excise the binary region commonly assume a limiting case whereby all material inflowing at $R=a$ is accreted, however, they find a reduction in the accretion rate of $\sim 25\%$ as material is stirred and slingshotted back towards the CBD. While total black hole accretion rates in our simulations are roughly equal to the mass flow through the cavity, many exceed it slightly, which is likely a result of the second relaxation phase we employ, during which accretion is prohibited and the well-defined mini-discs form.

We show inflow rates onto black holes in Fig.~\ref{fig:mdot_time} and similarly to previous works we find that they vary on times scales of $\sim t_{\rm bin}$ \citep[e.g.,][]{MacFadyenEtAl08, CuadraEtAl2009, RoedigEtAl2012, MunozLai16, LaiMunoz23, WangEtAl23}. While these fluctuations, driven by the binary orbit and induced accretion streams, are common among simulations, some additionally show long-term variability on a timescale of $\sim 5t_{\rm bin}$ associated with the formation of a lump at the inner edge of the CBD \citep{ShiEtAl12, MacFadyenEtAl08, D'OrazioEtAl13, MunozLai16, MirandaEtAl17, WangEtAl23, LaiMunoz23}, although this is only expected when gas can cool rapidly \citet{WangEtAl23}.

 Many simulations find that accretion occurs preferentially onto the secondary black hole and drives the binary mass ratio towards unity \citep[e.g.,][]{RoedigEtAl2012, FarrisEtAl14, DuffellEtAl20, MunozEtAl20, DittmannRyan21, SiwekEtAl23Acc}. In our simulations we do not observe this behaviour, instead finding that accretion occurs preferentially onto the primary black hole, which has important implications for spin alignment and GW emission. A key difference in our work is that we employ an inflow-regulated accretion model (described in Section~\ref{sec:bh_spin_model}), while the vast majority of other simulations employ some form of sink particle prescription, whereby gas residing within a fixed sink radius is removed from the simulation domain and added to the black hole. This change in approach may affect the accretion rate onto both black holes, however,  we expect the impact of this difference to be small given that our model will still remain dependent on the amount of mass in the vicinity of each black hole. Hence, the thermodynamic and viscous modelling of the gas likely plays  the critical role in understanding the accretion process.  Our chosen cooling prescription means that heat generated via compression or shocks is not immediately radiated away. This is particularly evident at the outer edge of the cavity where the disc material puffs up (see Fig.~\ref{fig:cbd_profiles}). Interestingly, several previous studies of hot/thick discs have found preferential accretion onto the primary instead of the secondary \citep{OchiEtAl05, HanawaEtAl10, YoungEtAl2015}, with \citet{YoungEtAl2015} specifically showing that the amount of gas moving from the secondary to primary Roche lobe is dependant on gas temperature, with more gas reaching the primary in hotter systems. Additionally, \citet{RoedigEtAl2012} showed an adiabatic equation of state results in less well-defined mini-discs and streams compared to the isothermal case. They further found that while the accretion rate onto the secondary black hole is independent of the sink radius for an isothermal gas, it scales with the sink radius in the adiabatic case, suggesting that for small enough sink radii, preferential accretion may be reversed.  As well as the gas thermodynamics, the viscosity likely also plays an important role. \citet{DittmannRyan23} recently showed that whether a constant kinematic viscosity or an $\alpha$-viscosity is used affects the level of preferential accretion on to the secondary, while \citet{DuffellEtAl20} found that the ratio of $\dot{M}_{2}/\dot{M}_{1}$ tends towards unity at low viscosity.  While we reserve a detailed study to future work, in comparing these previous findings with the characteristics of our simulations, we conclude
 that the lack of preferential accretion on the secondary is driven by a combination of slow cooling, and hence ``hot'' discs/cavity, the low intrinsic viscosity in our simulations, and to some extent the specific mass accretion prescription we have adopted. An important open question that future studies need to address is the likely range of visco-thermo-dynamical states of mini-discs representative of the entire population of black hole binaries.
 
 If accretion occurs preferentially onto the secondary black hole, it would lead to parsec-separation binary AGN being primarily powered by the secondary. A potential piece of evidence countering this situation comes from observations of large-scale radio jets that exhibit precession on timescales consistent with geodetic precession driven by a low-mass binary companion \citep{KrauseEtAl19}. Given the powerful nature of these sources, it would seem likely that the jet originates from accretion onto the primary black hole, which tentatively disfavours preferential secondary accretion. However, further observations, for example with LOFAR, SKAO, or precursor facilities, as well future GW detections with LISA are needed to definitively solve this open question.

Finally, LISA will allow us to build our understanding of the black hole populations spins \citep{Lisa17, LISA23}, as such it is important to understand how black hole spins evolve during CBD phases of their evolution. The sub-grid accretion disc model of \citet{fiacconi+18} allows us to self-consistently evolve the black hole spin. Several similar models have been developed for use in galaxy formation simulations \citep[e.g.,][]{DuboisEtAl14, fiacconi+18, BustmanteSpringel19, CenciEtAl21, SalaEtAl21, HuskoEtAl22, MassonneauEtAl23}, and have been used to determine feedback properties \citep[e.g.,][]{DuboisEtAl14, BeckmannEtAl19, BeckmannEtAl22, TalbotEtAl21, TalbotEtAl22, TalbotEtAl23, HuskoEtAl22, BollatiEtAl23binary, BollatiEtAl23feedback}. Simulations of binary evolution in large ($\sim 100$~pc) circumnuclear discs have tracked spin evolution using similar sub-grid models to ours \citep{BollatiEtAl23binary}. However, while previous simulations of CBDs have tracked the instantaneous rotational angular momentum of the accretor \citep[e.g.,][]{MunozEtAl19, MunozEtAl20, DittmannRyan21}, to the best of our knowledge ours is the first work to simulate parsec scale binaries and CBDs in order to self-consistently evolve sub-grid accretion discs and incorporate physical spin-alignment timescales. This allows us to not only evolve the spin magnitude but also its alignment. In general, the spins of black holes in systems in which the binary and CBD are coplanar gradually align with the global angular momentum. The picture is less clear cut, at least during early evolution, in systems where the binary and CBD are initially misaligned because the mini-disc angular momentum can be highly variable \citep[see also,][]{MoodyEtAl19}. This can inhibit black hole spins from aligning with the binary angular momentum over long timescales, with several of our simulations undergoing non-negligible periods during which the black hole spin and binary angular momentum become further misaligned (see Fig.~\ref{fig:spin_alignment_evo}). This being said, despite some early deviations, the relative angle between black hole spins does reduce during the course of the simulations, with $\Delta\theta_{\rm bh}^{\rm bh}$ differing with respect to the coplanar cases by roughly $\pm30\%$. As such, if the system evolves unperturbed over a sufficiently long timescale such that the mini-discs are able to align with the global angular momentum of the system, then spin-alignment should be expected \citep{MillerKrolik13, GerosaEtAl15}. However, as noted previously, the study we present here covers a restricted range of initial spin, binary and CBD orientations/properties, and the sub-grid accretion disc model does not capture all external processes driving the angular momentum evolution of the accretion disc, in particular the effect of the external tidal field of the gas and companion black hole, which could directly impact spin alignment \citep[see e.g.,][]{GerosaEtAl20}. Additionally, very few simulation works have studied initially misaligned systems and it would be pertinent to explore the long term spin evolution in systems covering a wide range of initial inclination angles and CBD properties in future works, including in the regimes where tearing/breaking of the CBD \citep{NixonEtAl13, AlyEtAl15} and/or accretion discs \citep{NixonEtAl11Alignment, NixonEtAl12SpinTearing, GerosaEtAl20, NealonEtAl22} occurs, and for cases of polar alignment of the CBD about the binary \citep{AlyEtAl15}. Further the scenario in which the secondary black hole is embedded within the CBD could provide interesting insights into spin evolution, with recent studies of stellar mass black holes embedded in AGN accretion discs finding that turbulence can result in stochastic accretion and spin evolution \citep{ChenLin23}. As we discuss next, spin-alignment has important implications for merger remnants and GW detection, as well as feedback physics (e.g., efficiency and direction).

\subsection{Implications for black hole mergers and gravitational wave detection}
\label{sec:discussion_gw}

Once at parsec scales, the evolution of the binary depends upon the environmental properties \citep{LISA23}. In sufficiently gas-rich environments a CBD may form and effectively torque the binary. As discussed above, simulations have found that interactions between the binary and CBD can lead to either expansion or shrinking of the binary \citep{CuadraEtAl2009, RoedigEtAl2012, RoedigSesana14, MirandaEtAl17, TangEtAl17, MoodyEtAl19, MunozEtAl19, MunozEtAl20, DuffellEtAl20, HeathNixon20, TiedeEtAl20, D'OrazioDuffel21, FranchiniEtAl2021, FranchiniEtAl22, SiwekEtAl23Orbit, WangEtAl23}, which makes it non-trivial to determine the rate of GW detections. Based on our simulations we expect massive, self-gravitating CBDs to provide a plausible channel through which a binary can shrink. However, it is worth highlighting that the CBDs in reality may be more dynamic with significant scope to improve modelling of the gas thermodynamics and include the effects of star formation as well as stellar and black hole feedback. Additionally, to reach merger it will likely require numerous individual accretion events and we find that retrograde events are likely more effective than prograde events at shrinking a binary \citep[see also,][]{NixonEtAl11Retrograde, DunhillEtAl14, TiedeDOrazio23}. 

As discussed in Section~\ref{sec:discussion_bhevo}, not only does the interaction of the binary with a CBD alter its orbital properties and provide a channel through which to bring the black holes closer together but also impacts the individual black hole properties that shape the eventual merger remnant properties, GW emission and the stochastic GW background. The GW strain amplitude scales with the binary chirp mass,  
\begin{equation}
\label{eq:chirp_mass}
\mathcal{M}=\frac{\left(M_{\bullet, 1}M_{\bullet, 2}\right)^{3/5}}{\left(M_{\bullet, 1}+M_{\bullet, 2}\right)^{1/5}}=\left[\frac{q}{\left(1+q\right)^2}\right]^{3/5}M_{\rm bin}\,,
\end{equation}
which for a circular binary scales as $\mathcal{M}^{5/3}$ \citep{Thorne1987}. As such, for a fixed binary mass, equal mass binaries provide the highest amplitude signals. Accretion can play an important role in determining the amplitude of the GW background both in terms of simply increasing the binary mass and in setting the final mass ratio \citep[e.g.,][]{SesanaEtAl09PTA, KelleyEtAl17, SiwekEtAl20}.

 The alignment (or misalignment) of black hole spins can directly impact the GW emission \citep{VecchioEtAl04, KleinEtAl09, KleinEtAl13, LangEtAl11, GerosaEtAl16}. Additionally, high magnitude misaligned spins can result in merger remnants with large recoil velocities \citep{CampanelliEtAl07, GonzalezEtAl2007, LoustoEtAl2011, LoustoEtAl13, LoustoEtAl19, SperhakeEtAl20}. This scenario could lead to black holes being ejected from their host galaxy \citep{RedmountRees89, VolonteriEtAl10, GerosaSesana15}, thus preventing or severely reducing possible future mergers in the galactic core and hence reducing the number of GW events \citep{Sesana07, SesanaEtAl09Recoils, BlechaLoeb08}. Considering both the binary-CBD and spin alignment timescales, analytical estimates by \citet{MillerKrolik13} suggest that spin alignment should occur in all but extreme mass ratio binaries. When considering coplanar systems, \citet{GerosaEtAl15}, predict that preferential accretion onto secondary black holes inhibits spin-alignment in low mass ratio systems, although it can proceed for $q\gtrsim 0.2$ systems \citep[see also, ][]{GerosaEtAl20}. Our simulations present somewhat different behaviour. The lack of preferential secondary accretion in our simulations means that efficient spin alignment can occur even for low $q$ systems and we note that longer timescales found in retrograde systems are likely a result of our inflow accretion model and a different gas cooling prescription compared to the majority of the literature. Secondly, as discussed in the previous section, additional complexity arises for systems in which there is initial misalignment between the binary and CBD. However, as long as the system evolves unperturbed such that the binary, CBD and mini-discs come into alignment, we would also expect spin alignment to occur. Additionally, if the black holes co-exist in a large-scale post-merger coherent flow, such as a nuclear gas disc, their spins may align prior to the CBD phase \citep[e.g.][]{BogdanocivEtAl07, DottiEtAl10}. Once the binary is at $\lesssim$~parsec separation and a cavity forms, the subsequent spin evolution will depend on whether the binary experiences coherent or chaotic/randomly orientated accretion. The former would likely drive and maintain high spins and their alignment. On the other hand, chaotic/randomly orientated accretion would result in misaligned accretion events \citep{KingPringle06, KingPringle07, NixonEtAl13, DunhillEtAl14} for which we believe long-term spin-alignment requires further study. It is important to stress that high recoil velocities also require high black hole spins. While spin up could be achieved via long term coherent accretion onto the binary (which is expected to lead to spin alignment), chaotic accretion is expected to spin down black holes in the long term due to the higher specific angular momentum of retrograde events \citep{KingPringle06, KingHofmann08}. This being said, for the system studied here, in which we assume the binary forms from two already highly spinning black holes, the timescale for binary spin down could be longer than the binary inspiral time, especially in the case of low mass ratio and/or retrograde systems, meaning that the black holes could still have non-negligible spins at merger. LISA will be able to put important constraints on the population of black hole spins in merging systems and therefore shed light on black hole accretion physics \citep{LISA23}.

\subsection{Caveats and future work}
\label{sec:discussion_future}

There are some caveats and subsequent areas of future development that we briefly highlight here. Firstly in terms of the numerical methods, in this work, we use a hierarchical time integration \citep{SpringelEtAl2021}, including hybrid force calculations that employs direction summation at small radii and a tree method at larger radii. As shown in Section~\ref{sec:torques}, while this gives broadly consistent results compared to a pure direct summation calculation and maintains stable binary orbits, small differences in gravitational torques, however, would suggest that in future works direct summation between the binary components and the gas component should be used \citep[although as in other works, to ease computational expense a tree can still be used for the gas self-gravity, e.g.,][]{CuadraEtAl2009, RoedigEtAl2012}. The black hole accretion model also differs from previous works in calculating mass inflow rates onto the sub-grid accretion disc as opposed to using a sink particle approach. A future study is important to understand the extent to which differences in these approaches affect both the binary and black hole evolution, and would allow the development of more physically realistic and robust accretion prescriptions and place limits on the required resolution to achieve this.

The parameter space when modelling CBDs and binaries is vast and there are many avenues to consider when expanding on the presented simulations. Based on the above discussion, as well as extending the range of binary properties, a key area to probe is the thermal modelling of the gas.  The ratio between accretion rates onto the secondary and primary black holes is very likely sensitive to the exact thermal state of the gas and hence the employed cooling prescription (see discussion in Section~\ref{sec:discussion_bhevo}). Many processes govern the radiative heating and cooling of the gas around a binary, however, fully self-consistent simulations employing full radiative-transfer would be prohibitively expensive for wide parameter surveys. As such, variations in the $\beta-$cooling, such as those studied by \citet{WangEtAl23} would provide a tractable and insightful avenue. Employing $\beta-$cooling within self-gravitating CBDs can result in a system that self-regulates its temperature/thickness (as long as $\beta$ is not so small as to drive fragmentation) and as such other parameters that would be useful to investigate are the total disc mass and viscosity \citep[see e.g.,][]{FranchiniEtAl2021}, including the regime in which disc breaking occurs \citep{NixonEtAl13, AlyEtAl15}.  Furthermore, in the currently employed $\beta$-cooling prescription (see Equation~(\ref{eq:cool})) the positions of gas cells is always calculated with respect to the binary centre of mass. However, one could alternatively treat gas within the mini-discs separately by instead using the distance to the corresponding binary component. If using a universal $\beta$ value, this would reduce cooling times of gas close to the secondary black hole in low $q$ systems, although we note that the choice of $\beta$ need not be the same around each black hole and in the CBD. Feedback from accreting black holes may also impact the binary evolution \citep{delValleVolonteri18, BollatiEtAl23binary} as well as the properties of gas feeding the binary from larger scales \citep{DehnenKing13, Zubovas15, BourneSijacki17, FaberDehnen18, TalbotEtAl23}. In future, we aim to include feedback prescriptions coupled to the sub-grid accretion disc model \citep[][Koudmani et al., in prep]{TalbotEtAl21, TalbotEtAl22, TalbotEtAl23} to explore this further.

Finally, there are some caveats related to the sub-grid modelling of the black hole-accretion disc system. Firstly we assume that the sub-grid accretion disc is a thin \citet{shakura+73} disc. Such discs are expected to exist for high Eddington ratios, however, the low Eddington ratios seen in our simulations suggest that a thick or truncated disc may be more appropriate \citep{YuanNarayan14, KoudmaniEtAl23}. In order to achieve an analytically tractable solution, the sub-grid accretion disc model of \citet{fiacconi+18} was derived under the small-warp approximation \citep[e.g.,][]{martin+07, perego+09, DottiEtAl13, fiacconi+18}, however, this may not be applicable when the black hole spin and accretion disc are initially highly misaligned. Additionally, depending on the disc properties, instead of attaining a steady warp the accretion disc may break and result in either reduced or enhanced rates of accretion \citep{NixonEtAl11Alignment, NixonEtAl12SpinTearing, GerosaEtAl20, NealonEtAl22}. Given that we resolve gas down to the scale of the sub-grid accretion disc, the mini-discs that we simulate should potentially also experience torquing by the spin of their host black hole. However, no such back reaction from the sub-grid model onto the resolved gas is included in our simulations and may affect the mini-disc behaviour, especially in the case of misaligned binaries. Finally, in a binary, each accretion disc is perturbed by the other black hole \citep{MartinEtAl09, TremaineDavis14, MillerKrolik13}, which can impact the spin alignment \citep{MillerKrolik13, GerosaEtAl20} and whether or not the accretion disc breaks \citep{NealonEtAl22}.

\section{Conclusions}
\label{sec:conclusion}

We have performed high-resolution simulations of $\sim $ parsec scale separation black hole binaries surrounded by gas-rich gravito-turbulent CBDs ($M_{\rm cbd}/M_{\rm bin}=0.1$; $M_{\rm bin} = 2\times 10^{6}$~M$_{\odot}$) including the effects of self-gravity. Employing a super-Lagrangian refinement techniques allows us to effectively resolve the streams and mini-discs that form within the cavity. By using a sub-grid model that mediates the black hole mass and spin evolution via a thin $\alpha-$disc prescription we have, for the first time, been able to self-consistently track the spin evolution of the individual black holes in such CBD simulations and to study their alignment. The full simulation suite includes twelve setups varying the initial binary mass ratio ($q=1$, $1/3$ or $1/10$), eccentricity ($e=0$ or $0.5$) and inclination angle ($i=0\degr$, $45\degr$ or $180\degr$). Our main findings are summarised as follows:
\begin{itemize}
    \item The evolution of the CBD, cavity, streams and mini-discs depends sensitively on the binary properties (Section~\ref{sec:cbd}). Higher mass ratio and eccentric binaries result in larger cavities, while retrograde binaries result in smaller cavities. Mini-disc sizes typically scale with the black hole mass (roughly filling the Roche lobe for prograde circular binaries), although eccentric binaries have truncated mini-discs.
    \item We find that the net torque experienced by binaries drives them towards merger (Sections~\ref{sec:binary_evo}), with low mass ratio binaries and retrograde binaries expected to shrink more rapidly. 
    \item For prograde binaries, the net gravitational torques experienced by the binary depend on the balance of net positive torques from gas at $R\lesssim a$ and net negative torques from gas at $R\gtrsim a$. For retrograde binaries, the net gravitational torque is dominated by negative torques from gas a $R\lesssim a$. Accretion torques are always positive and generally sub-dominant.
    \item Unlike many previous works, accretion in our simulations does not proceed preferentially onto the secondary black hole (Section~\ref{sec:bh_accretion}). This implies binaries to generally maintain a constant mass ratio up to merger, which for low mass ratio systems would result in smaller chirp masses, lower GW strain amplitudes and a weaker GW background. 
    \item We attribute the lack of preferential accretion onto the secondary black hole to a combination of using an adiabatic equation of state with slow $\beta-$cooling, low intrinsic viscosity and possibly due to differences between our inflow rate accretion model and the traditional sink particle model often employed in the literature. 
    \item Alignment of black hole spins and their accretion discs occurs on timescales of $\lesssim$ the binary inspiral time, with the majority of alignment times being much shorter (Section~\ref{sec:bh_spin_evo}). The lack of preferential secondary accretion means this is true even for the primary black hole in low mass ratio systems.
    \item In coplanar systems we expect black hole spins to align with each other and the global angular momentum of the system prior to merging. Additionally, while the mini-discs, which feed the black hole accretion discs that torque the black hole spin, do not align on short timescales, especially for low mass ratio systems, black hole spins do appear to align with each other in the long term. However, we caution that wider parameter space studies, the effects of external tidal fields on the sub-grid accretion disc, and scenarios including disc breaking/tearing are necessary to fully address the question of spin alignment efficiency in CBDs. Further, in galaxies with bulges or central nuclear star clusters, stellar scattering may speed up binary shrinking on parsec scales, meaning that spin alignment is only likely to be efficient on sub-parsec scales. An understanding of the regimes in which the spins of merging black holes align with each other and the binary angular momentum (or not) is critical in determining expected black hole recoil velocities, galaxy occupation fractions and GW event detection rates. 
\end{itemize}

\section*{Acknowledgements}
We thank the referee for constructive comments on our original manuscript. We additionally thank Alex Dittmann, Pedro Capelo, Martin Krause, Jim Pringle and Chris Nixon for useful discussions, feedback and comments. M.A.B. acknowledges support from a UKRI Stephen Hawking Fellowship (EP/X04257X/1). M.A.B., D.F. and D.S. acknowledge support by European Research Council Starting Grant 638707 ``Black holes and their host galaxies: coevolution across cosmic time''. M.A.B and D.S. additionally acknowledge support from the Science and Technology Facilities Council (STFC). The simulations were performed on the DiRAC Darwin Supercomputer hosted by the University of Cambridge High Performance Computing Service (http://www.hpc.cam.ac.uk/), provided by Dell Inc. using Strategic Research Infrastructure Funding from the Higher Education Funding Council for England and funding from the Science and Technology Facilities Council; the COSMA Data Centric system at Durham University, operated by the Institute for Computational Cosmology on behalf of the STFC DiRAC HPC Facility.This equipment was funded by a BIS National E-infrastructure capital grant ST/K00042X/1, STFC capital grant ST/K00087X/1, DiRAC Operations grant ST/K003267/1 and Durham University.

\section*{Data availability}
The data underlying this article will be shared on reasonable request to the corresponding author.

\bibliographystyle{mnras}
\bibliography{cbdisc}

\appendix

\section{Spin alignment timescales}
\label{apndx:spin_alignment}
\begin{figure*}
\begin{center}
\includegraphics[width=1.55\columnwidth]{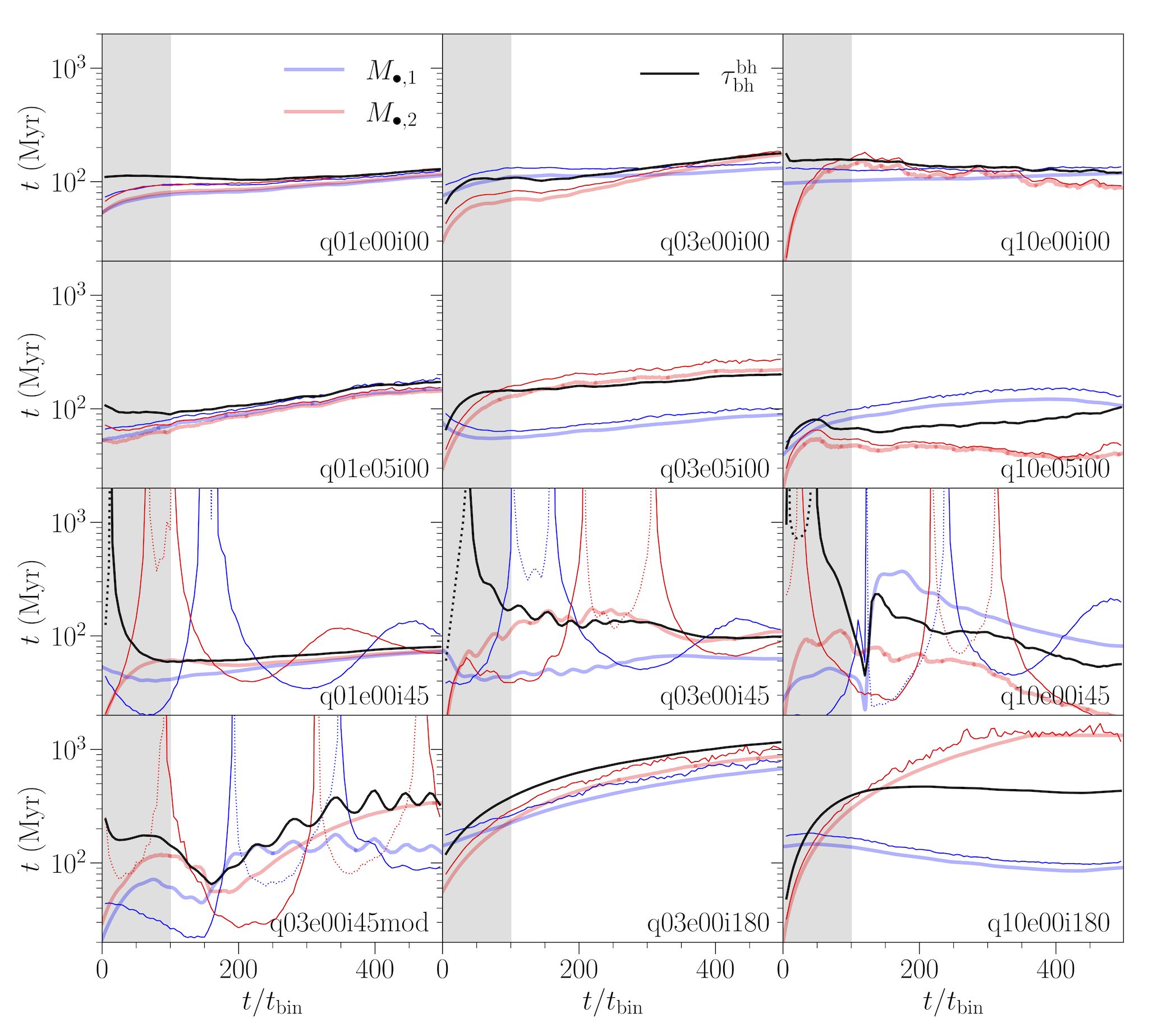}
\caption{Evolution of the black hole spin alignment timescales. Each panel represents a different binary, with the grey shaded regions illustrating the first $\sim 100$ binary orbits that are excluded from the mean timescale estimates used in Fig.~\ref{fig:spin_tal_q}. Blue and red lines represent the primary and secondary black hole in each system, respectively. Thick lines show the alignment timescale assumed by the \citet{fiacconi+18} model as given in Equation~(\ref{eq:alignment}), while thin lines show the alignment timescale between the black hole spin and binary angular momentum calculated directly from the simulation and given by $\tau_{\rm bh}^{\rm bin}=\theta_{\rm bh}^{\rm bin}/|\dot{\theta}_{\rm bh}^{\rm bin}|$, where theta is the relevant alignment angle. The timescale on which the black hole spins align with each other $\tau_{\rm bh}^{\rm bh}=\theta_{\rm bh}^{\rm bh}/|\dot{\theta}_{\rm bh}^{\rm bh}|$ is additionally shown by the black line. In all cases solid and dotted lines indicate when $\dot{\theta}$ is negative or positive, i.e. when the system is aligning or misaligning, respectively. For coplanar systems, the analytic alignment timescale between the black hole spin and its accretion disc matches well to the simulated alignment timescale between the black hole spin and binary angular momentum, as one would expect. However, it is clear that differences arise between these timescales in the case of systems that are initially misaligned, indicating that global alignment could initially be inhibited in such systems.}
\label{fig:spin_alignment_evo}
\end{center}
\end{figure*}

The spin alignment timescales presented in Fig.~\ref{fig:spin_tal_q} and discussed in section~\ref{sec:bh_spin_evo} represent the median expected alignment times between the black hole spin and its accretion disc angular momentum over $\sim 400$ binary orbits. In Fig.~\ref{fig:spin_alignment_evo} we instead show the evolution of the spin alignment timescale given by Equation~\ref{eq:alignment} for the primary (blue) and secondary (red) black hole for each run by the thick lines in each panel. However, the alignment timescale between the black hole spins and binary angular momentum measured directly from the simulation is potentially more relevant for considering misalignment triggered black hole recoils \citep{CampanelliEtAl07, GonzalezEtAl2007, LoustoEtAl2011, LoustoEtAl13, LoustoEtAl19, SperhakeEtAl20}, and is given by $\tau_{\rm bh}^{\rm bin}=\theta_{\rm bh}^{\rm bin}/|\dot{\theta_{\rm bh}^{\rm bin}}|$, where $\theta$ is the angle between the black hole spin and binary angular momentum. The evolution of $\tau_{\rm bh}^{\rm bin}$ is shown by the thin blue and red lines, for the primary and secondary black holes, respectively. Additionally, the timescale on which the black hole spins align with each other, $\tau_{\rm bh}^{\rm bh}=\theta_{\rm bh}^{\rm bh}/|\dot{\theta}_{\rm bh}^{\rm bh}|$, is shown by the black line in each panel. In all cases dotted lines represent periods during which $\theta$ is increasing, i.e., the spin and binary are becoming more misaligned. 

There are a couple of key points to note. Firstly, for the initially coplanar systems we find that the evolution of the analytic alignment timescales given by Equation~(\ref{eq:alignment}), the black hole spin-binary alignment timescale, $\tau_{\rm bh}^{\rm bin}$, and the spin-spin alignment timescales, $\tau_{\rm bh}^{\rm bh}$, follow each other very well, with only fairly minor differences. We also note that for these systems there is at most mild evolution of the alignment timescales. On the other hand, if we consider initially misaligned systems, the alignment timescales also undergo a richer and more variable evolution, with potentially significant differences between the three timescale estimates. In particular, $\tau_{\rm bh}^{\rm bin}$ exhibits more variation including periods of slow alignment and increasing misalignment, in some cases over significant periods of time, due to the relative evolution of both the black hole spin and binary angular momentum. This being said, while $\tau_{\rm bh}^{\rm bh}$ can also exhibit slow alignment and periods of increasing misalignment, this is confined to very early times and in general $\tau_{\rm bh}^{\rm bh}$ roughly follows the behaviour of $\tau_{\rm align}$ at later times.

\begin{figure*}
\begin{center}
\includegraphics[width=1.6\columnwidth]{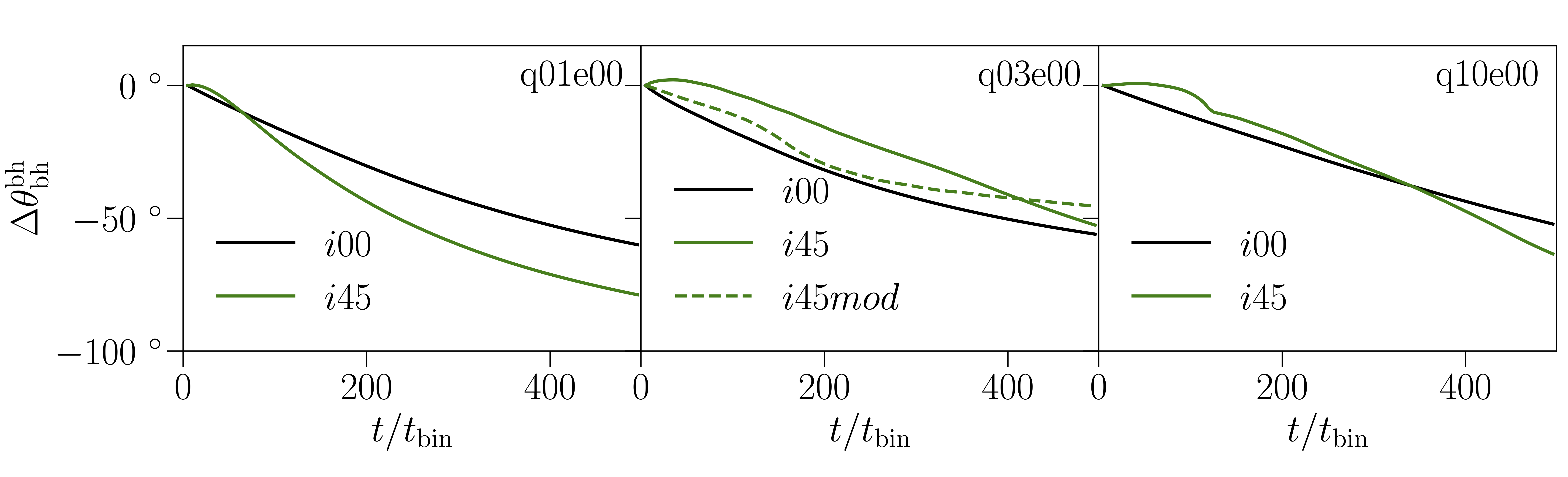}
\caption{Evolution of the change in angle between the spin of each black hole, $\Delta\theta_{\rm bh}^{\rm bh}$. Each panel represents a different binary mass ratio as labelled in the top right, noting that we only include prograde, initially circular binaries in order to compare initially misaligned systems (shown in green) with their coplanar (shown in black) counterparts. In general the evolution is similar for all binaries, although initially misaligned systems may initially experience periods in which $\Delta\theta_{\rm bh}^{\rm bh}$ increases. For a given mass ratio, after $500$~binary orbits the difference in $\Delta\theta_{\rm bh}^{\rm bh}$ between initially coplanar and initial misaligned systems is at most $\sim\pm 30\%$.}
\label{fig:spin_spin_theta}
\end{center}
\end{figure*}

To consider the latter point further, in Fig.~\ref{fig:spin_spin_theta} we show the evolution of the change in angle between the black hole spins, $\Delta\theta_{\rm bh}^{\rm bh}$, for all prograde, initially circular binaries. Each panel shows a different binary mass ratio with black and green lines showing initially coplanar and misaligned systems respectively. We see that in general the change in the spin-spin alignment angle is fairly similar for all runs, although initially misaligned systems can undergo initial short periods of growing misalignment. By the end of the simulations, the difference in $\Delta\theta_{\rm bh}^{\rm bh}$ between runs of a given mass ratio differ by at most $\pm30\%.$ Based on this behaviour, and assuming the system survives unperturbed for a sufficiently long time it is expected that all systems will eventually come into alignment \citep{MillerKrolik13, GerosaEtAl15}. Although we point the reader to Sections~\ref{sec:bh_spin_evo}, \ref{sec:discussion_bhevo} and \ref{sec:discussion_gw} for discussion of caveats and possible implications.

\end{document}